\documentclass[fleqn,usenatbib]{mnras}

\usepackage[justification=centering]{caption}
\usepackage{newtxtext,newtxmath}
\usepackage{mathrsfs,amsmath}
\usepackage[T1]{fontenc}
\usepackage{ulem}
\usepackage{xcolor}
\usepackage{CJKutf8} 

\usepackage{graphicx,xspace,lscape,subfigure,booktabs,multirow,float}
\usepackage[export]{adjustbox}

\newcommand{\toc}{\par\begin{center}\vspace{0.5mm}\begingroup\small\let\cleardoublepage\relax\let\clearpage\relax\mytoc\endgroup\vspace{0.5mm}\end{center}\par} 
 \makeatletter\newcommand\mytoc{\@starttoc{toc}}\makeatother 
\long\def\symbolfootnote[#1]#2{\begingroup%
\def\thefootnote{\fnsymbol{footnote}}\footnote[#1]{#2}\endgroup} 

\title[Merging Triple BHs in PopIII Clusters]{Merging Hierarchical Triple Black Hole Systems with Intermediate-mass Black Holes in Population III Star Clusters}

\author[Shuai Liu et al.]{
Shuai Liu (刘帅)$^{1,2}$,
Long Wang (王龙)$^{2}$\thanks{E-mail: wanglong8@mail.sysu.edu.cn}, 
Yi-Ming Hu (胡一鸣)$^{2}$\thanks{E-mail: huyiming@mail.sysu.edu.cn}, \medskip Ataru Tanikawa$^{3,4}$, \\ \Large{\rm{and Alessandro A. Trani$^{5,6,7}$}}
\\
$^{1}$School of Electronic and Electrical Engineering, Zhaoqing University, Zhaoqing 526061, People's Republic of China \\
$^{2}$School of Physics and Astronomy, Sun Yat-sen Univiersity, Zhuhai 519082, People's Republic of China \\
$^{3}$Department of Earth Science and Astronomy, College of Arts ans Sciences, The University of Tokyo, 3-8-1 Komaba, Meguro-ku, Tokyo 153-8902, Japan\\
$^{4}$Center for Information Science, Fukui Prefectural University, 4-1-1 Kenjojima, Matsuoka, Eiheiji-Town, Fukui, 910-1195, Japan \\
$^{5}$Niels Bohr International Academy, Niels Bohr Institute, Blegdamsvej 17, 2100 Copenhagen, Denmark \\
$^{6}$Research Center for the Early Universe, School of Science, The University of Tokyo, Tokyo 113-0033, Japan \\
$^{7}$Okinawa Institute of Science and Technology, 1919-1 Tancha, Onna-son, Okinawa 904-0495, Japan 
}

\date{Accepted XXX. Received YYY; in original form ZZZ}

\pubyear{2023}

\begin{document}
\begin{CJK}{UTF8}{gbsn}
	\label{firstpage}
	\pagerange{\pageref{firstpage}--\pageref{lastpage}}
	\maketitle

	\begin{abstract}

		Theoretical predictions suggest that very massive stars have the potential to form through multiple collisions and eventually evolve into intermediate-mass black holes (IMBHs) within Population III star clusters embedded in mini dark matter haloes.
		In this study, we investigate the long-term evolution of Population III star clusters, including models with a primordial binary fraction of $f_{\rm b}=0$ and 1, using the $N$-body simulation code \textsc{petar}.
		We comprehensively examine the phenomenon of hierarchical triple black holes in the clusters, specifically focusing on their merging inner binary black holes (BBHs), with post-Newtonian correction, by using the \textsc{tsunami} code.
		Our findings suggest a high likelihood of the inner BBHs containing IMBHs with masses on the order of $\mathcal{O}(100)M_{\odot}$, and as a result, their merger rate could be up to $0.1{\rm Gpc}^{-3}{\rm yr}^{-3}$. The orbital eccentricities of some merging inner BBHs oscillate over time periodically, known as the Kozai-Lidov oscillation, due to dynamical perturbations. Detectable merging inner BBHs for mHz GW detectors LISA/TianQin/Taiji concentrate within $z<3$. More distant sources would be detectable for CE/ET/LIGO/KAGRA/DECIGO, which are sensitive from $\mathcal{O}(0.1)$Hz to $\mathcal{O}(100)$Hz. Furthermore, compared with merging isolated BBHs, merging inner BBHs affected by dynamical perturbations from tertiary BHs tend to have higher eccentricities, with a significant fraction of sources with eccentricities closing to 1 at mHz bands. GW observations would help constrain formation channels of merging BBHs, whether through isolated evolution or dynamical interaction, by examining eccentricities.

	\end{abstract}




	\section{Introduction}\label{sec:intro}

	A class of black holes (BHs) with masses ranging from $10^{2}M_{\odot}$ to $10^{5}M_{\odot}$, known as intermediate-mass black holes (IMBHs), is speculated to exist in the gap between stellar-mass black holes (SBHs) and massive black holes (MBHs) in the Universe. IMBHs have attracted significant attention due to their potential role in explaining the formation of MBHs in the first hundreds of millions of years after the Big Bang \citep{Volonteri:2010wz, Wu2015, Banados2018, Greene:2012gk, Reines:2016kej, 2017IJMPD..2630021M, Inayoshi:2019fun, 2020ARA&A..58..257G,2021ApJ...907L...1W}, as well as the anomalies observed in dwarf galaxies (DGs), e.g., core-cusp and the number of DGs \citep{Silk:2017yai, Barai:2018iyd}. In particular, IMBHs could serve as seeds for the formation of MBHs through their coalescence and gas accretion. Moreover, the early feedback from IMBHs could quench star formation (SF), reduce the number of DGs, and impact the central density profile of DGs.

	Considerable efforts have been dedicated to the search for IMBHs so far, but concrete observations of them are scarce. Electromagnetic wave (EM) observations have yielded some candidates, but they can not be identified as IMBHs conclusively. Ultraluminous x-ray
	sources as IMBH candidates in external galaxies and active galactic nuclei (AGN), e.g., M82 \citep{PortegiesZwart:2004ggg}, Galaxy ESO 243-49 HLX-1 \citep{Farrell:2011rj}, NGC 2776 \citep{Mezcua:2015pra}, could potentially be explained as SBHs with super-Eddington accretion. Globular clusters, including M15 \citep{Gerssen:2002iq, Gerssen:2002sd, vanderMarel:2002ip} and G1 \citep{Gebhardt:2002in, Gebhardt:2005cy}, as well as $\omega$-Centauri \citep{2017MNRAS.464.2174B, Noyola:2008kt}, are also considered possible locations for IMBHs, but the velocity dispersion of center stars
	caused by potential IMBHs could alternatively be explained by a concentrated group of SBHs or neutron stars (NSs), a stellar-mass binary black hole (SBBH), or radial anisotropy configuration \citep{Hurley:2007as, 2017MNRAS.468.4429Z, 2019MNRAS.488.5340B}. Fortunately, the detection of gravitational
	waves (GWs) has brought a breakthrough in the search for IMBHs. The merger remnant of the SBBH called GW190521 \citep{LIGOScientific:2020iuh} detected by LIGO/Virgo is a BH with a mass of $\sim142M_{\odot}$, falling within the mass range of IMBHs. This marks the first direct observational
	evidence for the existence of IMBHs. Subsequently, several similar events, e.g., GW190426\_190642 \citep{LIGOScientific:2021usb},
	have also been detected. However, binary black holes (BBHs) consisting of IMBHs have not been detected directly by LIGO/Virgo/KAGRA yet. It has been suggested that the primary component of GW190521 may be an IMBH \citep{2020ApJ...904L..26F, Nitz:2020mga}, but this claim is still under debate. The challenge in the direct detection of IMBHs arises from the fact that the frequencies of GWs emitted by heavier BBHs with IMBHs are lower than the sensitive frequency bands of the current ground-based detectors. However, upcoming space-borne GW detectors, such as TianQin \citep{TianQin:2015yph}, Taiji \citep{Ruan:2018tsw}, LISA \citep{2017arXiv170200786A} and DECIGO \citep{Kawamura:2011zz},
	as well as the next generation ground-based GW detectors Cosmic Explorer (CE) \citep{Reitze:2019iox}
	and Einstein Telescope (ET) \citep{Punturo:2010zz}, are expected to
	have sensitivities at lower frequency ranges, making them capable of detecting BBHs with IMBHs across a wide range of redshifts \citep{Fregeau:2006yz, Amaro-Seoane:2006zlf, Amaro-Seoane:2007osp,Amaro-Seoane:2009iyc, Mapelli:2010ht, Gair:2010dx, Yagi:2012gb,Rasskazov:2019tgb, Jani:2019ffg,Emami:2019bss,2019arXiv190605864A, Datta:2020vcj, Arca-Sedda:2020lso,Deme:2020ewx,Fragione:2020rmf, Liu:2021yoy, Wang2022b, Fragione:2022avp, 2022MNRAS.517.1339G, Fragione:2022ams, Torres-Orjuela:2023hfd}.

	While searching for IMBHs, various possible explanations for their formation have been proposed,
	which could be divided into three categories roughly. The first one involves the mergers of stellar-mass objects in a dense environment. For example, successive or runaway mergers of SBHs in GCs could produce IMBHs with masses greater than $10^{3}M_{\odot}$ \citep[e.g.][]{Miller:2001ez, Gultekin:2004pm, 2015MNRAS.454.3150G, Mapelli:2021syv, Mapelli:2021gyv, 2021MNRAS.501.5257R, Mouri:2002mc, 2015MNRAS.454.3150G}. In environments with higher density, e.g., circumnuclear giant H II region \citep[e.g.][]{Taniguchi:2000mp} and nuclear star clusters \citep[e.g.][]{Antonini:2018auk, Fragione:2020nib, Kroupa:2020sru, Fragione:2021nhb, Rose:2021ftz}, massive IMBHs would form, because the gravitational potential from galaxies could prevent merger remnants with natal kick from escaping from clusters, where the natal kick could be from supernovae (SNe) and dynamical few-body interactions, as well as recoil kick from asymmetric GW radiation. Secondly, IMBHs could form in gaseous environments. In AGN disks around MBHs, IMBHs could form through mergers of stars due to mass migration and subsequent gas accretion \citep{McKernan:2012rf, McKernan:2014oxa}. IMBHs may also be formed through the direct gravitational collapse of metal-poor giant gas without the formation of stars at the galactic center \citep[e.g.][]{Mayer:2009qr, Mayer:2014nva}. Lastly, IMBHs could be formed through the evolution of very massive stars (VMSs) with low metallicities. In regions of starburst SF with high central densities, the successive mergers of massive stars could produce VMSs, which would then evolve to IMBHs \citep[e.g.][]{PortegiesZwart:2002iks, PortegiesZwart:2004ggg, PortegiesZwart:2005zp, Freitag:2005yd, Kremer:2020wtp, Gonzalez:2020xah}, or even collapse to IMBHs directly without pair-instability SNe \citep{Spera:2017fyx}.

	Population III (PopIII) star clusters are also potential formation sites for VMSs which could evolve to IMBHs \citep{Sakurai:2017opi}. Most PopIII stars are massive, and they could merge to VMSs or become IMBHs, due to their extremely low metallicities \citep{2016MNRAS.462.1307S, Chon:2021jlx, Latif:2021xad}. If PopIII clusters are embedded in mini dark matter haloes \citep[e.g.][]{2020MNRAS.492.4386S}, the haloes would prevent them from being disrupted by the galactic potential, allowing them to survive from the early stage of the Universe to the present \citep{Wang2022b}.

	In prior investigations regarding the formation of IMBHs within PopIII star clusters, \cite{Sakurai:2017opi} delved into the formation of VMSs within PopIII star clusters through the utilization of a zoom-in cosmological simulation. Furthermore, \cite{Reinoso:2018bfv} examined the impact of stellar collisions on the formation of VMSs by $N$-body simulations.

	\cite{Wang2022b} studied the long-term evolution of PopIII star clusters in mini dark matter haloes using the star-by-star $N$-body code \textsc{petar}, as well as the GW mergers of BBHs consisting of IMBHs. Several previous studies also investigated merging BBHs formed from isolated PopIII binaries \citep{2004ApJ...608L..45B, 2017MNRAS.471.4702B, 2014MNRAS.442.2963K, 2020MNRAS.498.3946K, 2021MNRAS.501L..49K, 2021MNRAS.504L..28K, 2021PTEP.2021b1E01K, 2016MNRAS.460L..74H, 2017MNRAS.468.5020I, 2021MNRAS.505.2170T, 2021ApJ...910...30T, 2023MNRAS.524..307S, 2023MNRAS.525.2891C} and PopIII star clusters \citep{2020MNRAS.495.2475L, 2020ApJ...903L..40L, 2021MNRAS.501..643L}. Merging BBHs in triple systems formed dynamically in PopIII clusters are not explored, although many papers suggested BBHs formed from isolated or dynamical PopI/II triple/quadruple systems for GW sources \citep{Antonini:2013tea, 2017ApJ...836...39S, 2018ApJ...863....7R, 2019MNRAS.486.4781F, 2020ApJ...898...99H, Trani:2021tan}.

	In this study, we first examine triple systems comprising IMBH or SBH components dynamically formed rather than being primordial in PopIII clusters, depending on the framework proposed by \cite{Wang2022b}.
	Nevertheless, it's important to acknowledge certain limitations in the models introduced by \cite{Wang2022b}.
	Firstly, these models do not incorporate primordial binaries.
	Although the characteristics of primordial binaries within PopIII clusters remain uncertain at present, observations of young clusters indicate a prevalence of multiple systems among OB stars \citep{2012Sci...337..444S,2013ARA&A..51..269D,2023A&A...674A..34G}, suggesting that this trend might extend to PopIII stars as well.
	Furthermore, in the models presented by \cite{Wang2022b}, binary mergers are handled using the orbital average method \citep{1963PhRv..131..435P,1964PhRv..136.1224P}. However, this method may not provide a robust description of their orbital evolution, especially when they exist within multiple systems \citep{Antonini:2013tea}.

	In this investigation, we conduct additional simulations, adhering to the initial conditions outlined in \citep{Wang2022b}, to enhance the statistical robustness of our results. Simultaneously, we undertake a series of simulations that incorporate primordial binaries.
	Furthermore, for triple BHs with merging inner BBHs (hereinafter referred to simply as ``merging triple BHs'') formed in the \textsc{petar} simulation, we evolve their orbits using the direct few-body numerical code including post-Newtonian (PN) correction \textsc{tsunami} \citep{2019ApJ...875...42T, Trani:2020hwc}, which can accurately simulate the orbital evolution of triple systems. Additionally, we study GWs from merging inner BBHs under perturbation. Note that henceforth, when discussing merging triple BHs in the context of \textsc{petar} simulation, it will be necessary to clarify that the inner BBHs are evolved to merge using the orbital average method, which may not accurately describe their actual merger processes.

	The rest of this paper is organized as follows: we present the initial conditions of PopIII clusters, and the $N$-body and few-body methods (codes) in Sec. \ref{sec:method}.
	The population properties of merging triple BHs and GWs emitted by their inner BBHs are
	investigated in Sec. \ref{sec:result}.
	Finally, we draw the conclusion in Sec. \ref{sec:conclusion}.
	Throughout the paper, geometrical units
	($G=c=1$) are used, unless otherwise specified, and the standard $\Lambda$CDM cosmological model \citep{Planck:2015fie} is adopted.

	\section{Method}\label{sec:method}

\begin{figure}
		\centering
		\includegraphics[width=0.4\textwidth]{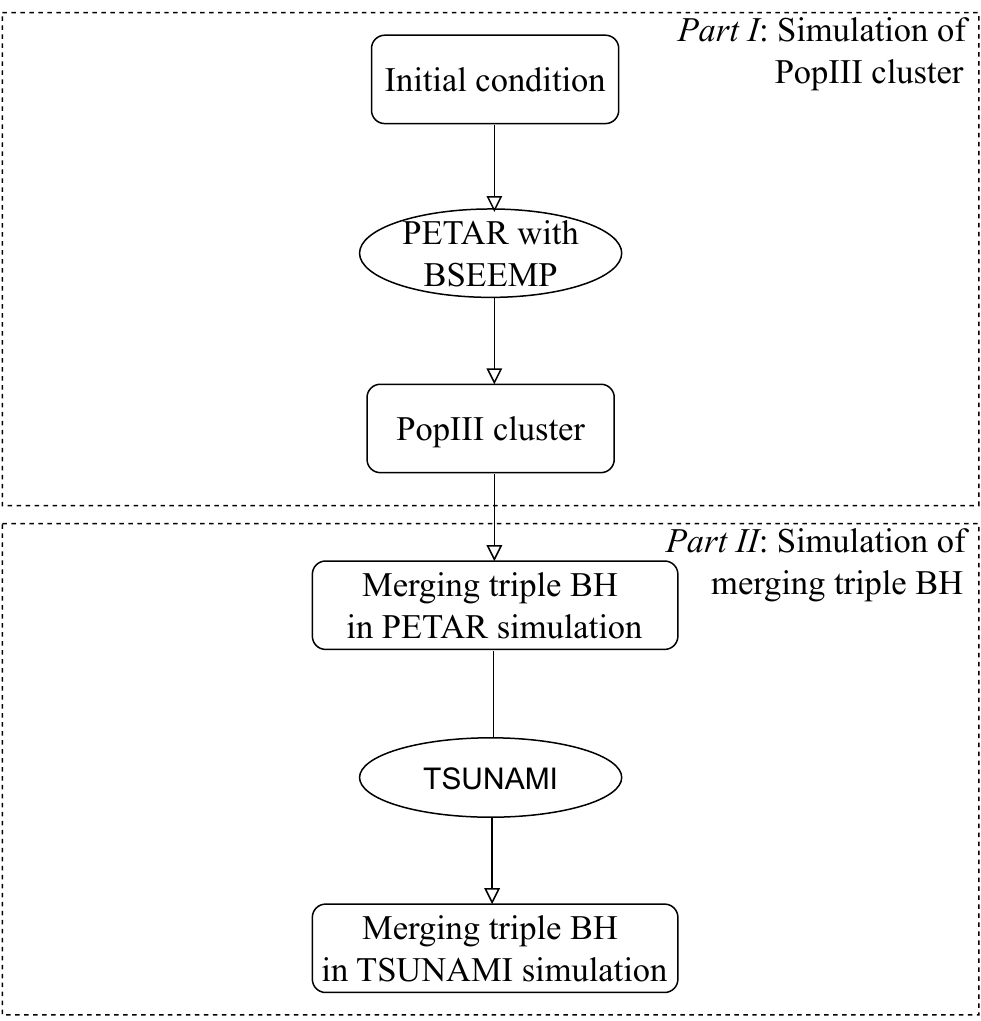}
		\caption{The flow chart outlines the complete process of evolving merging triple BHs in PopIII clusters, which consists of two main parts: simulations of PopIII clusters and simulations of merging triple BHs, respectively. Rectangles and ellipses represent target physical systems and methods (codes), respectively.}
		\label{fig:flow triple bh}
	\end{figure}

\begin{figure}
		\centering
		\includegraphics[width=0.4\textwidth]{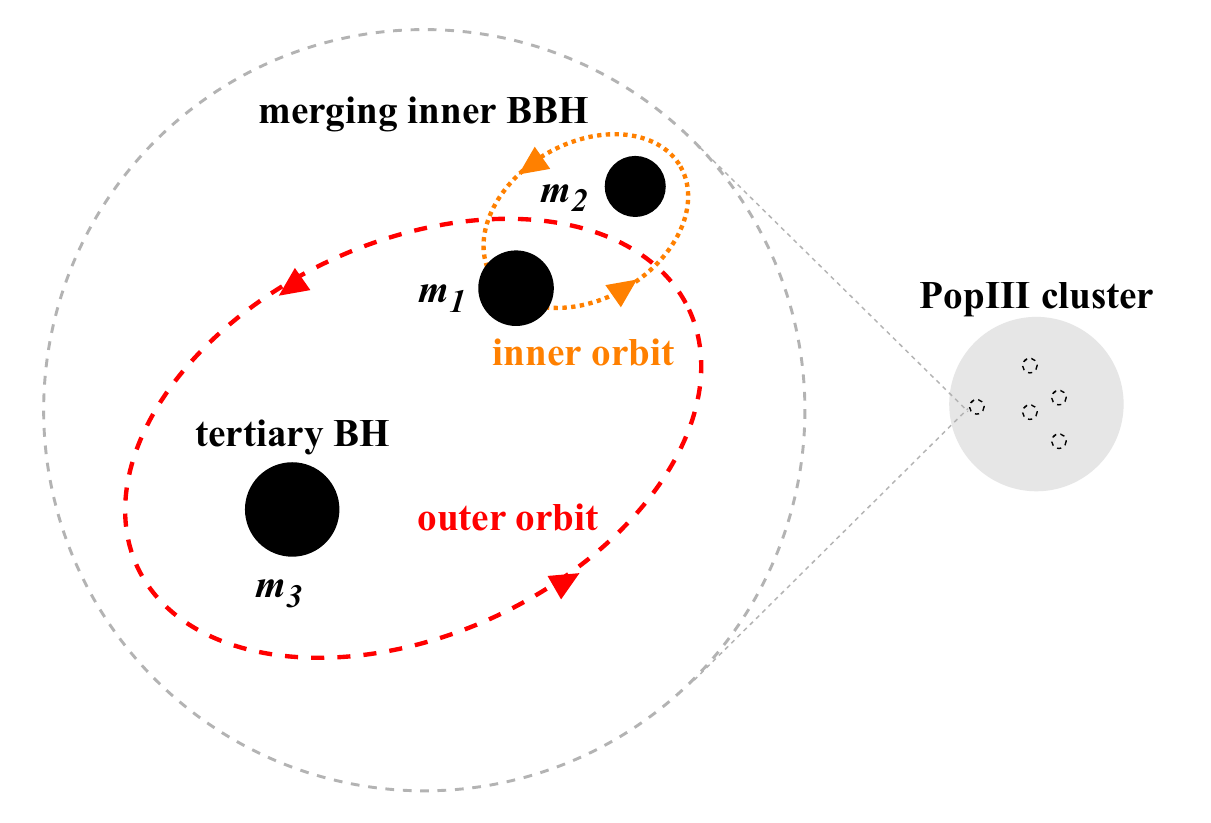}
		\caption{A sketch of merging triple BHs in PopIII clusters. The gray ball represents a PopIII cluster,
			in which small gray dashed circles are merging triple BHs. The enlarged version of a merging triple BH
			is in a gray large dashed circle, where a merging inner BBH with primary mass $m_{1}$ and
			secondary mass $m_{2}$ ($m_{2}\le m_{1}$) orbits around a tertiary BH with mass $m_{3}$. Orange dotted and red dashed circles are
			inner orbit with semimajor axis $a_{1}$ and outer orbit with semimajor axis $a_{2}$, respectively.}
		\label{fig:sketch triple bh}
	\end{figure}

	The complete process to track the long-term evolution of PopIII clusters and to evolve merging triple BHs within them is depicted in Fig. \ref{fig:flow triple bh}, where a schematic diagram of merging triple BHs is illustrated in Fig. \ref{fig:sketch triple bh}.
	It can be divided into two main parts: simulations of PopIII clusters using an integrated approach combining the $N$-body code \textsc{petar} with single and binary population synthesis code \textsc{bseemp} \citep{2020MNRAS.495.4170T}, as well as simulations of merging triple BHs using the high-accuracy method with the PN correction \textsc{tsunami} based on the results obtained from the cluster simulations.
	These two parts will be explained in the following two subsections in detail, respectively.

	\subsection{Simulation of PopIII cluster}

	\subsubsection{\textsc{petar}}

	We simulate the evolution of PopIII clusters using the high-performance $N$-body code \textsc{petar} \citep{2020MNRAS.497..536W}, which combines the particle-tree and particle-particle (P$^{3}$T) algorithm \citep{Oshino:2011ja} with the slow-down algorithmic regularization (SDAR) method \citep{2020MNRAS.493.3398W}.
	The P$^{3}$T component of the code handles long-range gravitational interactions and is implemented within the framework of \textsc{pentacle} \citep{2017PASJ...69...81I} and the Framework for Developing Particle Simulator \citep{Iwasawa:2016zci, 2020PASJ...72...13I}.
	Conversely, the SDAR method is employed to manage short-range interactions, ensuring precise and efficient treatment of binary orbital evolution and close encounters.
	Additionally, the impact of the galactic potential on PopIII cluster is implemented by \textsc{petar} using Galpy \citep{2015ApJS..216...29B}, with a detailed description provided in the following subsection.

	\subsubsection{\textsc{bseemp}}
	The single and binary stellar evolution in PopIII clusters is simulated by the fast population synthesis method \textsc{bseemp}, which could trace the
	stellar wind mass loss and BH formation of PopIII stars with the minimum metallicity $Z=2\times10^{-10}$.
	The details of star evolution in \textsc{bseemp} and the definition of production of star evolution are explained in \citep{Hurley2000,Hurley2002,2020MNRAS.495.4170T,2021ApJ...910...30T,Wang2022b}.
	For instance, the BHs within pair-instability
	mass gap, ranging from 60$M_{\odot}$ to 121$M_{\odot}$, are referred to as ``pair-instability BHs'' (PIBHs). The BHs with mass greater than $121M_{\odot}$
	and less than $60M_{\odot}$ are known as IMBHs and low-mass BHs (LBHs), respectively.

	\subsubsection{Initial condition}\label{subsubsec:initial condition}

	The initial conditions for the PopIII clusters in this study are derived from the long-term model ``NFWden\_long\_w9\_noms\_imf1'' without primordial binaries introduced in \citep{Wang2022b}. We select this model as it is expected to lead to the formation of VMSs, which eventually evolve to IMBHs with masses of up to $10^{3}M_{\odot}$.
	In order to improve statistical accuracy, we carry on 5 times more simulations (168 simulations) for the model ``NFWden\_long\_w9\_noms\_imf1'' than those performed by \cite{Wang2022b}. Each simulation is performed using different random seeds for generating initial positions, velocities and masses of stars.

	During the stellar evolution, the kick velocities of compact objects forming after SNe are also token into account. We adopt the rapid SNe model for remnant formation and material fallback \citep{Fryer:2011cx}, and the pulsation pair-instability SNe \citep{Belczynski:2016jno}. The kick velocity are assumed to follow a Maxwell velocity distribution with a dispersion of 265{\rm km/s}, based on the measurement of the proper motions of field NSs \citep{Hobbs:2005yx}. The fallback scenario assumes that some BHs have less or zero kick velocity due to mass fallback or failed SNe, resulting in their retention in star clusters. Meanwhile, the kick velocity of NSs formed by electron-capture supernovae are assumed to follow a Maxwell velocity distribution with a dispersion of {\rm 3 km/s}.

	Additionally, we include a new set of models that incorporate the presence of primordial binaries whose initial period, mass ratios, and eccentricities follow Sana's distributions \citep{2012Sci...337..444S}. The specific key parameters of these models are highlighted as follows.

	The initial mass of clusters $M_{\rm clu}$ is set to $10^{5}M_{\odot}$, similar to that of Model A, which is a fiducial model of star clusters embedded in mini-haloes models
	in \citep{2017MNRAS.472.1677S}. The initial half-mass radius $r_{\rm h}=1{\rm pc}$ is a typical value observed in star clusters. The
	central density model for clusters in \citep{1963MNRAS.125..127M, 1966AJ.....71...64K} is adopted, where the central concentration is determined by the ratio between the core
	radius $r_{\rm c}$ and the tidal radius $r_{\rm t}$ of clusters, denoted as $W$. The initial value of $W$ is set to $W_{0}=9$.

	We construct the initial mass function (IMF) of stars in PopIII clusters based on the hydrodynamic simulations \citep{2016MNRAS.462.1307S, 2021MNRAS.508.4175C, Latif:2021xad}. The IMF follows a single power-law profile:
	\begin{align}
		p(m)\propto m^{-\alpha}, m_{\rm min}<m<m_{\rm max},
	\end{align}
	where $\alpha=1$, $m_{\rm min}=1M_\odot$, and $m_{\rm max}=150M_\odot$. The IMF of PopIII star clusters would be more top-heavy, favoring the formation of more VMSs and more BHs, than the canonical IMF with the power index of $-2.35$ \citep[e.g.,][]{2001MNRAS.322..231K, 2003PASP..115..763C}.

	As the structure of the dark matter halo where the PopIII clusters embed is unclear to date, its potential is assumed to follow the model in \citep{1996ApJ...462..563N},
	\begin{align}
		\Phi=-\frac{M_{\rm vir}}{r[{\rm log}(1+C(z))-C(z)/(1+C(z))]{\rm log}(1+r/r_{\rm vir})},
	\end{align}
	where the virial mass $M_{\rm vir}=4\times10^{7}M_{\odot}$, the virial radius $r_{\rm vir}=280{\rm pc}$ and the concentration follows $C(z)=C(0)/(1+z)$, with the assumption that clusters evolve from $z=20$ to the present, and $C(0)=15.3$.

	The properties of primordial binaries in PopIII clusters
	are still uncertain, and the observations of the young SF region show that OB stars are mostly in binaries or high-order multiple systems \citep{2012Sci...337..444S,2013ARA&A..51..269D,2023A&A...674A..34G}, which could be applicable to PopIII clusters as well.
	In this study, we introduce a set of models featuring similar initial conditions but with a primordial binary fraction of $f_{\rm b}=1$, with the characteristics of these binaries aligning with observational constraints derived from \citep{2012Sci...337..444S}.

	We employ \textsc{petar} and to evolve 168 PopIII clusters with $f_{\rm b}=0$ and 1 for up to 12Gyr, respectively.

	\subsection{Simulation of merging triple BHs}
	\subsubsection{\textsc{tsunami}}
	Binary mergers in \textsc{petar} are handled by
	the orbital average method described in \citep{1963PhRv..131..435P,1964PhRv..136.1224P}, which may formally break down if the binary is orbited by the third object on a highly inclined orbit at a moderate distance \citep{Antonini:2013tea}. Therefore, we use \textsc{tsunami} to evolve orbits of merging triple BHs in the \textsc{petar} simulation. \textsc{tsunami} is a direct few-body numerical code \citep{2019ApJ...875...42T, Trani:2020hwc} used to evolve the dynamics of triple systems with high accuracy. It implements Mikkola's algorithmic regularization with a chain structure \citep{1999CeMDA..74..287M, 1999MNRAS.310..745M}, including 1PN and 2PN precession and 2.5PN GW radiation.

	\subsubsection{Initial condition}
	The orbital evolution of merging triple BHs formed in PopIII clusters can be traced by \textsc{petar}.
	The parameters (masses, positions and velocities) of the components of merging triple BHs are recorded by \textsc{petar} at various times between the first and last record times within the integrator time-step where inner BBHs are merging. The first record time refers to the moment when the triple BHs form in the \textsc{petar} simulation. The last record time represents the moment when the pericenters of the merging inner BBHs are $\sim100$ times the Schwarzschild radii of ($m_{1}+m_{2}$).
	We regard the parameters of merging triple BHs at the first and last record times in the \textsc{petar} simulation as the initial conditions of \textsc{tsunami} simulation, respectively.

	Due to the dense star environment of PopIII clusters, merging triple BHs might be perturbed through encounters
	with field objects on timescales shorter than their evolution time, which would alter their orbital properties
	significantly or even disrupt them. The timescale for collision with field objects is \citep{1987gady.book.....B}
	\begin{equation}
		T_{\rm coll} = 3.3\times10^{6}{\rm yr}\left(\frac{10^{2}{\rm pc}^{-3}}{n}\right)\left(\frac{100{\rm AU}}{a_{2}}\right)\left(\frac{100M_{\odot}}{m_{1}+m_{2}+m_{3}}\right)\left(\frac{\sigma}{1\rm km/s}\right),
	\end{equation}
	where $n$ is the number density of PopIII clusters. The collision time is estimated to be $\mathcal{O}$(1)~Myr, depending on the average number density of PopIII clusters $\langle n\rangle= \mathcal{O}(100)~{\rm pc^{-3}}$ \citep{Wang2022b}, the average semimajor axis of outer binaries of merging triple BHs $\langle a_{2}\rangle= \mathcal{O}(100)~{\rm AU}$, the average total mass of them $\langle m_{1}+m_{2}+m_{3}\rangle= \mathcal{O}(100)~M_{\odot}$, and the velocity dispersion $\sigma=\mathcal{O}(1)$km/s in the \textsc{petar} simulations\footnote{We select all the merging triple BHs in the \textsc{petar} simulation and make statistic on their semimajor axes, velocity dispersion and masses.}. Furthermore, we also estimate the timescale of dynamical processes which could disrupt merging inner BBHs, e.g., evaporation and flyby encounter , using Eq. (42) in \citep{1987gady.book.....B} and Eq. (17) in \citep{2023arXiv231217319W}. Both of these timescales exceed 10Gyr, which are much longer than $T_{\rm coll}$, indicating that they would not occur within $T_{\rm coll}$.

	Therefore, we evolve the merging triple BHs at the first and last record times in the \textsc{petar} simulation with \textsc{tsunami} for up to 10~Myr respectively, and only keep the triple BHs whose inner BBHs could merge eventually, i.e., merging triple BHs in the \textsc{tsunami} simulation. We compile statistics on the merger times of inner BBHs in the \textsc{tsunami} simulation and find that over 90\% and 95\% of them merge within 0.3 Myr and 0.5 Myr, respectively, which are significantly shorter than 10 Myr. This also indicates that setting the evolution time threshold to 10 Myr in the \textsc{tsunami} simulation is reasonable.

	\section{Result}\label{sec:result}

	In this section, we investigate population properties of merging triple BHs and GWs emitted by their inner BBHs within PopIII clusters, along with the impact of $f_{\rm b}$ on them.

	\subsection{Number of merging triple BHs in \textsc{petar} simulation}

\begin{figure}
		\centering
		\includegraphics[width=0.4\textwidth]{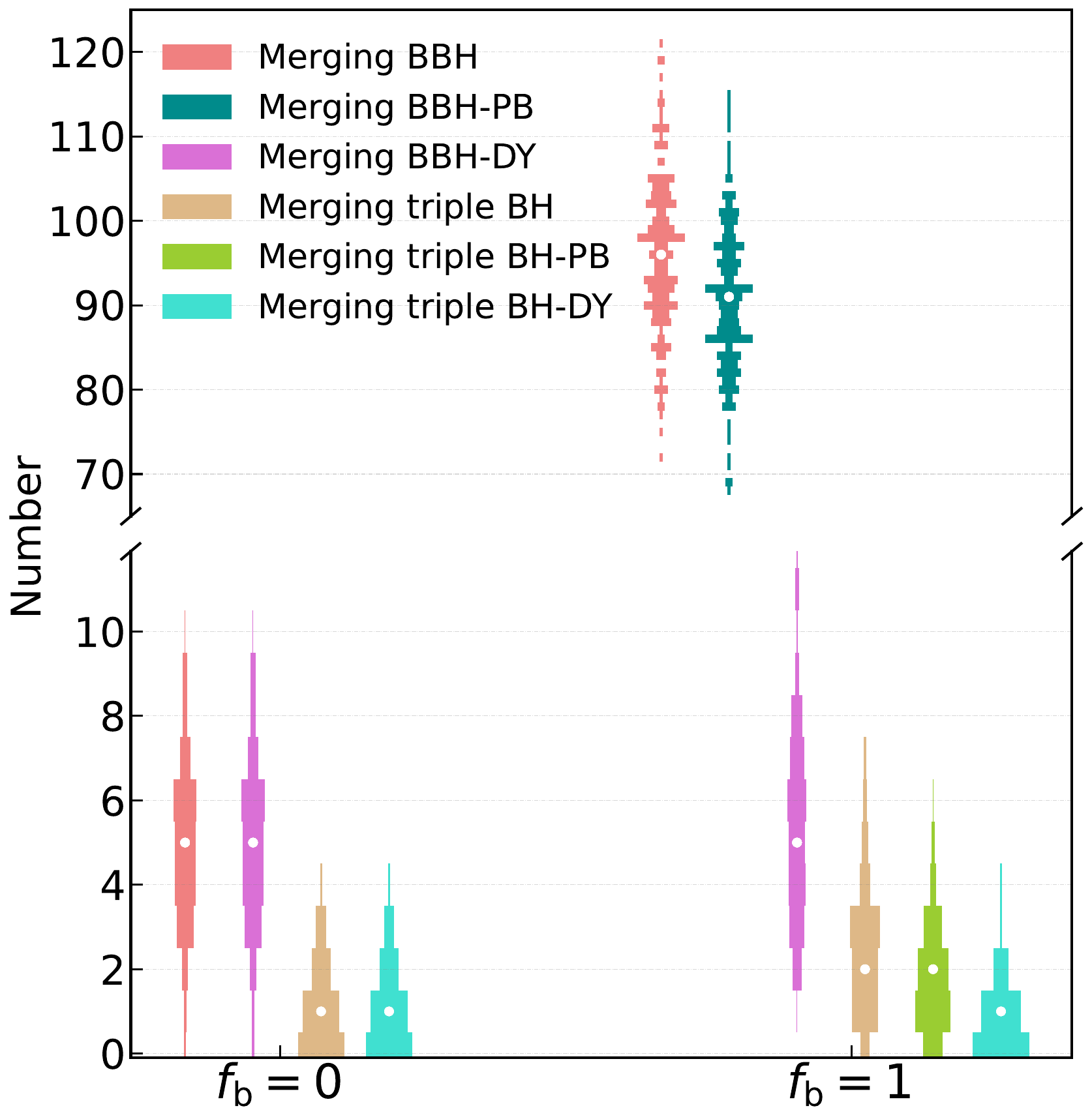}
		\caption{The number of merging triple BHs and merging BBHs during the evolution of PopIII clusters simulated by \textsc{petar}, which is denoted by box charts. Each box chart represents an entire set of simulations. The width of boxes represents the number fraction of merging triple BHs or BBHs within the values of the y-axis, similar to a histogram. The light coral color represents the total merging BBHs, of which ones formed by the evolution of primordial binaries and
			dynamical captures are denoted by \textit{suffixes} ``-PB'' and ``-DY'', respectively. The burly wood color
			represents the total merging triple BHs, the symbols representing the formation channels of their inner BBHs the same as those of merging BBHs.
			The white dots denote median values. The x-axis represents the cases with $f_{\rm b}$ of 0 and 1, respectively. Note that, in order to better present the chart, the scales located above and below the truncation of the y-axis are set to be different.}
		\label{fig:number binary triple bh}
	\end{figure}

	We investigate the number of merging triple BHs and merging BBHs in the \textsc{petar} simulation, as shown in Fig. \ref{fig:number binary triple bh}. In clusters without
	primordial binaries ($f_{\rm b}=0$), all the merging BBHs naturally form through dynamical processes. On average, there are 5
	merging BBHs in one cluster, with approximately one fifth of them occurring in merging triple BHs. If all the stars are paired initially, i.e.,
	$f_{\rm b}=1$, the average number of merging BBHs increases to $\sim100$, with the majority formed through the
	evolution of primordial binaries, and $\sim6$\% formed by dynamical captures. The average number of the merging triple BHs also increases to 2, with most of them having inner BBHs originating from primordial binaries and a few formed through dynamical captures.

	\subsection{Comparision between merging triple BHs in \textsc{petar} and \textsc{tsunami} simulation}

	In this subsection, we will compare merging triple BHs in the \textsc{petar} and \textsc{tsunami} simulations. In order to express convenience hereinafter, we refer to the merging triple BHs in the \textsc{petar} simulation as ``\textsc{petar} merging triple BHs''. It is also important to clarify here that \textsc{petar} merging triple BHs are those whose inner BBHs are evolved to merge using the orbital average method, which descriptions may deviate from their actual merger processes. We use the abbreviation just for the sake of convenience in the following comparisons.
	The merging triple BHs whose inner BBHs could evolve from the first and last record times of \textsc{petar} records to merge in the \textsc{tsunami} simulation are referred to ``\textsc{tsunami} merging triple BHs (the first record time)'' and ``\textsc{tsunami} merging triple BHs (the last record time)'', respectively.

\begin{figure}
		\centering
		\includegraphics[width=0.4\textwidth]{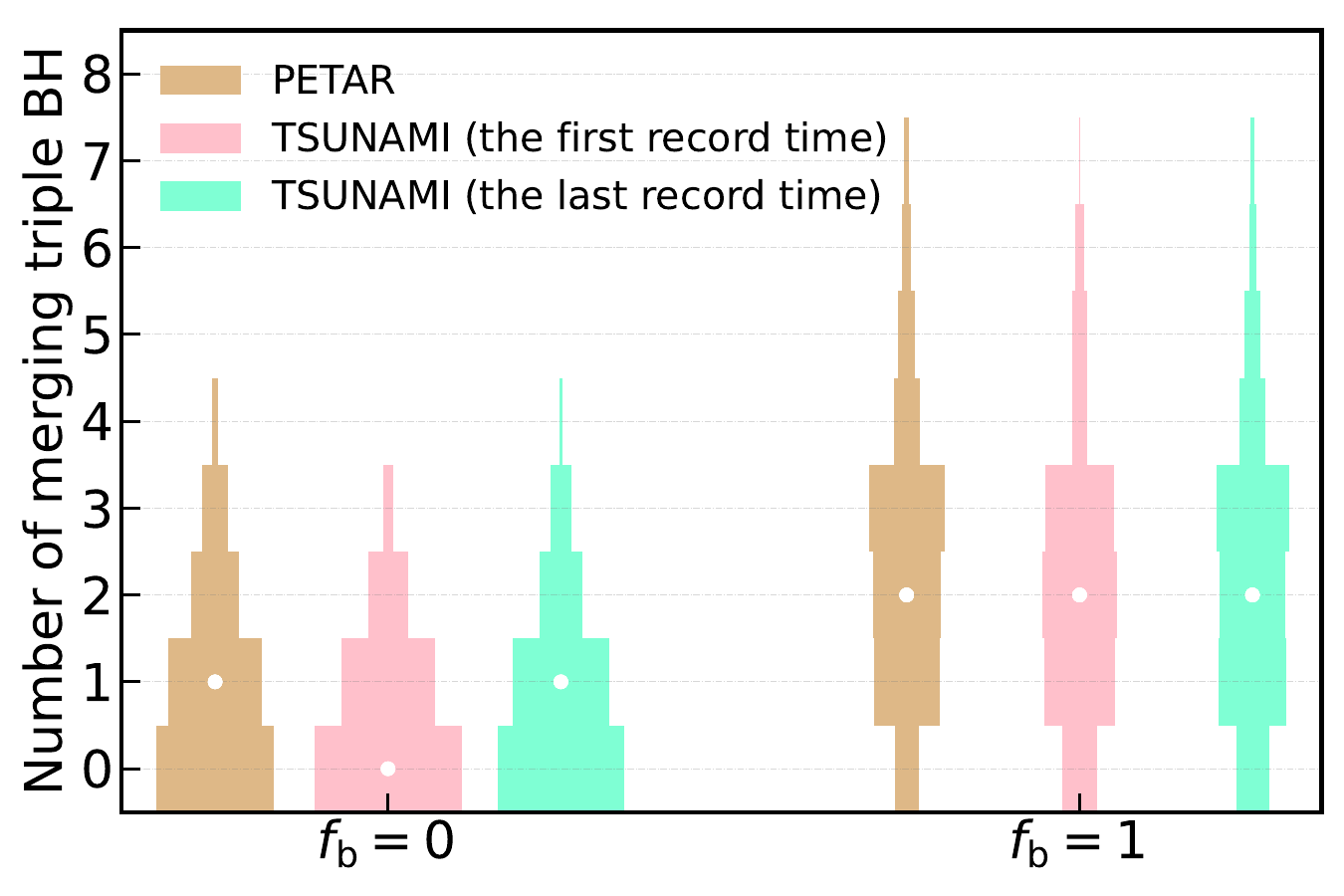}
		\caption{Comparison of the number of \textsc{petar} merging triple BHs and merging triple BHs in the \textsc{tsunami} simulation. Each box chart represents a complete set of simulations. The width of boxes indicates the number fraction of merging triple BH within the values on the y-axis, similar to the way a histogram displays data. The purple color represents \textsc{petar} merging triple BHs. The green and orange colors denote \textsc{tsunami} merging triple BHs (the first record time) and \textsc{tsunami} merging triple BHs (the last record time), respectively. The white dots denote median values. The x-axis represents cases with different values of $f_{\rm b}$.}
		\label{fig:number triple bh}
	\end{figure}

	Fig.~\ref{fig:number triple bh} provides a comparison among the numbers of the \textsc{petar} merging triple BHs ($N_{\mathrm{tbh,p}}$), the \textsc{tsunami} merging triple BHs (the first record time; $N_{\mathrm{tbh,t0}}$), and the \textsc{tsunami} merging triple BHs (the last record time; $N_{\mathrm{tbh,tf}}$).
	For models with $f_{\rm b}=0$ and 1, we find that $N_{\mathrm{tbh,t0}}\lesssim N_{\mathrm{tbh,tf}} \lesssim N_{\mathrm{tbh,p}}$. This will be explained in detail as follows.

	The \textsc{petar} merging triple BHs used as the evolved objects in the \textsc{tsunami} simulation are those that have been ascertained to undergo mergers in the \textsc{petar} simulation. Consequently, if the \textsc{tsunami} results show divergent orbital evolution for these triple BHs, some of them may not ultimately merge in the \textsc{tsunami} simulation. When \textsc{tsunami} adopts the parameters of merging triple BHs at the last record time from \textsc{petar} as its initial conditions, the inner BBHs would already be in the advanced stages of merging in the \textsc{tsunami} simulation.Consequently, these results lead to $N_{\mathrm{tbh,tf}}$ being greater than $N_{\mathrm{tbh,t0}}$ and comparable to $N_{\mathrm{tbh,p}}$.

	In the following, we will further study the difference between the \textsc{petar} merging triple BHs and \textsc{tsunami} merging triple BHs (the first record time) providing more evolution information. Hereinafter, we only analyze \textsc{tsunami} merging triple BHs (the first record time) and they are simply referred to ``\textsc{tsunami} merging triple BHs''.

	We investigate the evolution results of \textsc{petar} merging triple BHs in the \textsc{tsunami} simulation, as listed in Table \ref{tab:stable and unstable triple BHs}, and investigate their dynamical stability, which could lead to the difference between the \textsc{petar} and \textsc{tsunami} simulations. In particular, stable triple systems have a constant semimajor axis of inner binaries on a secular timescale, unstable triple systems, in contrast, would experience chaotic energy exchange, leading to one body escaping from the systems over a short timescale. The stability could be determined by the following criterion 
	\citep{2001MNRAS.321..398M}
	\begin{align}\label{eq:stability}
		\frac{a_{2}}{a_{1}}>\frac{3.3}{1-e_{2}}\left[\frac{2}{3}\left(1+\frac{m_{2}}{m_{1}+m_{2}}\right)\frac{1+e_{2}}{(1-e_{2})^{1/2}}\right]^{2/5}\times(1-0.3i_{\rm
			mut}/\pi),
	\end{align}
	where $i_{\rm mut}$ is the angle between inner and outer orbital planes. In the \textsc{petar} simulation, more than 50\% merging triple BHs are stable in both cases with $f_{\rm b}=0$ and 1. In the \textsc{tsunami} simulation, almost all the stable \textsc{petar} merging triple BHs could also merge. However, more than half of unstable \textsc{petar} merging triple BHs have different evolution results (merge, break, and neither merge nor break) in the \textsc{tsunami} simulation. This is an inevitable consequence of the chaotic nature of triple systems. Small differences in the initial conditions or in the integration algorithm are bound to create differences in the evolution \citep{2022ApJ...939...81H,2023MNRAS.tmp.2576P}. This is especially true for meta-stable or unstable triples, where the extreme chaotic behavior can amplify small differences between two neighboring solutions, until their macroscopic outcomes diverge \citep{2022ApJ...938...18L,2022A&A...659A..86P, Trani:2024xwq}. Fortunately, it has been shown that, despite different final outcomes for an individual simulation can arise from small differences in accuracy, algorithms, or machine architectures, the statistical outcome of many realizations is independent of these factors \citep{1991PASJ...43L...9S,2020MNRAS.493.3932B}. 

	\begin{table*}
		\centering
		\caption{Evolution results of \textsc{petar} merging triple BHs in the \textsc{tsunami} simulation and their stabilities.}
		\label{tab:stable and unstable triple BHs}
		\setlength{\tabcolsep}{1.0mm}{
			\begin{tabular}{ccccccccc}
				\toprule\toprule  
				Code                              & $f_{\rm b}$                        & Merge (stable) & Merge (unstable) & Break (stable) & Break (unstable) & Neither merge nor
				break (stable)                    & Neither merge nor break (unstable) &                                                                                                       \\
				\midrule  
				\multirow{2}{*}{\textsc{petar}}   & 0                                  & 101 (60.5\%)   & 66 (39.5\%)      & --             & --               & --                & --        \\
				\multirow{2}{*}{}                 & 1                                  & 346 (84.6\%)   & 63 (15.4\%)      & --             & --               & --                & --        \\
				\midrule
				\multirow{2}{*}{\textsc{tsunami}} & 0                                  & 99 (59.3\%)    & 19 (11.4\%)      & 1 (0.6\%)      & 47 (28.1\%)      & 1 (0.6\%)         & 0 (0.0\%) \\
				\multirow{2}{*}{}                 & 1                                  & 339 (82.9\%)   & 26 (6.4\%)       & 1 (0.2\%)      & 36 (8.8\%)       & 6 (1.5\%)         & 1 (0.2\%) \\
				\bottomrule\bottomrule 
			\end{tabular}}
	\end{table*}

\begin{figure}
		\centering
		\subfigure[]{
			\centering
			\includegraphics[width=0.39\textwidth]{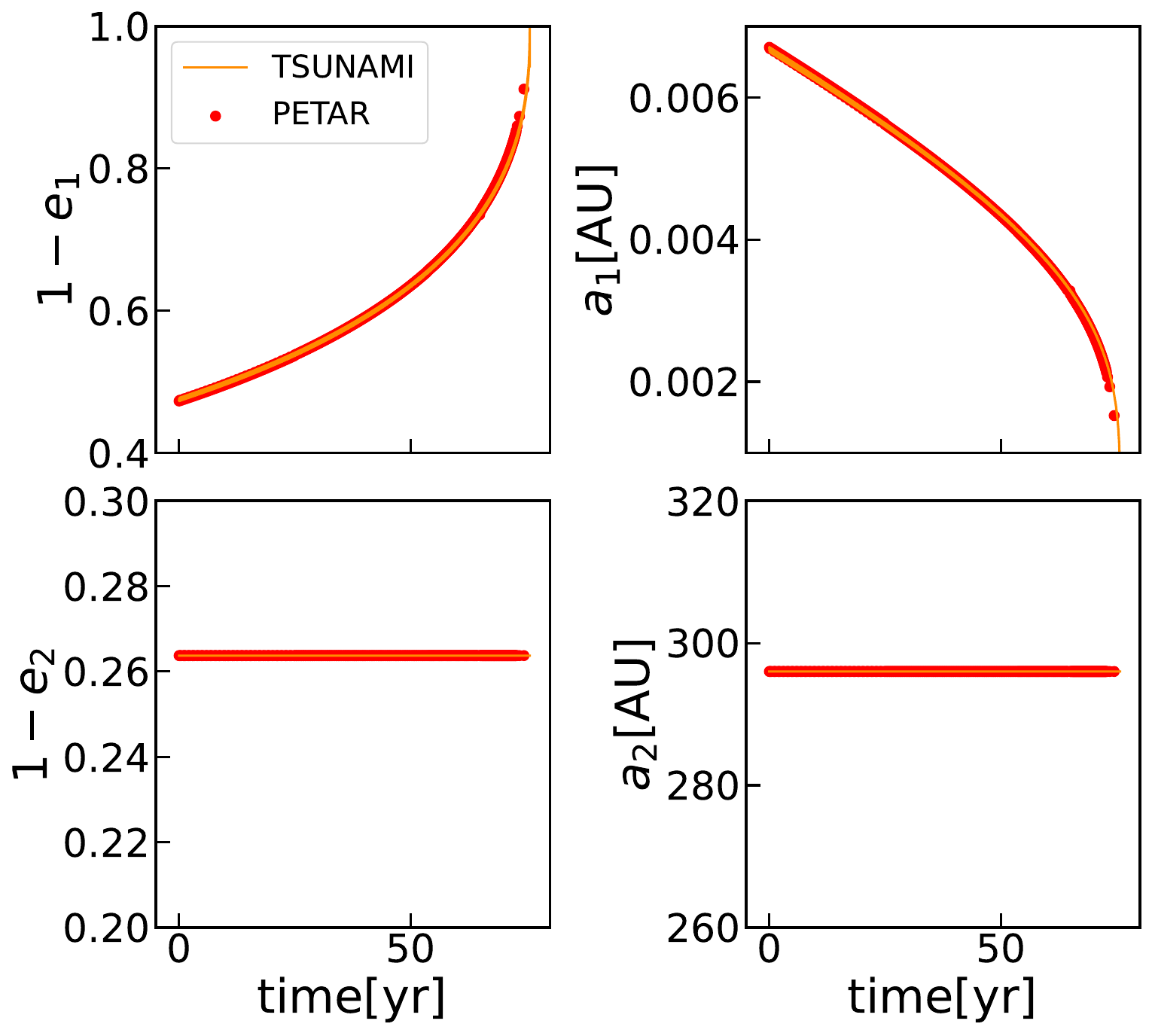}}
		\hspace{-10mm}
		\subfigure[]{
			\centering
			\includegraphics[height=0.34\textwidth, width=0.4\textwidth]{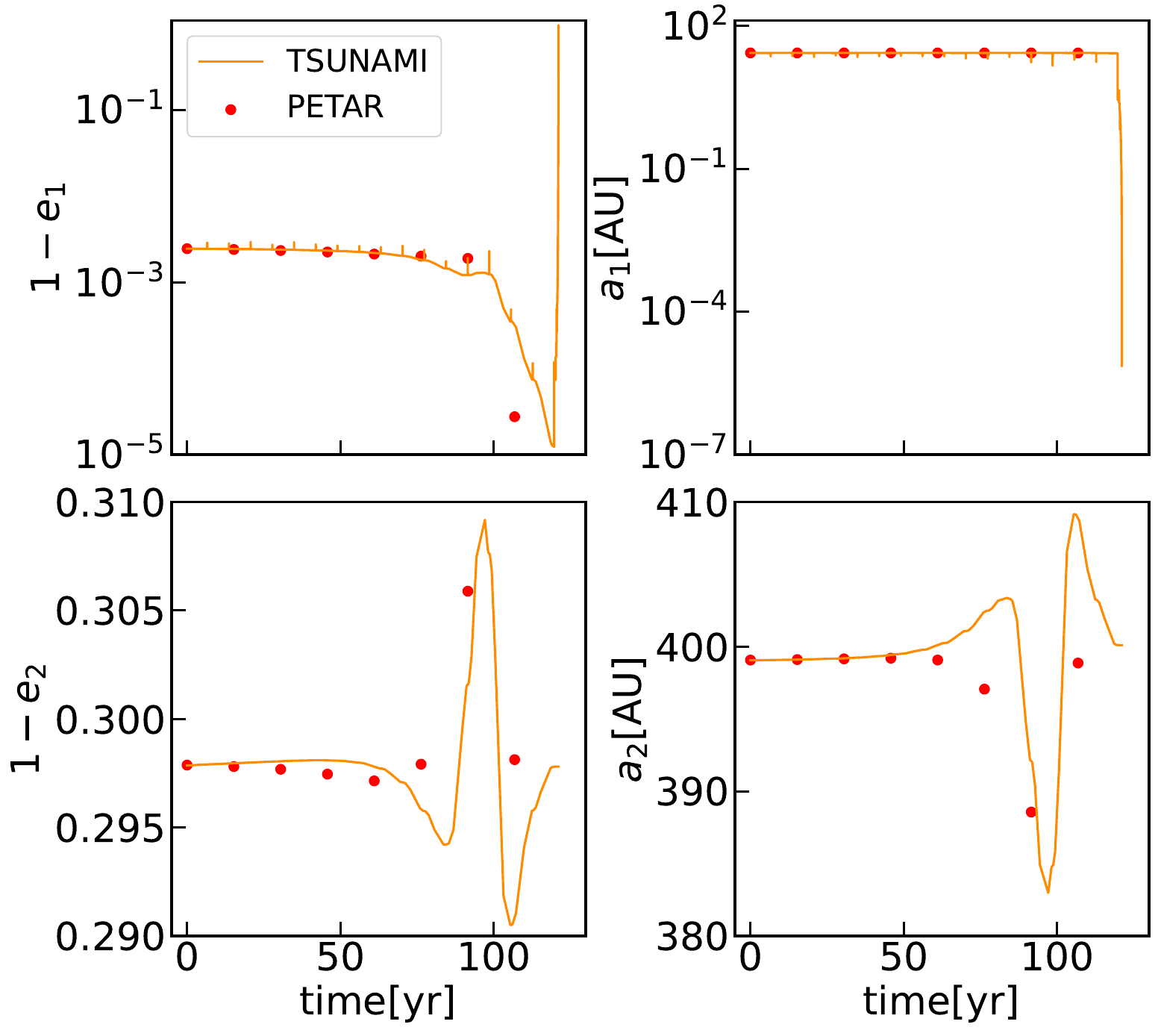}}
		\caption{The evolution of orbital parameters of (a) one stable merging triple BH with $m_{1}=200.1M_{\odot}, m_{2}=53.3M_{\odot}, m_{3}=44.8M_{\odot}$ and (b) one stable merging triple BH with $m_{1}=350.6M_{\odot}, m_{2}=49.2M_{\odot}, m_{3}=55.6M_{\odot}$ in both \textsc{petar} and \textsc{tsunami} simulations. The results from \textsc{petar} and \textsc{tsunami} simulations are plotted as red dots and orange solid lines, respectively. The x-axes are shifted by the time recorded.}
		\label{fig:triple bh merge petar tsunami}
	\end{figure}

	We compare orbital evolutions of several representative events in both \textsc{petar} and \textsc{tsunami} simulations. Two stable merging triple BHs are shown in Fig. \ref{fig:triple bh merge petar tsunami}.
	In the first example (a), there are no discernible perturbations from the tertiary BHs.
	In the second example (b), significant perturbations from the tertiary BHs affect both the inner and outer BBHs.
	These two examples illustrate scenarios where either GW radiation or dynamical perturbations predominantly influence the orbital evolution of the inner binaries.
	To illustrate this distinction, we can employ an analytical criterion proposed by \citep{Antonini:2013tea}:
	\begin{equation}\label{eq:criterion dynamic GW}
		\ell_{1}=\ell_{\rm GW},
	\end{equation}
	where $\ell_{1}=\sqrt{1-e_{1}}$ and $\ell_{\rm GW}$ are the defined dimensionless angular momenta of the inner BBHs and the critical angular momentum below which GW energy loss dominates its evolution, respectively.

	For the example (a), $\ell_{1}<\ell_{\rm GW}$, the inner BBH decouples from the third BH and evolves as an isolated binary approximately.
	The orbital evolutions using these two codes consistently agree with each other, implying
	that the GW effects from the orbital averaged method in \textsc{petar} and the direct numerical integration including the PN correction method in \textsc{tsunami} are practically identical for (almost) isolated binaries.

	However, in the case of example (b), where $\ell_{1}>\ell_{\rm GW}$, while the orbital evolution using both codes appears similar in the early stages, they exhibit significant differences as time progresses.
	These differences between the results suggest that when the dynamical influence from the third BHs on the inner BBHs cannot be ignored, the numerical three-body integration including the PN correction method and the orbital-averaged method may not agree with each other anymore, which is consistent with the conclusion from \citep{Antonini:2013tea}.

\begin{figure}
		\centering
		\includegraphics[width=0.4\textwidth]{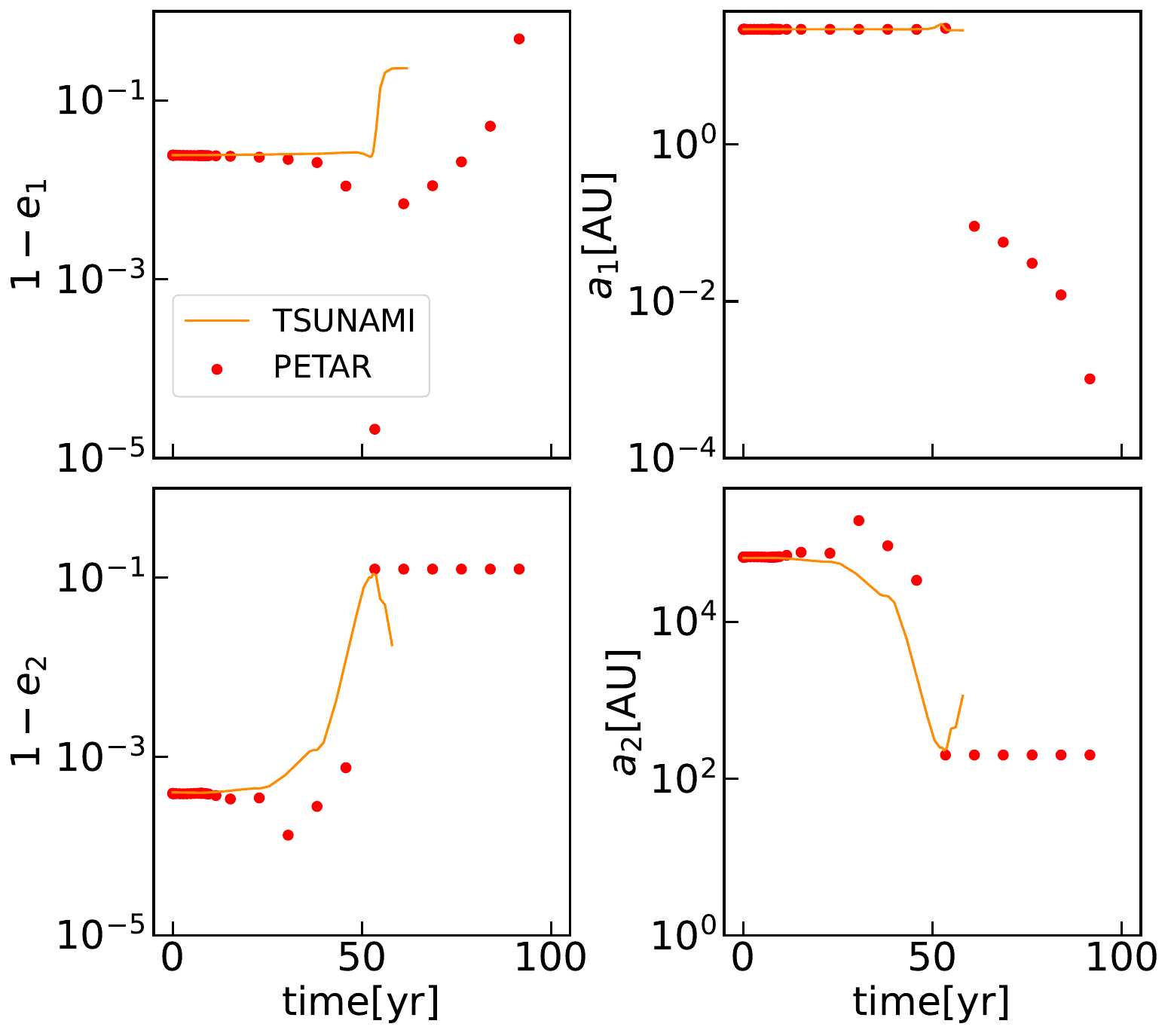}
		\caption{The evolution of orbital eccentricity (upper panel) and semimajor axis (lower panel) of an unstable triple BH with ($m_{1}=56.1M_{\odot}, m_{2}=55.8M_{\odot}, m_{3}=33.7M_{\odot}$) that the inner BBHs could merge in the \textsc{petar} simulation but break apart in the \textsc{tsunami} simulation. The results from \textsc{petar} and \textsc{tsunami} are plotted in red dots and orange solid lines, respectively. The x-axes are shifted by the time recorded.}
		\label{fig:triple bh merge petar break tsunami}
	\end{figure}

	An unstable triple BH whose inner BBH could merge in the \textsc{petar} simulation but break in the \textsc{tsunami} simulation is also shown in Fig. \ref{fig:triple bh merge petar break tsunami}. In the early stage of evolution, the evolutions between the \textsc{petar} and \textsc{tsunami} simulations are nearly identical. However, with the evolution, the differences between these two simulations become more and more significant. In particular, at about 50 yr, when the inner BBH is near the periapsis of outer BBH, the inner BBH starts to merge in the \textsc{petar} simulation. In contrast, the inner BBH gradually escapes from the gravitational influence of the third BH and the outer BBH breaks apart eventually in the \textsc{tsunami} simulation. This divergence in outcomes is a result of the difference between the two methods and the instability of triple BHs triggered near the outer periapsis.

\begin{figure*}
		\centering
		\includegraphics[width=0.9\textwidth, height=0.7\textwidth]{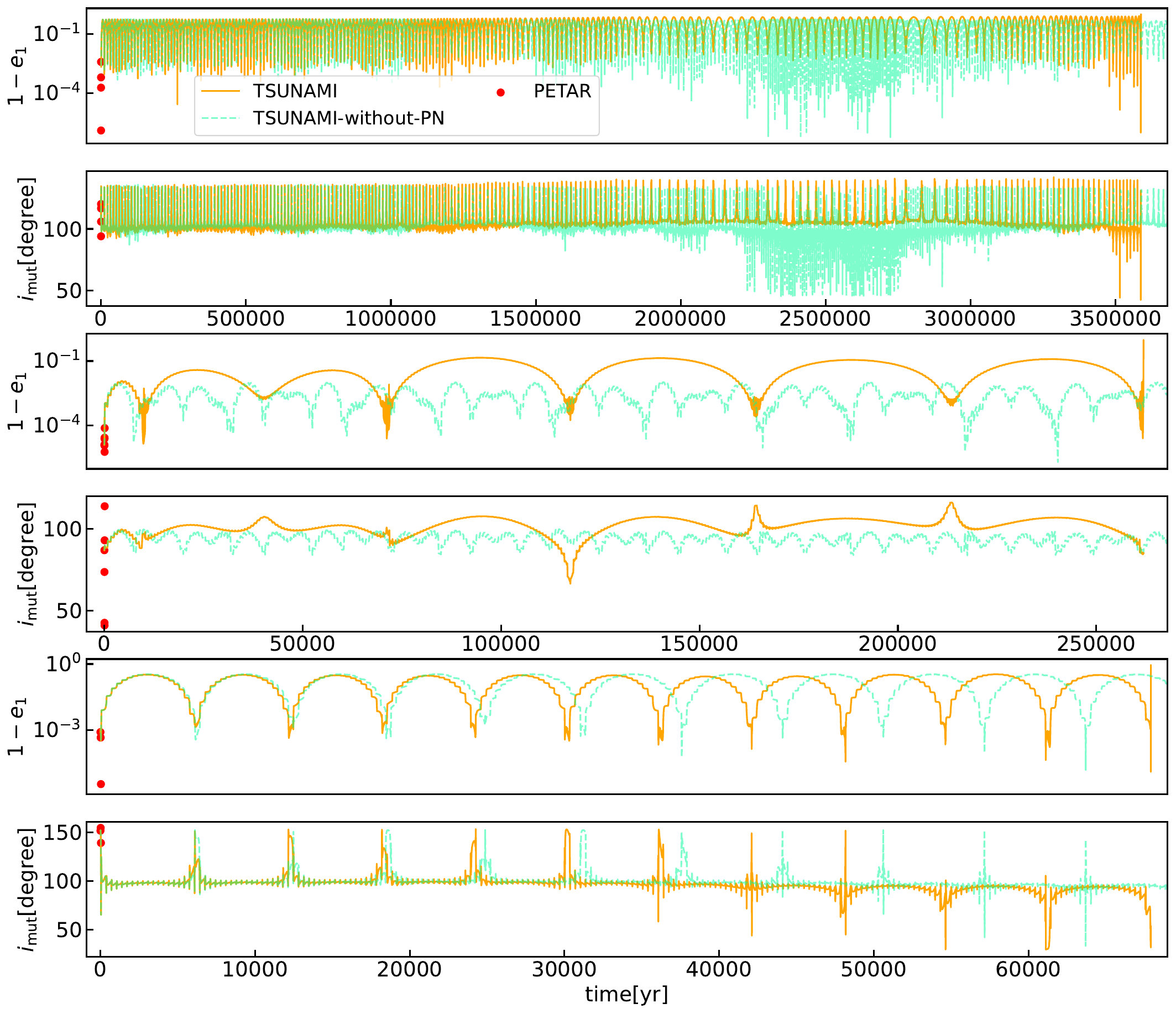}
		\caption{The orbital parameter evolutions ($e_{1}$ and $i_{\rm mut}$) of three merging triple BHs with the KL oscillation, having the component masses and initial orbital parameters of $(m_{1}, m_{2}, m_{3}, a_{1}, a_{2}, e_{1}, e_{2}, i_{\rm mut})$ = $(50.58M_{\odot},
				41.97M_{\odot}, 33.44M_{\odot}, 41.85{\rm AU}, 395.42{\rm AU}, 0.996, 0.613, 105.89^{\circ})$, $(153.19M_{\odot}, 432.74M_{\odot}, 53.60M_{\odot}$ $, 197.57{\rm AU}, 691.06{\rm AU}, 0.99999, 0.415, 86.95^{\circ})$, $(56.35M_{\odot},$
			$101.98M_{\odot}, 53.60M_{\odot}, 23.12{\rm AU}, 291.21{\rm AU}, 0.9992, 0.628,$ $151.67^{\circ})$ in the \textsc{tsunami} simulation, are shown from the upper to lower panels, respectively. Each pair of adjacent panels corresponds to one event. The results in the \textsc{petar} and \textsc{tsunami} simulations are plotted in red dots and orange solid lines, respectively. Additionally, the result in the \textsc{tsunami} simulation without the PN correction is plotted in a gray dashed line (due to the lack of PN correction, the evolution time becomes much longer. For the convenience of comparison, a part of the evolution is truncated). The x-axes are shifted based on the time recorded.}
		\label{fig:kozai effect}
	\end{figure*}

	Triple systems may have interesting Kozai-Lidov (KL) oscillation, i.e., $e_{1}$ and $i_{\rm mut}$ oscillate quasi-periodically with time \citep{Kozai:1962zz,1962P&SS....9..719L,2016ARA&A..54..441N}. The orbital evolution of several merging triple BHs with KL oscillation in the \textsc{tsunami} simulation, along with them in the \textsc{petar} simulation, is shown in Fig. \ref{fig:kozai effect}.
	The orbital elements, including $e_{1}$ and $i_{\rm mut}$ are almost the same at early stages of
	evolution in
	both \textsc{petar} and \textsc{tsunami} simulations. However, as the merging triple BHs evolve, $e_{1}$ increases rapidly, leading to the merger of the inner BBHs within a short time in the \textsc{petar} simulation (note that the KL oscillation could also occur in the \textsc{petar} simulation, but due to the quick merger of the inner BBH and low time resolution, it is challenging to observe). On the other hand, in the \textsc{tsunami} simulation, the merging triple BH exhibits quasi-periodic oscillations of $e_{1}$ and $i_{\rm imut}$ over a very long period. For example, the last merging triple BH in the plot evolves about $7\times10^{4}$ years in total, with a period of
	oscillation for $e_{1}$ being $\sim7000$ years, which is consistent with the KL timescale analytically obtained by \citep{2015MNRAS.452.3610A, Trani:2021tan}
	\begin{align}
		T_{\rm KL}=P_{1}\frac{m_{1}+m_{2}}{m_{3}}\left(\frac{a_{2}}{a_{1}}\right)^{3}(1-e_{2}^{2})^{3/2},
	\end{align}
	where $P_{1}=2\pi\sqrt{a_{1}^{3}/(m_{1}+m_{2})}$ is the Kepler period of the inner BBHs.
	Furthermore, we explore the effect of turning off the PN correction in the \textsc{tsunami} simulation and observe the orbital evolution of these events. While $e_{1}$ and $i_{\rm mut}$ undergo quasi-periodical oscillations, their behavior is distinct. Specially, at certain parameter regions, $e_{1}$ in the \textsc{tsunami} simulation without PN correction can be either smaller or larger than those in the \textsc{tsunami} simulation, which consists of the prediction that the PN effect can suppress or excite $e_{1}$ \citep{2013ApJ...773..187N}.

	Based on the above comparison, \textsc{tsunami} (direct three-body numerical integration including the PN correction) provides a more realistic evolution of merging triple BHs. Therefore, in the following analysis, we focus on \textsc{tsunami} merging triple BHs, unless explicitly stated otherwise.

	\subsection{Population property of merging triple BH}\label{subsec:population property}
\begin{figure}
		\centering
		\subfigure{
			\centering
			\includegraphics[width=0.4\textwidth]{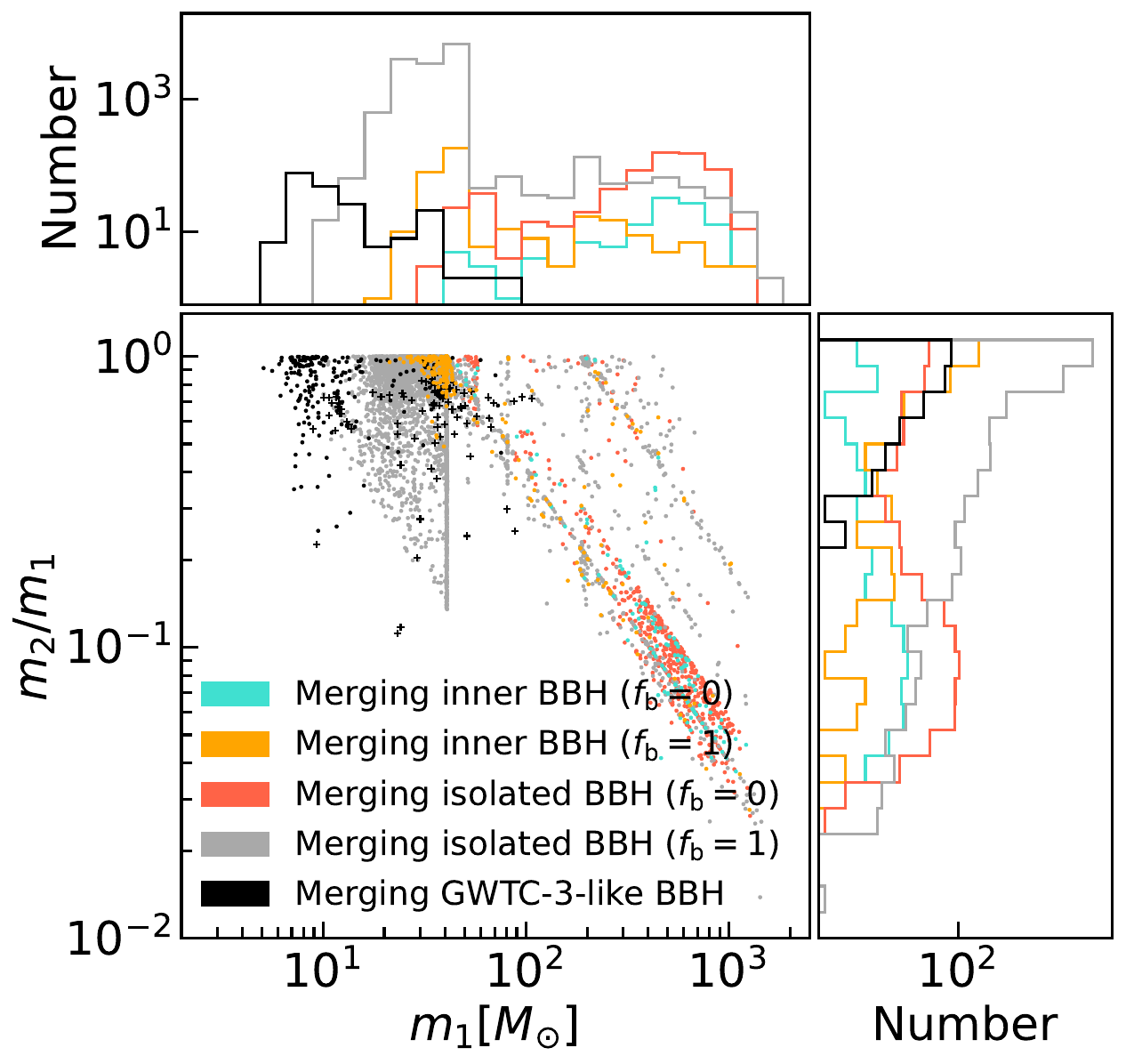}}
		\hspace{-50mm}
		\subfigure{
			\centering
			\includegraphics[width=0.4\textwidth]{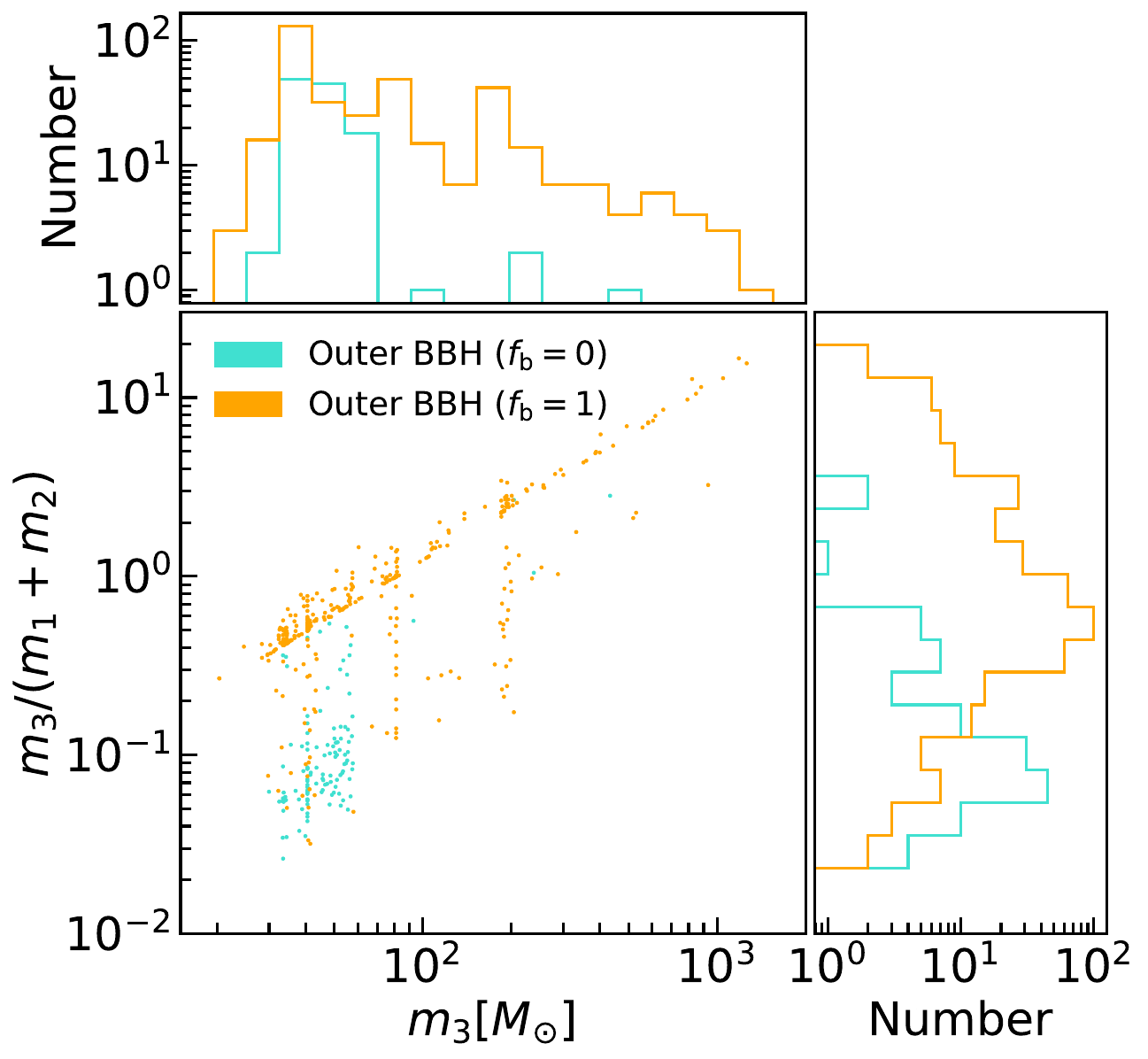}}
		\caption{\emph{Upper:} the distribution of the primary mass $m_{1}$ and mass ratio $m_{2}/m_{1}$.
		\emph{Bottom:} the distribution of the tertiary mass $m_{3}$ and mass ratio $m_{3}/(m_{1}+m_{2})$.
		The blue and orange colors denote merging triple BHs in the cases with $f_{\rm b}=0$ and 1, respectively. In the corresponding
		cases, merging isolated BBHs are represented by red and gray colors, respectively. The black color denotes merging BBHs inferred by LVKC based on
		GWTC-3.}
		\label{fig:mass and mass ratio}
	\end{figure}

	We investigate the mass distribution of merging triple BHs and compare it with that of merging isolated BBHs, as
	shown in Fig. \ref{fig:mass and mass ratio}. In the case of $f_{\rm b}=0$, both merging inner BBHs and merging isolated BBHs have primary masses $m_{1}$ of $\mathcal{O}(100)M_{\odot}$, which are indicative of IMBHs. Their inner BBHs and isolated BBHs tend to be incomparable, meaning the majority have mass ratio $q_{1}=m_{2}/m_{1}$ smaller than 0.2. In contrast, in clusters with primordial binaries ($f_{\rm b}=1$), most primary masses $m_{1}$ are typically $\mathcal{O}(10)M_{\odot}$, and component masses tend to be comparable. We can understand the difference as follows.

\begin{figure}
		\centering
		\includegraphics[width=0.3\textwidth, height=0.4\textwidth]{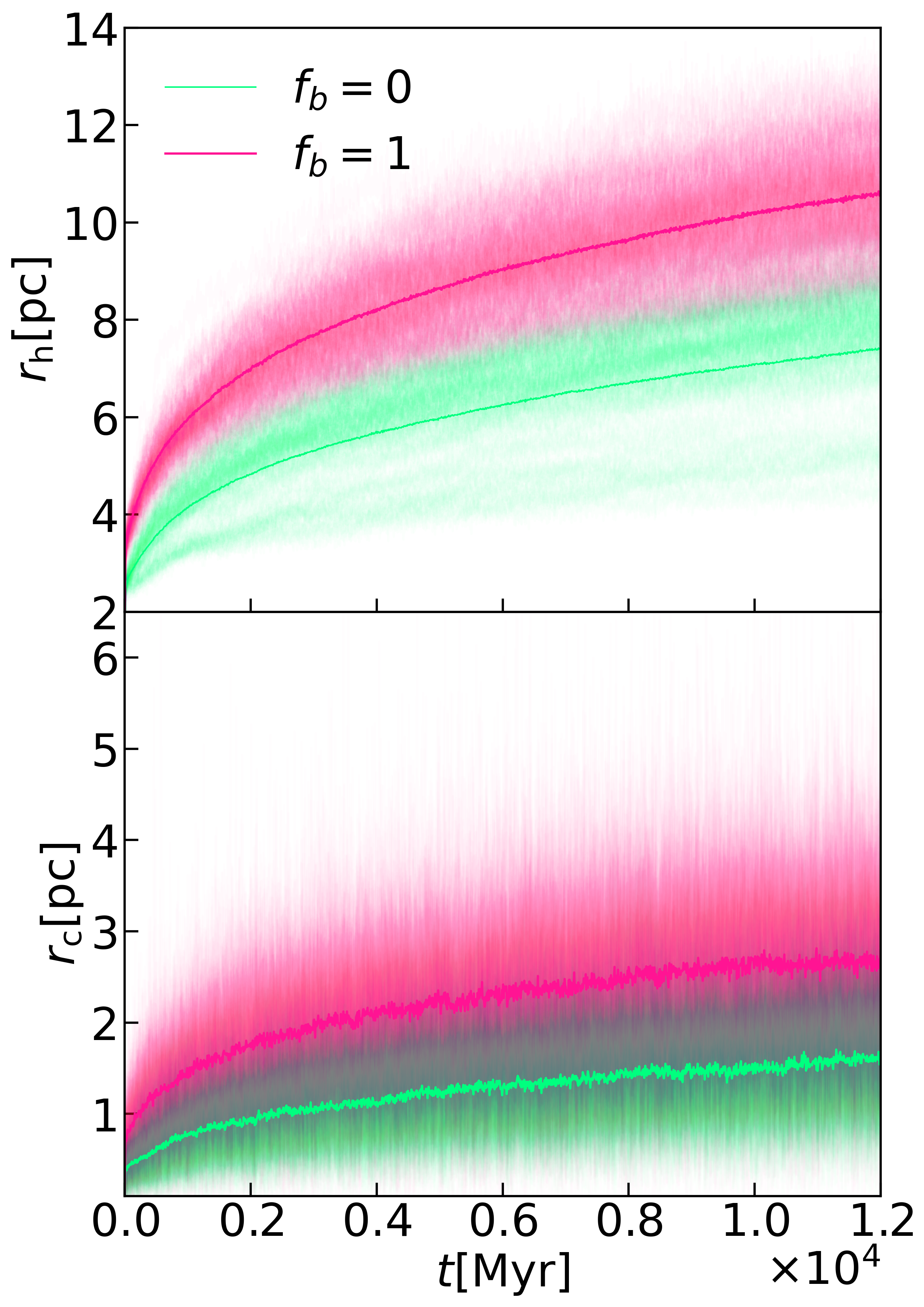}
		\caption{\emph{Upper:} The evolution of $r_{\rm h}$ of PopIII clusters. \emph{Lower:} The evolution of $r_{\rm c}$ of PopIII clusters. The cases with $f_{\rm b}=0$ and 1 are plotted in spring green and deep pink colors, respectively. The light colors represent 168 simulations, while the deep colors are mean values of them.}
		\label{fig:rh rc}
	\end{figure}

\begin{figure}
		\centering
		\includegraphics[width=0.4\textwidth, height=0.3\textwidth]{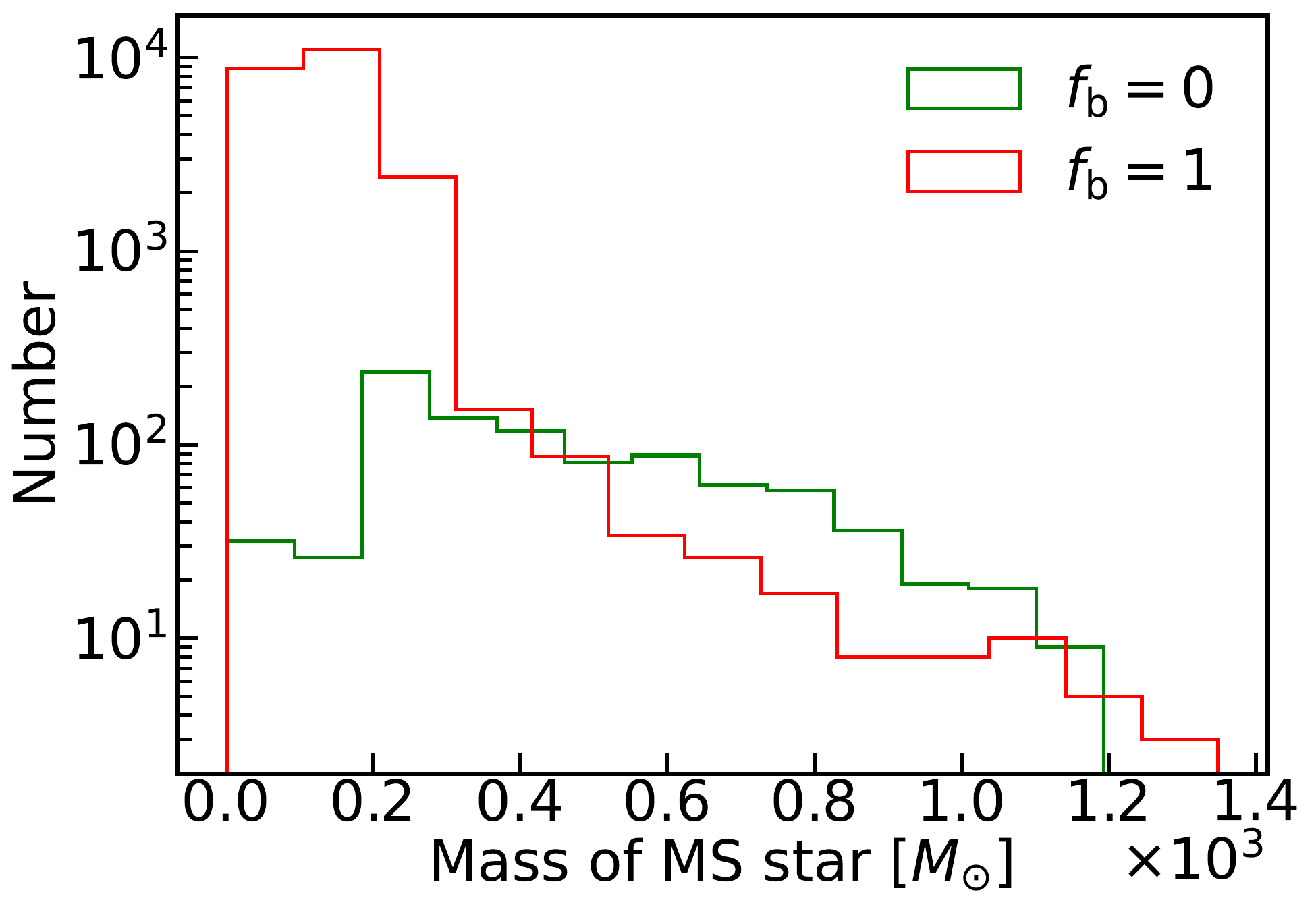}
		\caption{The mass distribution of main sequence stars from binary mergers. The y-axis represents the number of formed main sequence (MS) stars. The cases with $f_{\rm b}=0$ and 1 are plotted in green and red colors, respectively. }
		\label{fig:distrbution main sequence star}
	\end{figure}

	In the model with $f_{\rm b}=0$, $r_{\rm h}$ and $r_{\rm c}$ are smaller, as shown in Fig. \ref{fig:rh rc}, due to the absence of binary heating \citep{Wang:2021fxs}. This lead to an environment with higher number density of objects. Therefore, compared to the case with $f_{\rm b}=1$, both the number and proportion of VMSs heavier than $400M_{\odot}$ via multiple collisions, leading to the potential evolution into heavier BHs \citep{Wang2022b}, become greater in the case with $f_{\rm b}=0$, as shown in Fig. \ref{fig:distrbution main sequence star}. In clusters with primordial binaries, most merging inner BBHs and merging isolated BBHs originate from primordial binaries directly, as shown in Fig. \ref{fig:number binary triple bh}, which are expected to evolve into lighter BHs. In addition, BHs formed from the evolution of primordial binaries inherit the property of comparable mass, which results in a higher occurrence of mass ratios close to 1.

	The mass distribution of merging BBHs inferred by LVKC based on GWTC-3 is also presented\footnote{The data is from \href{https://zenodo.org/record/7843926}{https://zenodo.org/record/7843926}}.
	Compared to merging inner BBHs and merging isolated BBHs in PopIII clusters, they appear to be much lighter, which could be attributed to the observation selection effect of LIGO/Virgo. Since LIGO/Virgo are more sensitive at high frequency bands (hundreds of hertz), only merging comparable light BBHs with low redshifts are expected to be observed. Therefore, merging BBHs inferred by LVKC are biased towards lower masses. Regarding the outer BBHs of merging triple BHs, most of tertiary BHs $m_{3}$ are $\mathcal{O}(10)M_{\odot}$ in both cases with $f_{\rm b}=0$ and 1. However, the outer BBHs prefer to be comparable in the case with $f_{\rm b}=1$, because primordial binaries with comparable masses dominate the formation of merging triple BHs as well.

	\begin{figure}
		\centering
		\subfigure{
			\centering
			\includegraphics[width=0.4\textwidth]{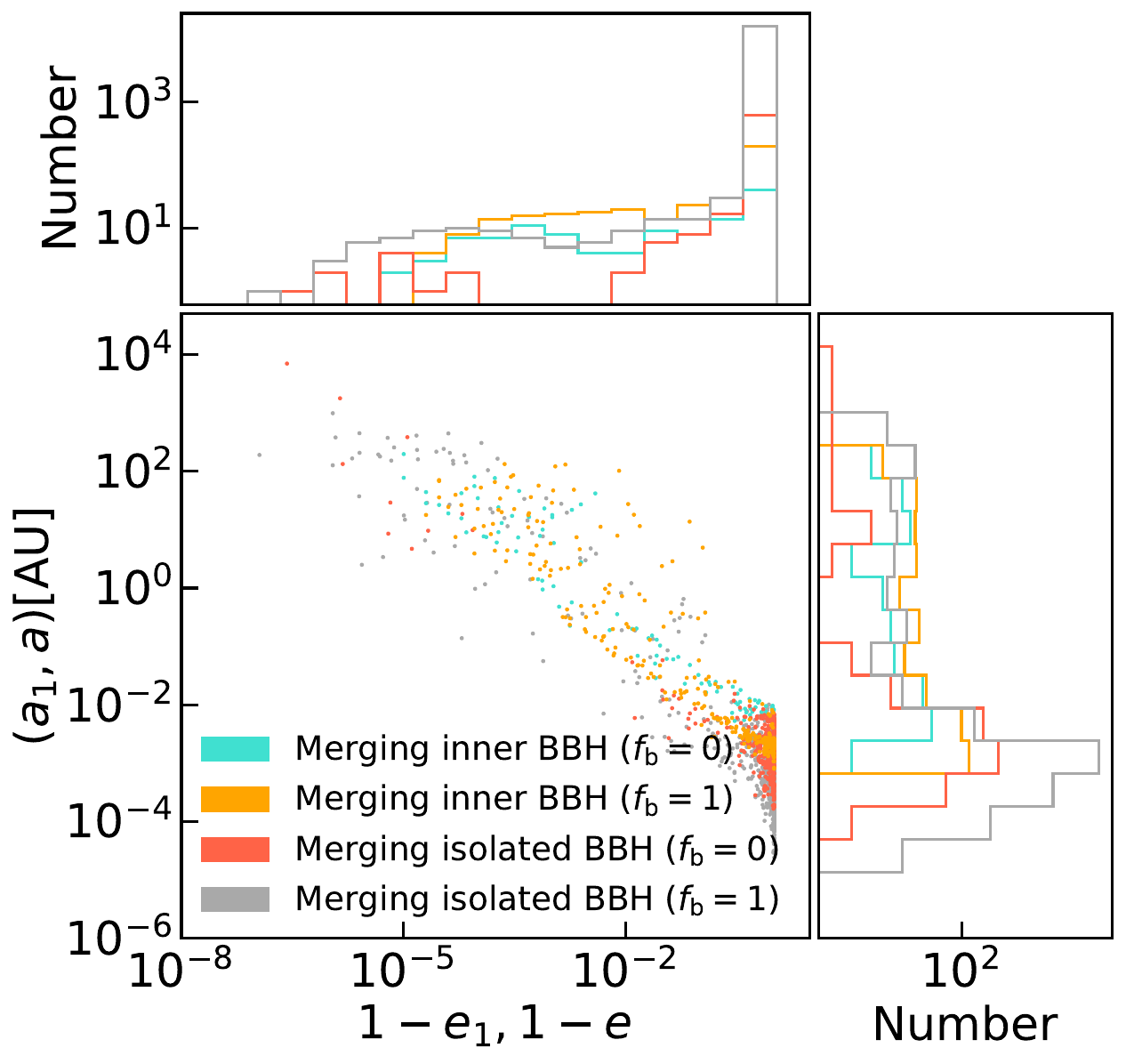}}
		\hspace{-50mm}
		\subfigure{
			\centering
			\includegraphics[width=0.4\textwidth]{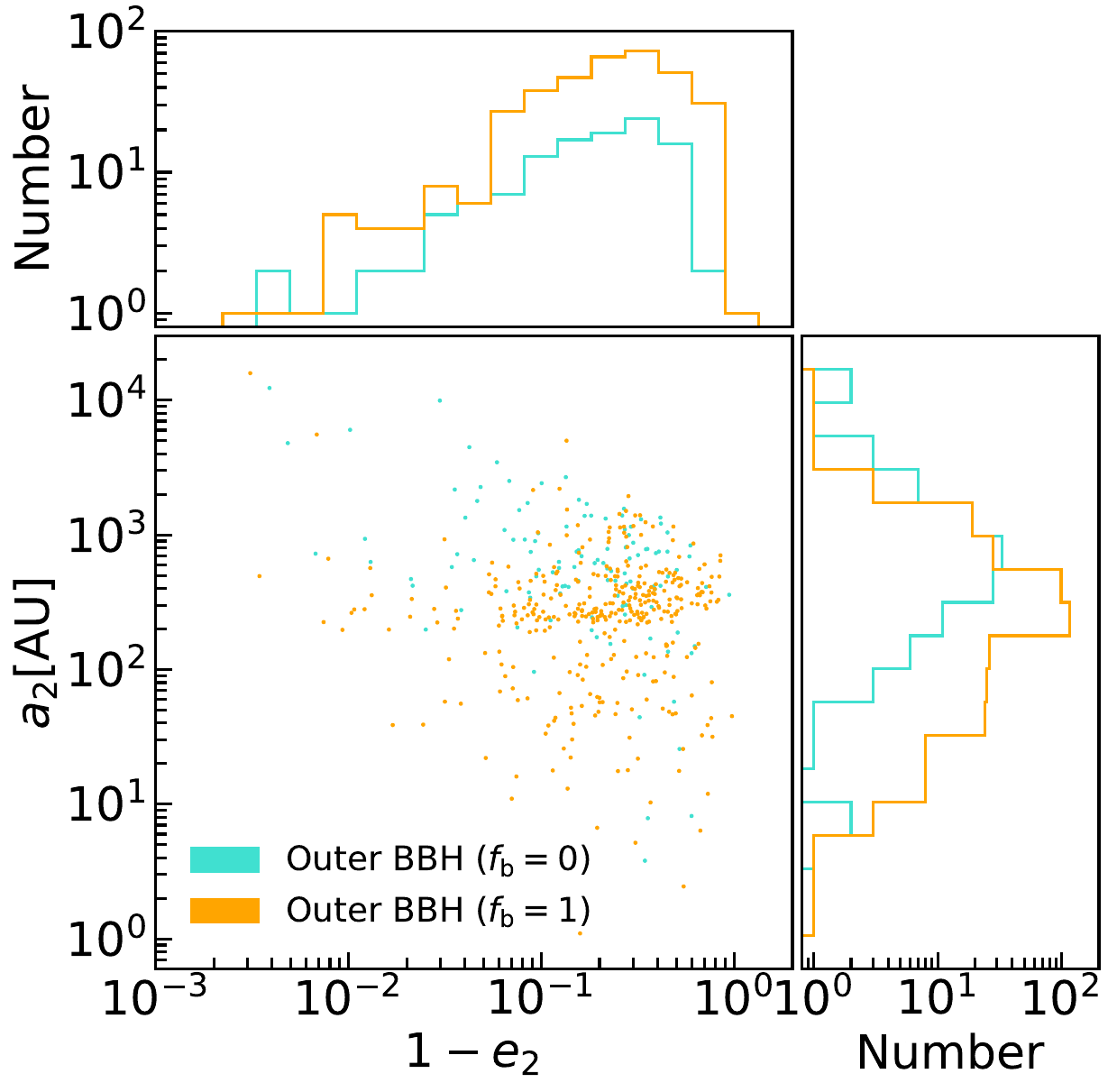}}
		\caption{\emph{Upper:} the distributions of the orbital eccentricity $1-e_{1}$ and the semimajor axis $a_{1}$ of the merging inner BBHs at their first record time. The distributions of $1-e$ and $a$ for merging isolated BBHs at their formation time, which is when BBHs appear during the evolution of binary stars, are also plotted.
		\emph{Bottom:} the distribution of the orbital eccentricity $1-e_{2}$ and the semimajor axis $a_{2}$ of the outer BBHs of merging triple BHs at their first record time.
		The blue and orange colors denote the merging triple BHs in the
		cases with $f_{\rm b}=0$ and 1, respectively. In the corresponding cases, the merging
		isolated BBHs are represented by red and gray colors, respectively.}
		\label{fig:ecc semi}
	\end{figure}

	\begin{figure}
		\centering
		\includegraphics[width=0.35\textwidth]{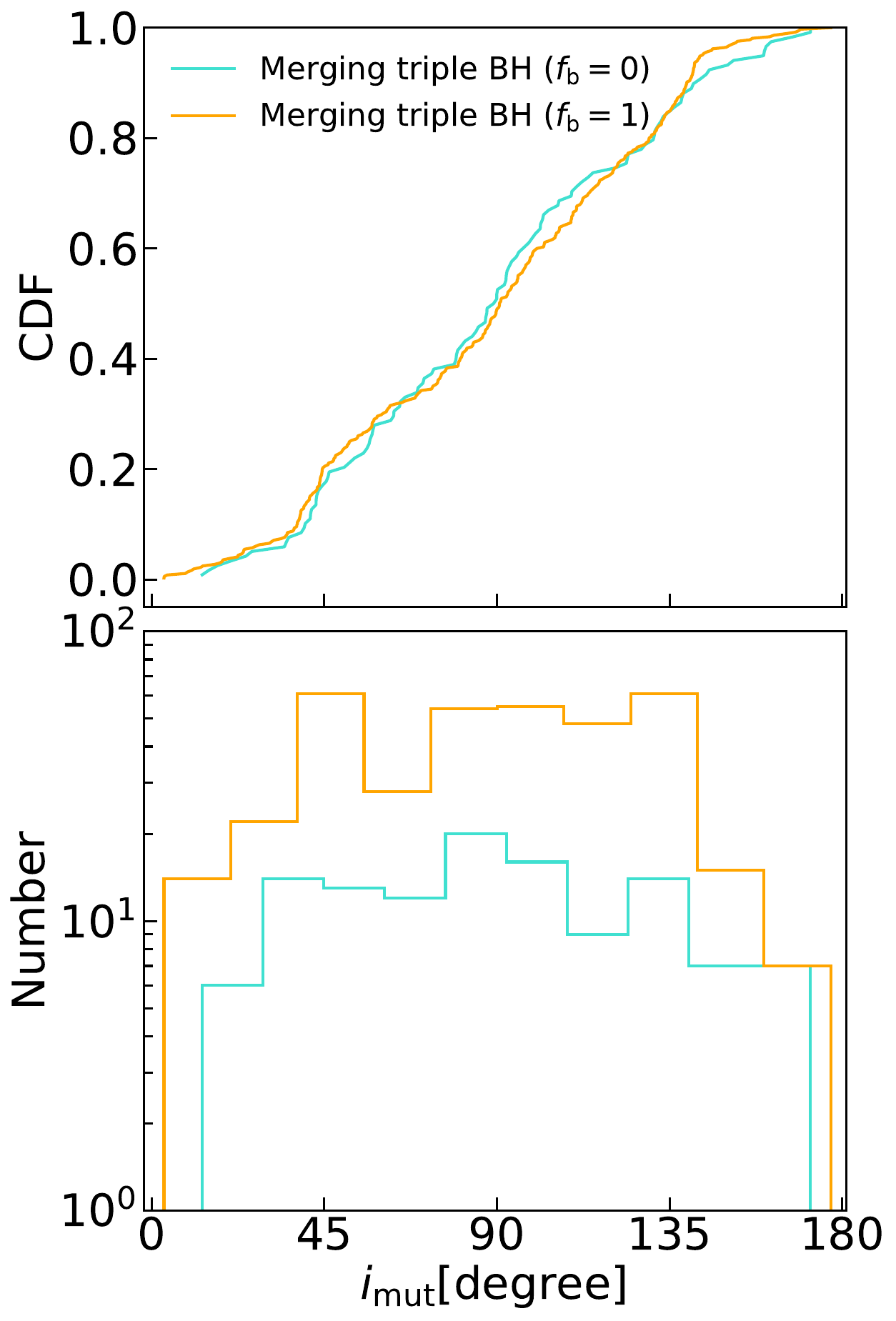}
		\caption{The distribution of angles $i_{\rm mut}$ between the inner and outer planes of merging triple BHs at the first record time. The lower and upper panels display the distribution of $i_{\rm mut}$ and the corresponding normalized cumulative distribution function (CDF). The blue and orange colors correspond to the cases with $f_{\rm b}=0$ and 1, respectively.}
		\label{fig:imut}
	\end{figure}

	The distribution of orbital eccentricity and semimajor axis of merging triple BHs and merging isolated BBHs is shown in Fig. \ref{fig:ecc semi}. In both of cases with $f_{\rm b}=0$ and 1, due to the interaction with the tertiary BHs, merging inner BBHs tend to have
	significantly higher eccentricities, some of which could be as low as $(1-e_{1})<10^{-3}$, even could reach $(1-e_{1})\sim10^{-5}$. In contrast, merging isolated BBHs tend to have steeper distributions of $(1-e)$, indicating that the proportion of sources with higher eccentricities is smaller than that of merging inner BBHs, as they lack the perturbation from the third BHs. Furthermore,
	since all the merging isolated BBHs are formed through dynamical captures in the case with $f_{\rm b}=0$, the proportion of more
	eccentric sources is higher than that in the case with $f_{\rm b}=1$, where all the sources are formed from the evolution of primordial binaries, inherently tending to have small eccentricities. In both cases with $f_{\rm b}=0$ and 1, the outer BBHs of merging triple BHs prefer to have eccentricities $0.6<e_{2}<0.9$, due to the gravitational capture process between the inner BBHs and the third BHs. The semimajor axes of the outer
	orbits tend to be much larger than those of the inner orbits, otherwise the latter would be disrupted by the tidal interaction. In addition, most merging triple BHs have large $i_{\rm mut}$ ranging from 45$^{\circ}$ to 135$^{\circ}$, as shown in Fig. \ref{fig:imut}. Note that in the above comparison, the orbital parameters of merging isolated BBHs are considered at their formation time, which is when BBHs form during the evolution of binary stars. However, for the merging triple BHs, their orbital parameters are considered at the first record time, as the formation time of triple systems can not be defined well. Nevertheless, this comparison can still highlights qualitatively the distinctions in orbital parameters between merging inner BBHs and isolated BBHs. A more fair comparison of orbital parameters at the same frequency will be further explored in the following subsection.

	We also study the evolutionary dominance of inner BBHs of merging triple BHs at their first record time, as listed in Table \ref{tab:dy and gw domination}. In the case with $f_{\rm b}=0$, the orbital evolution of more than 50\% of inner BBHs are dominated by the GW driven, while the evolutionary dominance is the dynamical interaction between them and the tertiary BHs in the case with $f_{\rm b}=1$. This discrepancy can be explained as follows. The inner BBH in the case with $f_{\rm b}=0$ are more eccentric than those in the case with $f_{\rm b}=1$, as shown in Fig. \ref{fig:ecc semi}. This makes that $\ell_{1}$ is more likely to be smaller than $\ell_{\rm GW}$ in the case with $f_{\rm b}=0$.

	\begin{table}
		\centering
		\caption{ The evolutionary dominance of inner BBHs of merging triple BHs at the first record time.}
		\label{tab:dy and gw domination}
		\setlength{\tabcolsep}{3.2mm}{
			\begin{tabular}{ccc}
				\toprule\toprule  
				$f_{\rm b}$ & Dynamical interaction ($\ell_{1}>\ell_{\rm GW}$) & GW driven ($\ell_{1}<\ell_{\rm GW}$) \\
				\midrule  
				0           & 22.0\%                                           & 78.0\%                               \\
				1           & 61.6\%                                           & 38.4\%                               \\
				\bottomrule\bottomrule 
			\end{tabular}}
	\end{table}

	\subsection{Merger rate of inner BH-IMBH of merging triple BH}
	\begin{table*}
		\centering
		\caption{Average numbers and corresponding merger rates of inner BBHs of merging triple BHs and merging isolated BBHs over redshift per 12 Gyr in cases with $f_{\rm b}=0$ and 1. These values are represented by $n(\mathcal{R}_{\rm lower}-\mathcal{R}_{\rm upper})$, where $n$ represents the average number, $\mathcal{R}_{\rm lower}$ and $\mathcal{R}_{\rm upper}$ in the unit of ${\rm Gpc}^{-3}{\rm yr}^{-1}$ are lower and upper limits of the average merger rate, respectively. The symbol BBH represents all the
			merging inner or merging isolated BBH, where IMBH-BH means binaries containing at least one IMBH, IMBBH denotes binaries
			composing of two IMBH, PIBH-LBH represents binaries consisting of PIBH and LBH, BLBH denotes binaries including two LBH. Some numbers are inconsistent due to rounding.}
		\label{tab:average number of merger}
		\setlength{\tabcolsep}{4.5mm}{
			\begin{tabular}{ccccccccc}
				\toprule\toprule  
				Type                              & $f_{\rm b}$ & BBH             & IMBH-BH        & IMBBH             & PIBH-LBH          & LBBH              \\
				\midrule  
				\multirow{2}{*}{Merging inner}    & 0           & 0.7 (0.02-0.1)  & 0.6 (0.02-0.1) & 0.03 (0.001-0.01) & 0.03 (0.001-0.01) & 0.05 (0.001-0.01) \\
				\multirow{2}{*}{}                 & 1           & 2.2 (0.1-0.4)   & 0.4 (0.01-0.1) & 0.1 (0.003-0.02)  & 0.1 (0.002-0.02)  & 1.6 (0.04-0.3)    \\
				\midrule
				\multirow{2}{*}{Merging isolated} & 0           & 3.9 (0.1-0.7)   & 3.4 (0.1-0.6)  & 0.2 (0.004-0.03)  & 0.1 (0.003-0.02)  & 0.4 (0.002-0.01)  \\
				\multirow{2}{*}{}                 & 1           & 93.6 (2.5-15.6) & 2.7 (0.1-0.5)  & 0.9 (0.03-0.2)    & 0.6 (0.02-0.1)    & 90.2 (2.4-15.0)   \\
				\bottomrule\bottomrule 
			\end{tabular}}
	\end{table*}

\begin{figure}
		\centering
		\includegraphics[width=0.5\textwidth, height=0.5\textwidth]{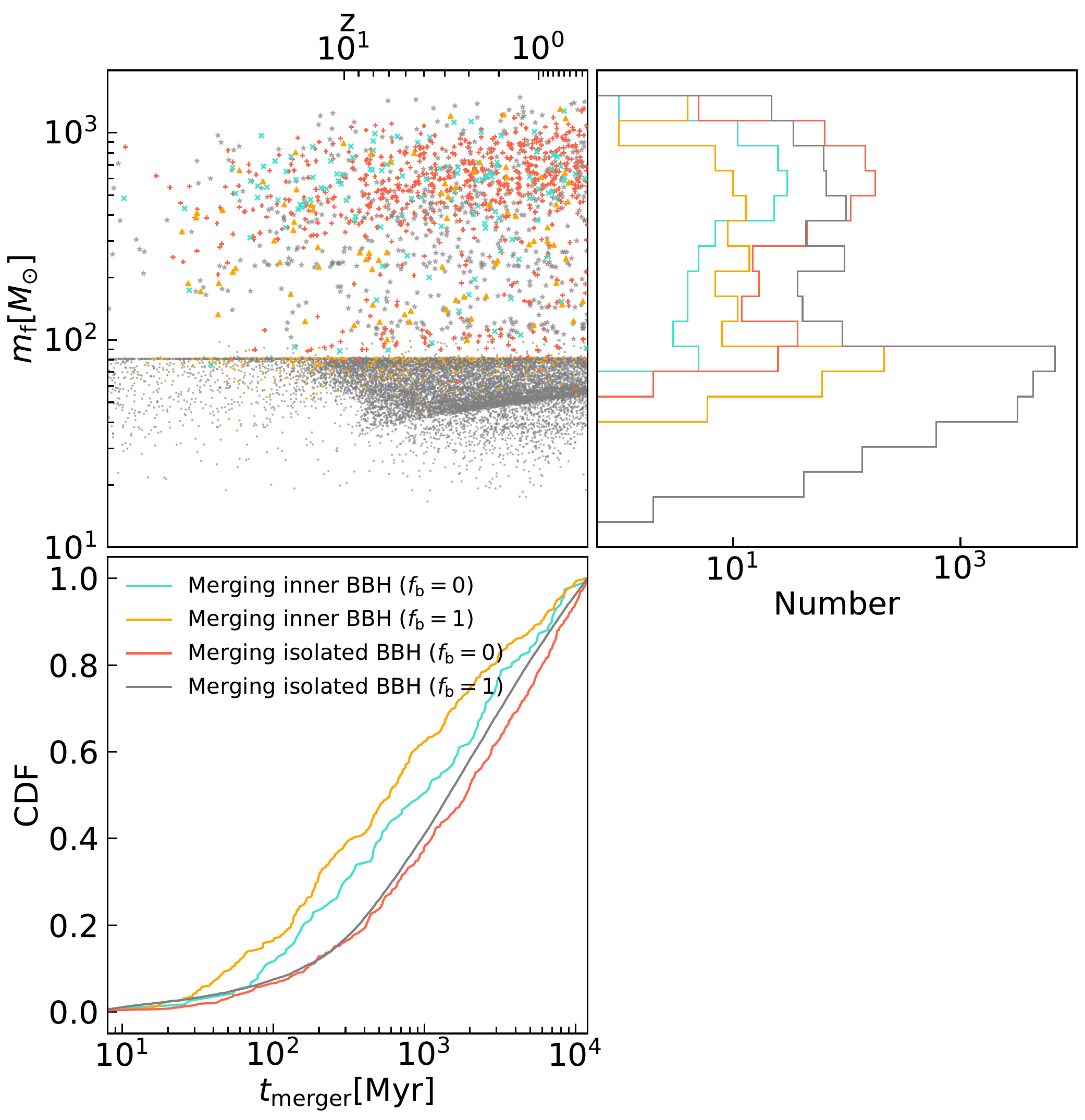}
		\caption{\emph{Upper:} the merger remnant masses ($m_{\rm f}$) of inner BBHs of merging triple BHs and merging isolated BBHs in the cases with $f_{\rm b}=0$
			and 1, which are plotted in different colors and shapes, respectively. The merging inner BBHs and merging isolated BBHs from primordial binaries are marked with
			dots. \emph{Lower:} the normalized CDF of merger times ($t_{\rm merger}$) of inner and isolated BBHs in the cases with different values of $f_{\rm b}$. The upper x-axis denotes the redshift $z$ corresponding to $t_{\rm merger}$. The merger times of inner BBHs here are equal to the first record time of inner BBHs recorded by \textsc{petar} plus their evolution time to mergers in the \textsc{tsunami} simulation. The merger time of isolated BBHs refer to the points during the evolution of PopIII clusters simulated by \textsc{petar} when the mergers occur.}
		\label{fig:mf cdf tmerger}
	\end{figure}

	The merger remnants and merger times of inner BBHs of merging triple BHs are shown in Fig. \ref{fig:mf cdf tmerger}. In both cases of $f_{\rm b}=0$ and 1, the inner BBHs begin to merge at about 10Myr . The total number of mergers increases logarithmically with time approximately, following the standard distribution of merger time $\propto t^{-1}$ \citep[e.g.][]{Tremou:2018rvq} until the present Universe, with about 80\% occurring before 3000Myr ($z>2$). However, the values of $f_{\rm
				b}$ affect the mass spectrum of merger remnants significantly. Specifically, when $f_{\rm b}=0$, most remnants would be IMBHs with $\mathcal{O}(100)M_{\odot}$. As $f_{\rm b}$ increases to 1, the remnants become lighter, with most of them being $\mathcal{O}(10)M_{\odot}$. We can understand the difference depending on Fig. \ref{fig:mass and mass ratio} and the explanation on it.

	For comparison, the merging isolated
	BBHs in cases with $f_{\rm b}=0$ and 1 are also investigated. Their mergers follow the trend of inner BBHs of merging triple BHs roughly, but the number of them increases relatively slowly
	during the period of 200-5000Myr. This is because the isolated BBHs, without perturbation from other BHs, take more time to undergo mergers. For the influence of $f_{\rm b}$ on the remnant masses of merging isolated BBHs, it exhibits a similar trend to that of merging inner BBHs.

	The merging inner and merging isolated BBHs can be categorized based on their component masses. The average number $n$ of such mergers and their corresponding merger rates $\mathcal{R}$ averaged over redshift are listed in Table \ref{tab:average number of merger}, where $\mathcal{R}$ can be estimated as below
	\begin{align}
		\mathcal{R}=\frac{n}{M_{\rm clu}/{\rm \rho_{\star}}\times T_{\rm evo}}
	\end{align}
	where $M_{\rm clu}=10^{5}M_{\odot}$ is the total stellar mass of PopIII clusters. The stellar mass formed per volume $\rho_{\star}$ during a period of time could be determined using a fitting formula based on star formation rate (SFR) density \citep{2022ApJ...926...83T,2020MNRAS.492.4386S}. The lower and upper limits of $\rho_{\star}$, derived by integrating the formula over the period from 100 to 500Myr (corresponding to a redshift range from 20 to 10) and considering re-ionization history, are $3.2\times10^{4}M_{\odot}{\rm Mpc}^{-3}$ and $2\times10^{5}M_{\odot}{\rm Mpc}^{-3}$ \citep{2020MNRAS.492.4386S,2021ApJ...919...41I}, respectively.

	In the case with $f_{\rm b}=0$, the merger rate of inner BBH reaches up to $0.1{\rm Gpc}^{-3}{\rm yr}^{-1}$, which is slightly larger than that of inner IMBH-BH. The upper merger rates of inner IMBBH, inner PIBH-LBH and inner LBBH are $0.01{\rm Gpc}^{-3}{\rm yr}^{-1}$. As $f_{\rm b}$ increases to 1, the merger rates increase by several to dozens of times, as expected, except for that of inner IMBH-BH. The slight decline in the merger rate of inner IMBH-BH could be attributed to the fact that most inner BBHs consist of comparable components with dozens of solar masses in the case with $f_{\rm b}=1$, as shown on the upper panel in Fig. \ref{fig:mass and mass ratio}. For merging isolated BBHs with different components, the upper merger rates are several to dozens of times those of corresponding inner BBHs in the case with the same $f_{\rm b}$. Depending on these above results, the merger rate of inner and isolated IMBBH could make a contribution to that of IMBBH constrained by LIGO/Virgo/KAGRA collaboration (LVKC) \citep{LIGOScientific:2021tfm, LIGOScientific:2019ysc}. If we regard PIBH-LBH and LBBH as GW190521-like sources and stellar-mass binary BHs (SBBHs) respectively, their merger rates could contribute to or explain those inferred by LVKC \citep{LIGOScientific:2020iuh,2020ApJ...900L..13A, KAGRA:2021duu}.

	\subsection{GW from inner IMBH-BH of merging triple BH}
	Merging eccentric BBHs are accompanied by the emission of GWs consisting of different harmonics, whereas only the second one is included in the case of circular orbits. The characteristic amplitude of the $n$ order ($n$th) harmonic of inspiral GWs is given, following \citep{2019PhRvD..99f3003K}, by
	\begin{align}\label{eq:hcn}
		h_{cn} = \sqrt{\frac{2}{3\pi^{4/3}}}\frac{\mathcal{M}^{5/6}}{D}f_{n}^{-1/6}\left(\frac{2}{n}\right)^{1/3}\sqrt{\frac{g(n,e)}{F(e)}},
	\end{align}
	where $f_{n}$ is the frequency of the $n$th harmonic, $\mathcal{M}$ and $D$ are the chirp mass and distance of the merging BBHs, respectively. Among these harmonics, the ($n_{\rm peak}=f_{\rm peak}/f_{\rm orb}$)th order one focused in the following has the maximum radiation power, where $f_{\rm orb}$ is the orbital frequency calculated by Kepler's third law. The peak frequency $f_{\rm peak}$ can be obtained by \citep{2021RNAAS...5..275H}
	\begin{align}
		f_{\rm peak} & = \frac{\sqrt{m_{1}+m_{2}}}{\pi} \nonumber                                         \\
		             & \times\frac{1-1.01678e+5.57372e^{2}-4.9271e^{3}+1.68506e^{4}}{[a(1-e^{2})]^{1.5}},
	\end{align}
	where $a$ and $e$ are the semimajor axes and orbital eccentricities of merging BBHs, respectively. As the merging BBHs
	evolve in their orbits, they undergo successive stages of merger and ringdown after the inspiral stages, rendering Eq. (\ref{eq:hcn}) unsuitable. Due to the circularization of GW radiation, the orbital eccentricities of BBHs decrease and approach zero as they enter the merger and ringdown phases, making the ($n_{\rm peak}=2$)th harmonic become dominant. The GWs from the merger and ringdown phases could be described by PhenomD waveform \citep{Husa:2015iqa,Khan:2015jqa}. Considering the significant effect of cosmic expansion on the motion of distant merging BBHs, some physical quantities should
	be replaced with their redshifted counterparts: $D\rightarrow D_{L}$(luminosity distance), $\mathcal{M}\rightarrow(1+z)\mathcal{M}$,
	$f_{n}\rightarrow f_{n}/(1+z)$, in both Eq. (\ref{eq:hcn}) and PhenomD waveform. Furthermore, the inspiral phase of merging BBHs can last much longer than the observation time of GW detectors, thus the fraction of mission lifetime
	sources spent within given frequency bins should be considered. Therefore, when calculating the characteristic strain of inspiral
	GWs, it should be multiplied by the square root of min$[1, \dot{f}_{n}(T_{\rm obs}/f_{n})]$
	\citep{2018MNRAS.481.4775D,2007ApJ...665L..59W,2005CQGra..22S.363S}, where $T_{\rm obs}=5$ years is adopted as a fiducial value here.

	We calculate the $f_{\rm peak}$ and corresponding GWs from inner BBHs of merging triple BHs, and compare them with those from merging isolated BBHs depending on Eq. (\ref{eq:hcn}) and PhenomD waveform model. It should be noted that since calculating the orbital eccentricities and semimajor axes of merging inner
	BBHs with \textsc{tsunami} will be not accurate anymore, when they begin to enter the bands of ground-based GW detectors like LIGO, because in this regime the PN approximation stops holding. Thus, we stop calculating them once $a$ reaches $\mathcal{O}(10^{-4})$AU. At this stage, GW radiation starts to
	dominate the evolution of merging inner BBHs and the perturbation from the tertiary BHs can be neglected safely, so we continue evolving the orbital of merging inner BBHs until $a$ approaches zero alternatively by the averaged orbital Eqs. (\ref{eq:a_dot e_dot}), which describes the evolution of merging isolated BBHs \citep{1963PhRv..131..435P,1964PhRv..136.1224P}
	\begin{subequations}\label{eq:a_dot e_dot}
		\begin{align}
			\frac{{\rm d}a}{{\rm d}t} & = -\beta\frac{F(e)}{a^{3}},                                                         \\
			\frac{{\rm d}e}{{\rm d}t} & = -\beta\frac{19}{12}\frac{e}{a^{4}(1-e)^{5/2}}\left(1+\frac{121}{304}e^{2}\right),
		\end{align}
	\end{subequations}
	where $\beta=64m_{1}m_{2}(m1+m2)/5$. As for the merging isolated BBHs, since they are not perturbated throughout their whole evolution, we evolve them with Eqs. (\ref{eq:a_dot e_dot}) until $a$ approaches to zero.

	\begin{figure*}
		\centering
		\includegraphics[width=0.7\textwidth, height=0.8\textwidth]{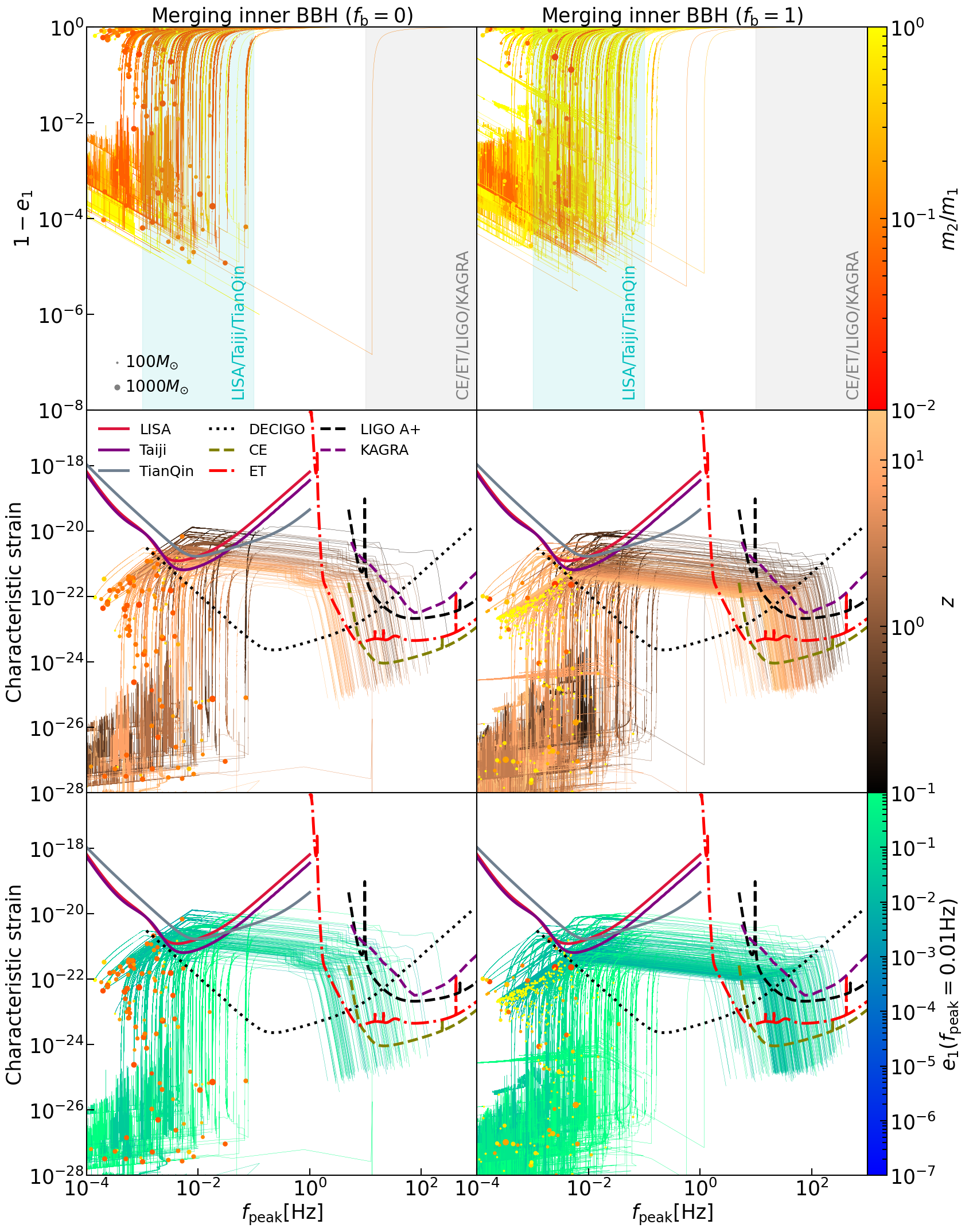}
		\caption{Inner BBHs of merging triple BHs. \emph{Upper}: the evolution of inner orbital eccentricity $e_{1}$ with $f_{\rm peak}$,
		and the color bar denotes the inner mass ratio $m_{2}/m_{1}$. The cyan and gray shadow areas represent the frequency ranges where the space-borne GW detectors LISA/TianQin/Taiji and the ground-based GW detectors CE/ET/LIGO/KAGRA are sensitive. \emph{Middle}: the characteristic strains of GWs with $f_{\rm peak}$, and
		the color bar represents the redshift of sources. \emph{Lower:} the characteristic strains of GWs with $f_{\rm peak}$, with the color bar representing the eccentricities at $f_{\rm peak}=0.01$Hz of sources. The size of dots scales with the total mass of the merging inner BBHs, the characteristic strains of noise of GW detectors\protect\footnotemark[5] \citep{Kawamura:2011zz,Robson:2018ifk,Wang:2019ryf,Michimura:2020xnj,LIGOScientific:2016wof,Hild:2010id} are plotted in different
		colors. The columns from left to right are cases with $f_{\rm b}=0$ and 1, respectively.}
		\label{fig:ecc hcn fpeak inner}
	\end{figure*}
	\footnotetext[3]{The power spectrum density of LIGO A+ is from
		\href{https://dcc.ligo.org/LIGO-T1800042/public}{https://dcc.ligo.org/LIGO-T1800042/public}}

	\begin{figure}
		\centering
		\includegraphics[width=0.4\textwidth]{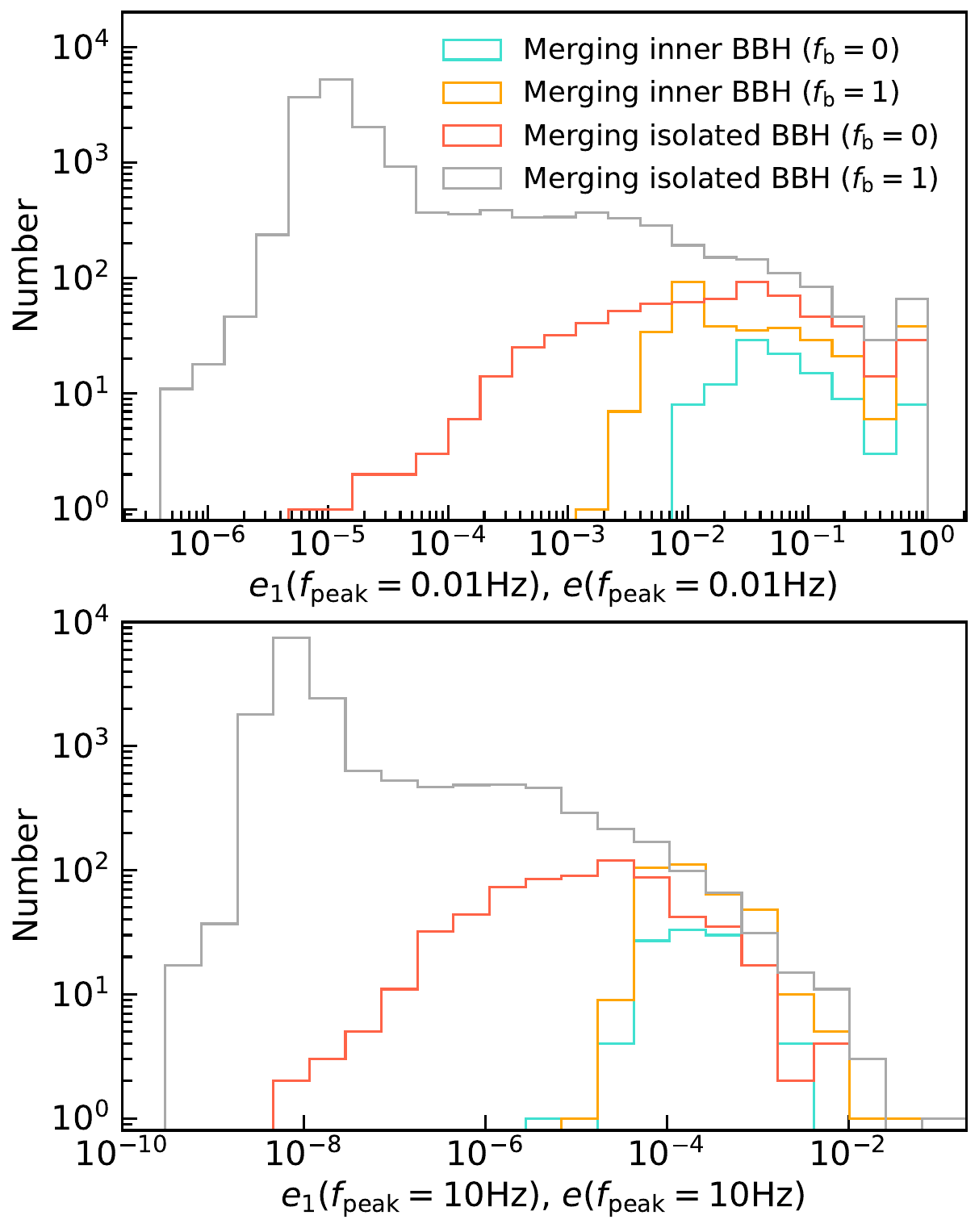}
		\caption{\emph{Upper}: the distribution of orbital eccentricity $e_{1}$($e$) of merging inner (isolated) BBHs at $f_{\rm peak}=0.01$Hz. \emph{Lower}: the distribution of $e_{1}$($e$) of merging inner (isolated) BBHs at $f_{\rm peak}=10$Hz. The distributions in the cases with $f_{\rm b}=0$ and 1 are plotted in
			different colors, respectively.}
		\label{fig:ecc at fpeak}
	\end{figure}

	\begin{figure*}
		\centering
		\includegraphics[width=0.8\textwidth]{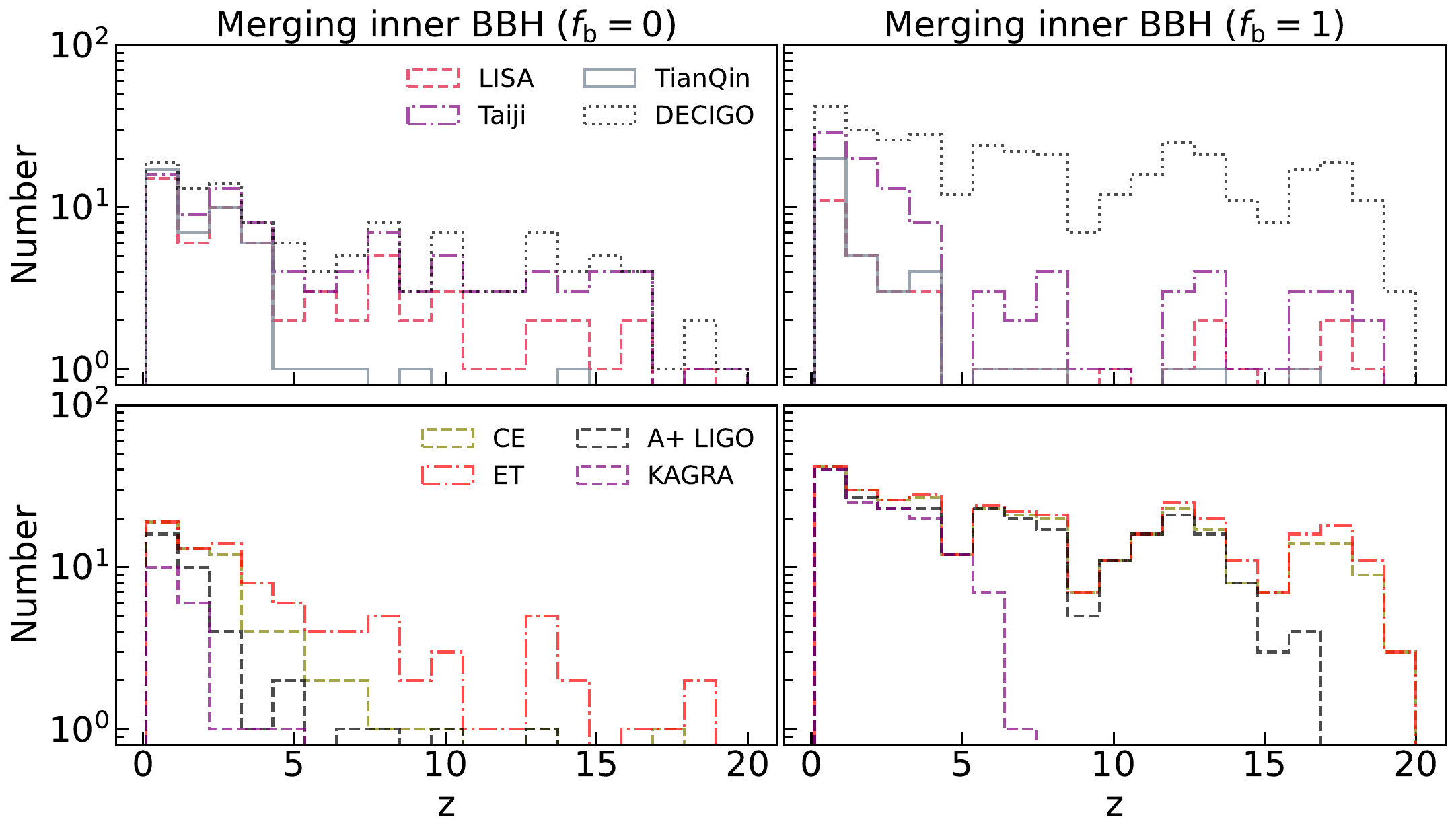}
		\caption{The redshift distributions of detectable merging inner BBHs are depicted for various GW detectors, distinguished by different colors and line styles. The left and right columns represent cases where $f_{\rm b}=0$ and 1, respectively. The upper panels showcase the space-borne detectors LISA/TianQin/Taiji/DECIGO, while the lower panels feature the ground-based detectors CE/ET/LIGO/KAGRA. Detectable sources are characterized by GW characteristic strains exceeding those of detector noise within a certain frequency range.}
		\label{fig:zz detectable inner}
	\end{figure*}

	\begin{figure*}
		\centering
		\includegraphics[width=0.8\textwidth]{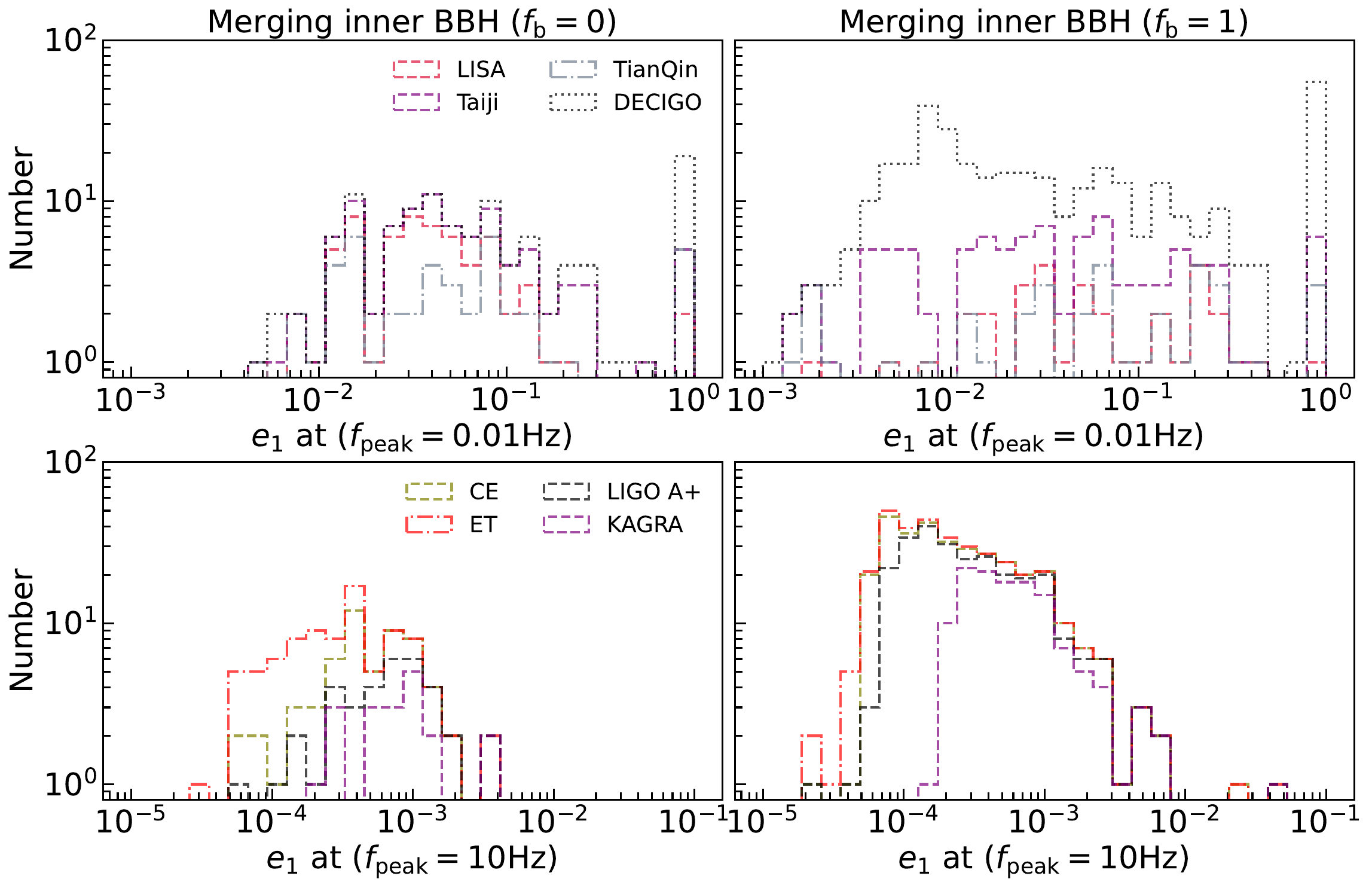}
		\caption{The distributions of orbital eccentricity at $f_{\rm peak}=0.01$Hz and 10Hz for detectable merging inner BBHs across various GW detectors are illustrated using different colors and line styles. The left and right columns correspond to cases where $f_{\rm b}=0$ and 1, respectively. The upper panels depict the space-borne detectors LISA/TianQin/Taiji/DECIGO, while the lower panels represent the ground-based detectors CE/ET/LIGO/KAGRA. Detectable sources are those for which the characteristic strains of GWs exceed those of detector noise within a certain frequency range.}
		\label{fig:e1 at fpeak detectable inner}
	\end{figure*}

	\begin{figure}
		\centering
		\includegraphics[width=0.4\textwidth]{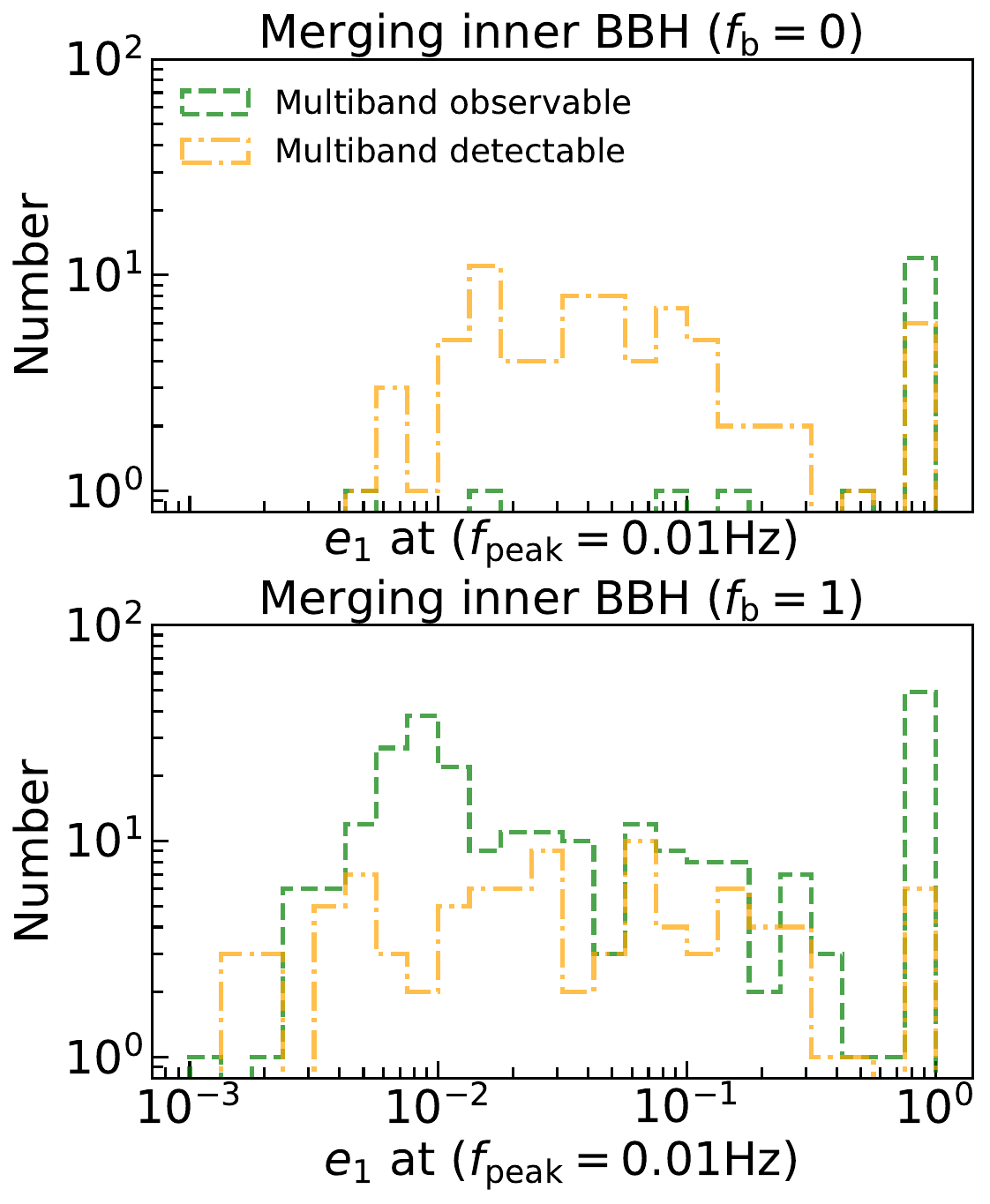}
		\caption{The eccentricity distributions at $f_{\rm peak}=0.01$Hz of multiband observable and multiband detectable merging inner BBHs are shown, represented by green and orange dashed lines, respectively. The multiband observable sources refer to merging inner BBHs that are undetectable by mHz space-borne detectors but are detectable by at least one ground-based detectors. The multiband detectable sources are those detectable by both at least one mHz detectors and at least one ground-based detectors. The upper and lower panels represent cases where $f_{\rm b}=0$ and 1, respectively.}
		\label{fig:multiband inner}
	\end{figure}


	\begin{figure*}
		\centering
		\includegraphics[width=0.7\textwidth, height=0.8\textwidth]{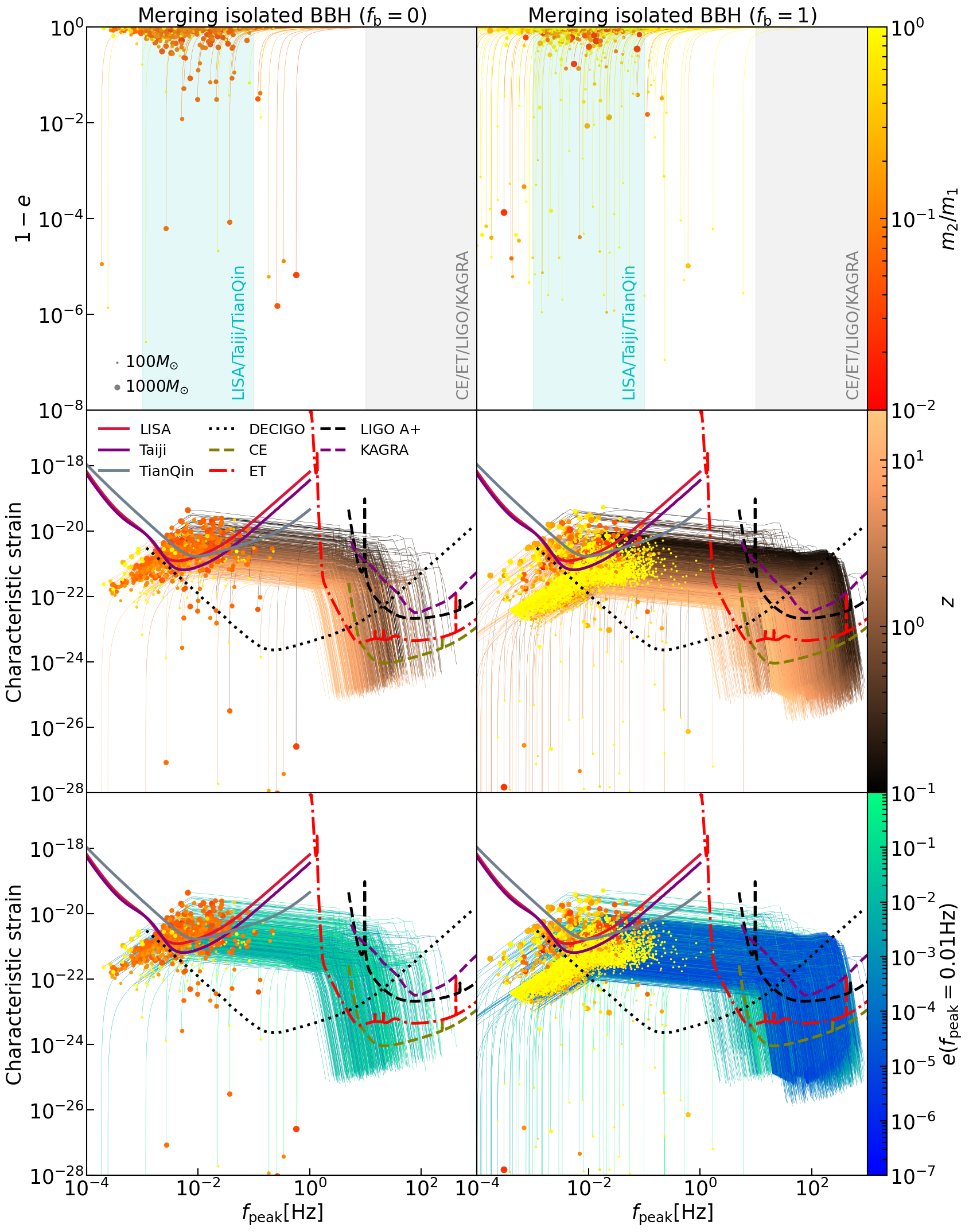}
		\caption{Merging isolated BBHs. \emph{Upper}: the evolutions of orbital eccentricity $e$ with $f_{\rm peak}$,
		and the color bar denotes mass ratio $m_{2}/m_{1}$. The cyan and gray shadow areas represent the frequency ranges where the space-borne GW detectors LISA/TianQin/Taiji and the ground-based GW detectors CE/ET/LIGO/KAGRA are sensitive. \emph{Middle}: the characteristic strains of GWs with $f_{\rm peak}$, with
		the color bar representing the redshift of the sources. \emph{Lower:} the characteristic strains of GWs with $f_{\rm peak}$, with
		the color bar representing the eccentricities at $f_{\rm peak}=0.01$Hz of sources. The size of the dots is scaled with the total mass of the sources, and the characteristic strains of the noise of GW detectors are plotted in different
		colors. The columns from left to right represent the cases with $f_{\rm b}=0$ and 1, respectively.}
		\label{fig:ecc hcn fpeak isolated}
	\end{figure*}

	\begin{figure*}
		\centering
		\includegraphics[width=0.8\textwidth]{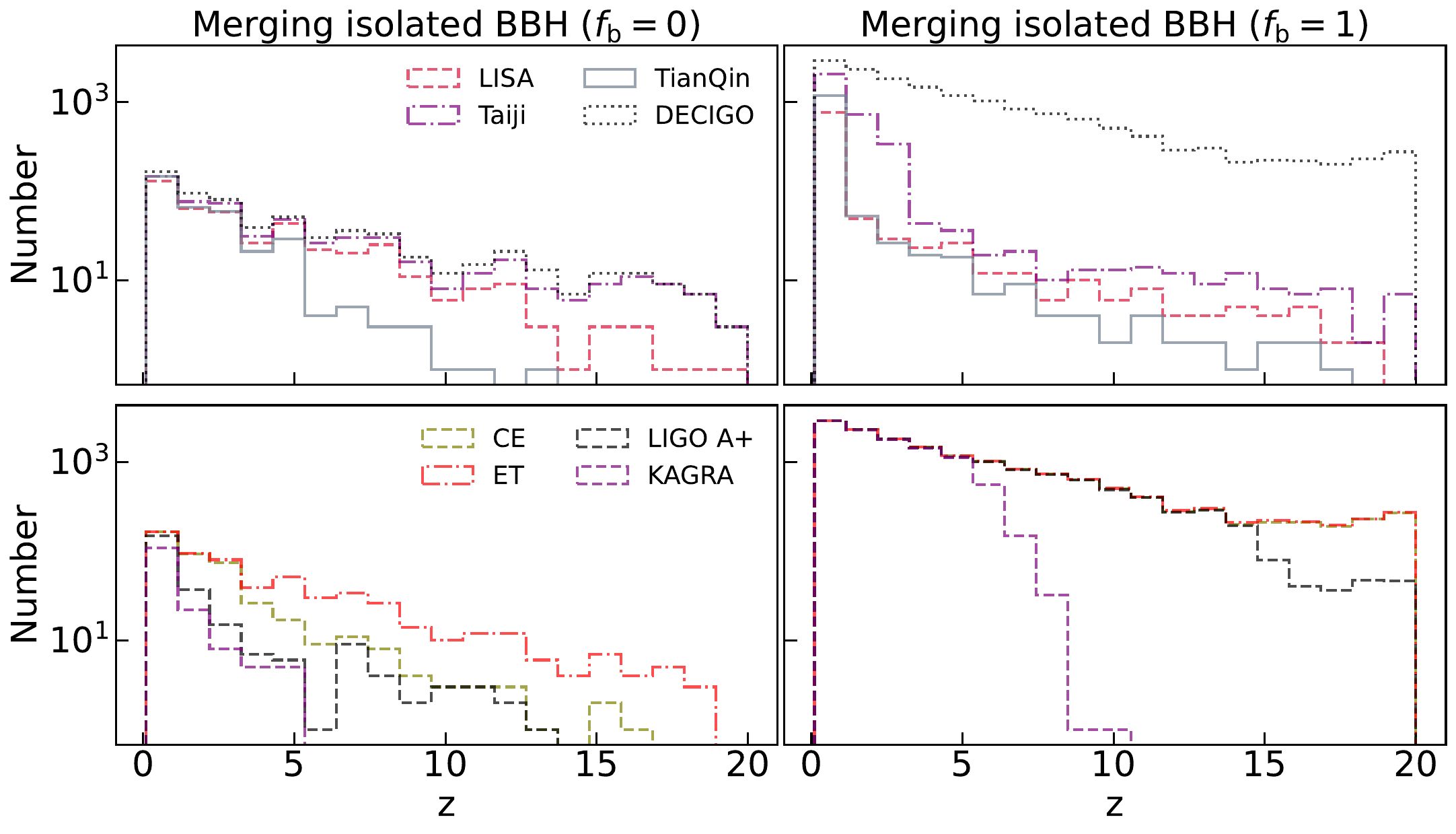}
		\caption{The redshift distributions of detectable merging isolated BBHs for various GW detectors are represented in different colors and line styles. The left and right columns correspond to the cases of $f_{\rm b}=0$ and 1, respectively. The upper panels depict the space-borne detectors LISA/TianQin/Taiji/DECIGO, while the lower panel show the ground-based detectors CE/ET/LIGO/KAGRA. The detectable sources are those whose characteristic GW strains exceed the detector noise strains within a certain frequency range.}
		\label{fig:zz detectable isolated}
	\end{figure*}

	\begin{figure*}
		\centering
		\includegraphics[width=0.8\textwidth]{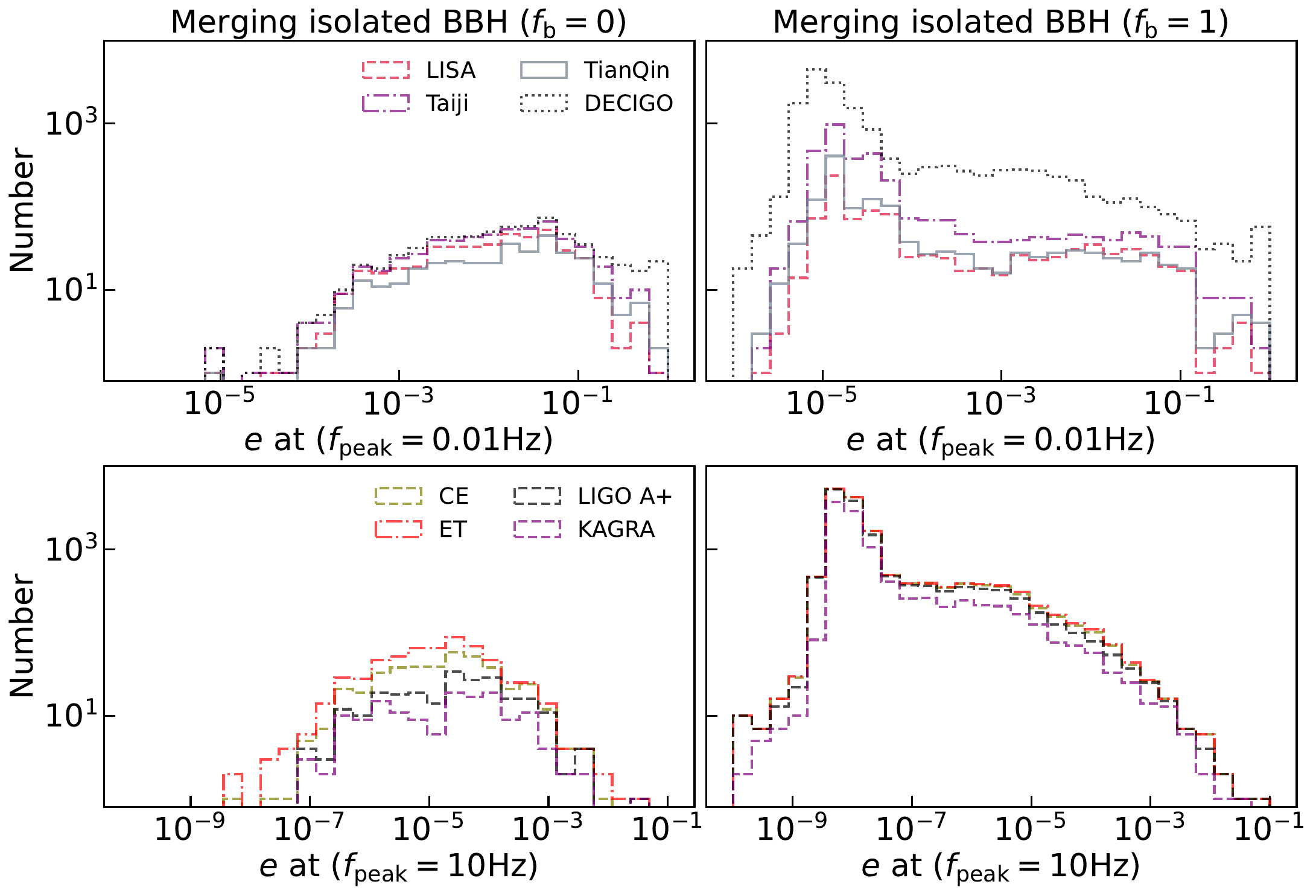}
		\caption{The distribution of orbital eccentricity at $f_{\rm peak}=0.01$Hz and 10Hz of detectable merging isolated BBHs for various GW detectors are represented in different colors and line styles. The left and right columns correspond to the cases of $f_{\rm b}=0$ and 1, respectively. The upper panels depict the space-borne detectors LISA/TianQin/Taiji/DECIGO, while the lower panel show the ground-based detectors CE/ET/LIGO/KAGRA. The detectable sources are those whose characteristic GW strains exceed the detector noise strains within a certain frequency range.}
		\label{fig:ecc at fpeak detectable isolated}
	\end{figure*}

	\begin{figure}
		\centering
		\includegraphics[width=0.4\textwidth]{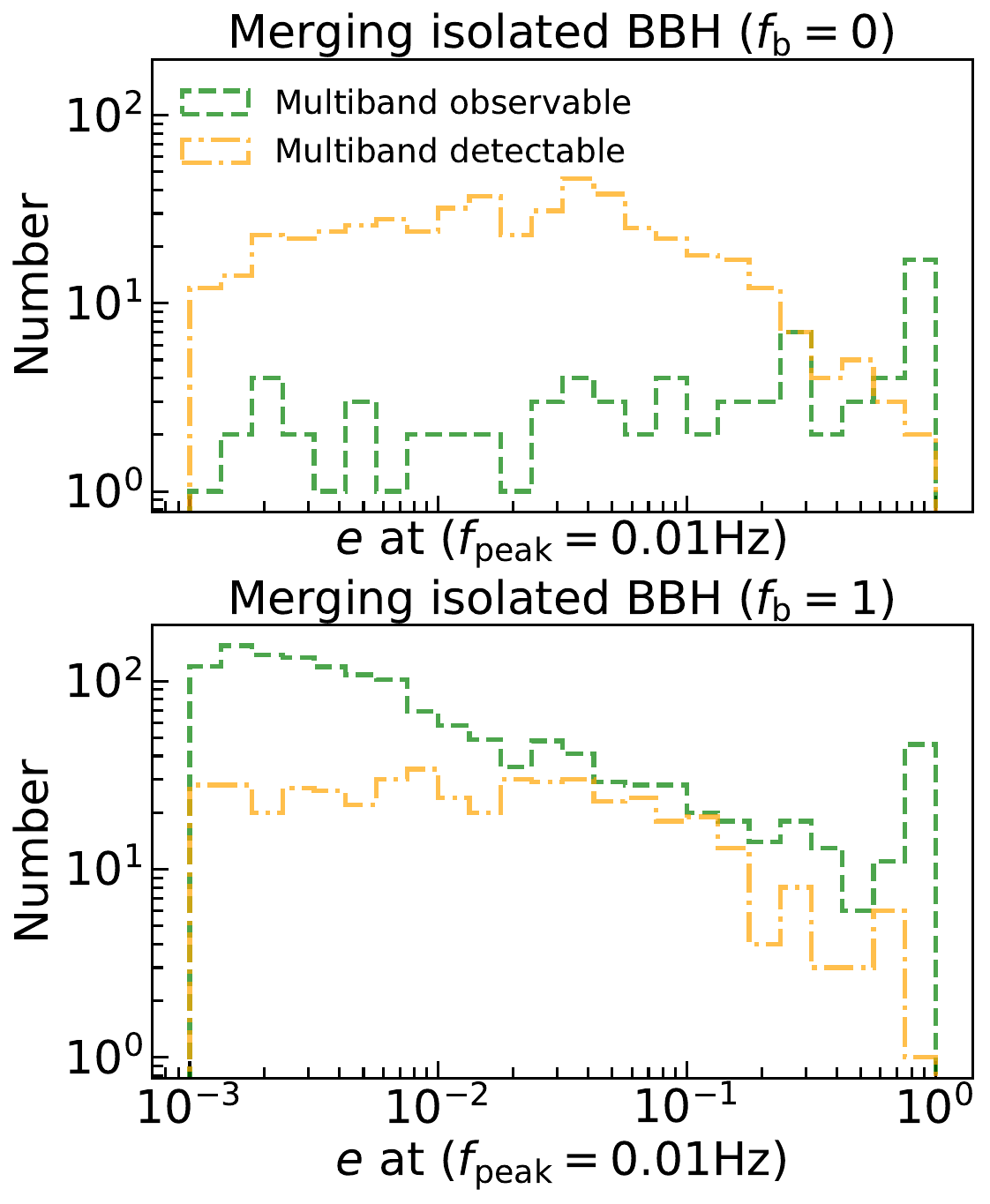}
		\caption{The distributions of eccentricities at $f_{\rm peak}=0.01$Hz of multiband observable and multiband detectable merging isolated BBHs, represented by green and orange dashed lines, respectively. The multiband observable sources denote the merging isolated BBHs are not detectable by mHz space-borne detectors but are detectable by at least one ground-based detectors. The multiband detectable sources are those detectable by both at least one mHz space-borne detectors and at least one ground-based detectors. The upper and lower panels correspond to the cases of $f_{\rm b}=0$ and 1, respectively.}
		\label{fig:multiband isolated}
	\end{figure}

	The evolution of $e_{1}$ with $f_{\rm peak}$ of all the inner BBHs, including inner IMBH-BH, within merging triple BHs, is shown on the upper panel in Fig. \ref{fig:ecc hcn fpeak inner}. In the cases with both $f_{\rm b}=0$ and 1, $e_{1}$ of some sources always decreases, similar to those shown on the upper panel in Fig. 10 in \citep{Wang2022b}\footnote[4]{In Fig. 10, when calculating the trajectory of $a$ and $e$ to depict the evolution of $e$ and GW strains with peak frequency for all merging BBHs using Eqs. (\ref{eq:a_dot e_dot}), it is assumed that the merging BBHs remain unaffected by perturbations from other objects.}. This occurs because these sources decoupled from the third BHs, as seen in the example (a) in Fig. \ref{fig:triple bh merge petar tsunami}, and their orbital evolution is dominated by GW radiation. However, the evolution of $e_{1}$ of the remaining sources displays peaks, oscillations and sharp turning points at lower frequencies before they evolve as isolated binary systems approximately. This is attributed to the oscillations of their orbital elements caused by the dynamical interactions between them and the third BHs before their GW radiation becomes dominant, as observed in the example (b) in Fig. \ref{fig:triple bh merge petar tsunami} and other sources in Fig. \ref{fig:kozai effect}. Most merging inner BBHs have detectable orbital eccentricities at the mHz space-borne detectors LISA/TianQin/Taiji sensitive bands. The distribution of $e_{1}$ at $f_{\rm peak}=0.01$Hz where these mHz detectors are sensitive is plotted on the upper panel in Fig. \ref{fig:ecc at fpeak}. Most of $e_{1}$ are larger than 0.01, and a fraction of sources have $e_{1}$ closing to 1.
	However, when the merging BBHs are at $\sim10$Hz where the ground-based CE/ET/LIGO/KAGRA become sensitive, they can be regarded as quasicircular due to the circularization of GW radiation, with most $e_{1}$ smaller than $10^{-3}$, as shown on the lower panel in Fig. \ref{fig:ecc at fpeak}.

	The characteristic strains of GWs from merging inner BBHs and those of different detector noise, with the color bar denoting their redshifts, are shown on the middle panel in Fig. \ref{fig:ecc hcn fpeak inner}. In both cases with $f_{\rm b}=0$ and 1, the characteristic strains of GWs emitted by the merging inner BBHs decoupled from the third BHs are similar to those on the medium panel in Fig. 10 in \citep{Wang2022b}\footnote[5]{Fig. 10 in \citep{Wang2022b} did not account for the ringdown effect in calculating strains. Consequently, our results provide a more accurate representation of signals in the high frequency range. Additionally, the strain curves for detectors in their Fig. 10 omitted the factor $f_{\rm peak}^{0.5}$, leading to an overestimation of the lower limit of strain in the high-frequency range. We have rectified this in our study.}. For the merging inner BBHs that are affected by the tertiary BHs, the characteristic strains of GWs from them also display peaks, oscillations and sharp turning points at lower frequency ranges, before they are decoupled and emit GWs as approximately isolated merging BBHs. Depending on the characteristic strains of GWs and the noise of detectors, certain sources, including merging inner IMBH-BH, are detectable, i.e., the GW strains are higher than those of detector noise within certain frequency ranges. The redshift distributions of detectable sources are shown in Fig. \ref{fig:zz detectable inner}. In both cases of $f_{\rm b}=0$ and 1, for the space-borne detectors LISA/TianQin/Taiji/DECIGO, the majority of detectable sources are located at $z<3$. Particularly, the number of detectable sources diminishes for Taiji, LISA and TianQin sequentially at high redshifts due to their reduced sensitivities at low frequency ranges. Compared with these detectors, the Deci-Hz space-borne detector DECIGO would have the most detectable sources, as it is sensitive over wider redshift range. For the detectors CE/ET/LIGO/KAGRA, the detectable sources also tend to concentrate at $z<3$. CE/ET would observe more sources at higher redshift than LIGO/KAGRA, because of their enhanced sensitivities at lower frequency ranges. The detectable sources in the case of $f_{\rm b}=1$ are more than those in the case of $f_{\rm b}=0$, as there would be more triple BHs formed when primordial binaries exist, as illustrated in Fig. \ref{fig:number binary triple bh}.

	The characteristic strains of GWs from merging inner BBHs, with the color bar representing their eccentricities $e_{1}$ at $f_{\rm peak}=0.01$Hz, are shown on the lower panel in Fig. \ref{fig:ecc hcn fpeak inner}. In both cases of $f_{\rm peak}=0$ and 1, the eccentricities at $f_{\rm peak}=0.01$Hz of detectable sources for LISA/TianQin/Taiji/DECIGO concentrate between $\mathcal{O}(10^{-3})$ and $\mathcal{O}(10^{-1})$, along with a significant fraction of detectable sources with $e_{1}\sim1$, as shown on the upper panel in Fig. \ref{fig:e1 at fpeak detectable inner}. Among the mHz detectors LISA/TianQin/Taiji, the more sensitive one (e.g., Taiji) would have more detectable sources, including those with higher eccentricities. This is because detectors with better sensitivities would detect GWs with lower strains caused by higher eccentricities. For CE/ET/LIGO/KAGRA, the $e_{1}$ at $f_{\rm peak}=10$Hz of most detectable sources ranges between $\mathcal{O}(10^{-5})$ and $\mathcal{O}(10^{-3})$, as the sources themselves have small eccentricities, as shown on the lower panel in Fig. \ref{fig:ecc at fpeak}. CE/ET would have more detectable sources than LIGO/KAGRA, due to their better sensitivities at low frequency range. We also explore the sources that are not detectable by mHz detectors but detectable by at least one ground-based detectors (multiband observable sources). Their eccentricity distributions at $f_{\rm peak}=0.01$Hz are illustrated in Fig. \ref{fig:multiband inner}. In both cases of $f_{\rm b}=0$ and 1, a significant fraction of multiband observable sources have eccentricities $e_{1}\sim1$. This can be explained by the fact that high eccentricities suppress the strains of GWs so that they are lower than those of the mHz detector noise. The eccentricities of such sources at mHz band could not be identified by LISA/TianQin/Taiji alone, but they would be constrained via multiband observation, e.g., LISA+ET. This could be performed by tracking sources in ground detected catalogs back to data streams from mHz detectors. If the a source is not a significant trigger in LISA/TianQin/Taiji, it could have large eccentricity at mHz band. Furthermore, the eccentricities at $f_{\rm peak}=0.01$Hz of sources detectable by both at least one mHz detector and at least one ground-based detector are also shown in Fig. \ref{fig:multiband inner}. The majority of them have eccentricities ranging from $\mathcal{O}(10^{-3})$ to $\mathcal{O}(10^{-1})$. There is also a fraction of multiband detectable sources with eccentricities closing to 1, because of their low redshifts or large masses, as illustrated in the middle panel of Fig. \ref{fig:ecc hcn fpeak inner}.

	\subsection{GW from merging isolated BBH}
	For comparison, we also plot the evolution of $e$ with $f_{\rm peak}$ of all the merging isolated BBHs in both cases with $f_{\rm b}=0$ and 1, as well as the corresponding characteristic strain of GWs emitted from them in Fig. \ref{fig:ecc hcn fpeak isolated}. Due to the lack of perturbations from the tertiary BHs, there are no peaks, oscillations and sharp turning points observed in the evolution of $e$ and GW strains with $f_{\rm peak}$.

	The distributions of $e$ at 0.01Hz and 10Hz of merging isolated BBHs are also plotted in Fig. \ref{fig:ecc at fpeak}. The distribution of $e$ in the case with $f_{\rm b}=0$ is peaked at values similar to those of merging inner BBHs, along with a larger fraction of smaller eccentricities, because all the merging isolated BBHs are formed by dynamical captures in this case, resulting in larger orbital eccentricities. However, $e$ in the case with $f_{\rm b}=1$ is peaked at significantly smaller values. This is attributed to the prevalence of merging isolated BBHs originating from primordial binaries, which tend to exhibit orbits with lower eccentricities.

	The redshift $z$ distributions of detectable merging isolated BBHs for various detectors are plotted in Fig. \ref{fig:zz detectable isolated}. These distributions are similar to those of merging inner BBHs detectable, because of the selection effect of detectors, i.e., the sources with redshifts that make their GW strains larger than detector noise become detectable.

	The distributions of $e$ of detectable merging isolated BBHs at $f_{\rm peak}=0.01$Hz and 10Hz for various detectors are illustrated in Fig. \ref{fig:ecc at fpeak detectable isolated}. In comparison with the case of merging inner BBHs, these distributions of $e$ show a shift towards smaller values of $e$. Particularly at $f_{\rm peak}=0.01$Hz, there is a notable absence of eccentricities closing to 1. The distributions of $e$ of multiband observable and multiband detectable merging isolated BBHs at $f_{\rm peak}=0.01$Hz and 10Hz across various detectors are also plotted in Fig. \ref{fig:multiband isolated}. These distributions of $e$ also indicate a tendency towards smaller eccentricities, with an increased fraction of low eccentricities,  compared with the case of merging inner BBHs. This trend above can be attributed to the fact that merging isolated BBHs without perturbation from tertiary BHs tend to have smaller eccentricities, as depicted in Fig. \ref{fig:ecc at fpeak}.


	\section{Conclusion}\label{sec:conclusion}

	In this study, we track the long-term evolution of PopIII star clusters embedded in mini dark matter haloes, including models with a primordial binary fraction of $f_{\rm b}=0$ and 1, using the $N$-body code \textsc{petar} which binary mergers via GW radiation are dealt with the orbital average method. To obtain a more accurate evolution of triple BHs, we utilize the \textsc{tsunami} code with PN effect to evolve the merging triples found in the \textsc{petar} simulation.
	Then, we compare the dynamical evolution results of the merging triple BHs between these two methods, and investigate the orbital properties and GW radiation of merging triple BHs in detail.

	In the \textsc{petar} simulation, when $f_{\rm b}=0$, the inner BBHs of all the merging triple BHs are formed by dynamical capture, and the average number of the merging triple BHs in one PopIII cluster is 1. As $f_{\rm b}$ increases to 1, almost all the merging triple BHs have inner BBHs formed by the evolution of primordial binaries, with the average number of the merging triple BHs becoming $2$. In both cases with $f_{\rm b}=0$ and 1, the \textsc{tsunami} simulation yields a comparable number of merging triple BHs. Specially, the number of stable merging triple BHs, which account for the vast majority of all the merging triple BHs, are almost equal in these two methods. For unstable merging triple BHs, however, the \textsc{tsunami} results are significantly less. This discrepancy can be attributed to the difference between these two methods and the instabilities of merging triple BHs.

	The mass distribution of merging triple BHs is affected by the values of $f_{\rm b}$ significantly. In the case with $f_{\rm b}=0$, most primary components $m_{1}$ of inner BBHs are
	IMBHs with $\mathcal{O}(100)M_{\odot}$, and the inner BBHs prefer to be unequal. The tertiary BHs $m_{3}$ concentrate at
	$\mathcal{O}(10)M_{\odot}$, and the outer BBHs also tend to be unequal. When $f_{\rm b}=1$, however, $m_{1}$ becomes lighter, most of
	them are dozens of solar masses, and the inner BBHs tend to be comparable. For the outer BBHs, most $m_{3}$ are still dozens of solar
	masses, but the mass ratio tends to be 1. The mass distribution of merging isolated BBHs follows trends similar to that of the inner BBHs. Unlike the mass distribution of merging triple BHs, the distribution of orbital parameters is not dependent on $f_{\rm b}$ significantly. At
	the early evolutionary time of the merging triple BHs, both inner and outer orbits prefer to be largely eccentric ($\sim0.9$), larger than those of the merging isolated BBHs. Furthermore, the orbital evolution of more than half of inner BBHs is dominated by GW driven at early evolutionary time in the case with $f_{\rm b}=0$, whereas the dominance of the orbital evolution of most inner BBHs is the dynamical interaction between them and the third BHs.

	In both cases of $f_{\rm b}=0$ and 1, the inner BBHs of merging triple BHs could merge from $\sim10$Myr to the present Universe, with about 80\% occurring at the redshift of $z>2$. In the case with $f_{\rm b}=0$, most merger remnants of inner BBHs are IMBHs with hundreds of solar masses. The upper merger rates of inner and isolated BBH could be $0.4{\rm Gpc}^{-3}{\rm yr}^{-1}$ and $15.6{\rm Gpc}^{-3}{\rm yr}^{-1}$, respectively. Specially, the upper merger rates of inner IMBH-BH and inner IMBBH are $0.1{\rm Gpc}^{-3}{\rm yr}^{-1}$ and $0.01{\rm Gpc}^{-3}{\rm yr}^{-1}$ respectively, which are about one-tenth those of the isolated cases. The upper merger rates of inner and isolated IMBBH could make a contribution to that constrained by GW observations. The upper merger rates of inner and isolated PIBH-LBH are $0.02{\rm Gpc}^{-3}{\rm yr}^{-1}$ and $0.1{\rm Gpc}^{-3}{\rm yr}^{-1}$ respectively, contributing to or explaining that of GW190521 inferred by LVKC. Furthermore, inner and isolated LBBH could also make a contribution to SBBHs detected by GWs.

	The inner BBHs of merging triple BHs tend to have significant orbital eccentricities $e_{1}$ within space-borne detector bands in both cases of $f_{\rm b}=0$ and 1. Specially, most of them have $e_{1}$ between $10^{-2}$ and $10^{-1}$, with a significant fraction of them having $e_{1}\sim1$ at $f_{\rm peak}=0.01$Hz where the mHz space-borne detectors LISA/TianQin/Taiji are more sensitive. When the merging inner BBHs reach at $f_{\rm peak}=10$Hz where the ground-based detectors CE/ET/LIGO/KAGRA become sensitive, the residual $e_{1}$ concentrates between $\sim10^{-5}$ and $\sim10^{-3}$. The merging inner BBHs, including merging inner IMBH-BH and PIBH-LBH, detectable for various space-borne and ground-based detectors concentrated at $z<3$. At $f_{\rm peak}=0.01$Hz, the eccentricities of most detectable sources for space-borne detectors are at $e_{1}<0.1$, with a significant fraction of them having $e_{1}\sim1$. Some sources with small eccentricities at $f_{\rm peak}=0.01$Hz could be detectable by both at least one mHz space-borne detectors and at least one ground-based detectors (multiband detectable). While some sources with high eccentricities at $f_{\rm peak}=0.01$Hz would not be detectable by mHz space-borne detectors but detectable by at least one ground-based detectors, whose eccentricities at mHz could be constrained via multiband observation, e.g, LISA+ET.

	The redshift $z$ distributions of merging isolated BBHs detectable by various GW detectors are similar to those of detectable merging inner BBHs, because of the selection effect of detectors. Compared with merging inner BBHs, the distributions of eccentricity $e$ at $f_{\rm peak}=0.01$Hz and 10Hz of merging isolated BBHs exhibittails with lower values. In particular, most merging isolated BBHs would have lower eccentricities when $f_{\rm b}=1$. Consequently, the distributions of $e$ at $f_{\rm peak}=0.01$Hz and 10Hz of detectable merging isolated BBHs, as well as of multiband detectable and multiband observable merging isolated BBHs, shift towards smaller values, with no significant fraction of sources having $e\sim1$. The difference in eccentricities between merging inner BBHs and merging isolated BBHs, along with relevant GW detection results, are expected to help us understand and constrain the formation channels of merging BBHs in PopIII clusters.

	In this work, GW detection results of merging inner BBHs and isolated BBHs are studied qualitatively by comparing characteristic strains of GWs and detector noise. Quantitative and more precise estimations through signal-to-noise ratio calculations will be explored in the future.

	\section*{Acknowledgements}
	We thank the valuable comments and suggestions from referee. We thank Hang Yu for discussions during the preparation of this manuscript. We also thank the helpful discussion from Shu-Tao Yang. This work has been supported by the fellowship of China Postdoctoral Science Foundation (Grant No. 2021TQ0389), the
	Natural Science Foundation of China (Grants No. 12173104, No. 12261131504), and Guangdong Major Project of Basic and Applied Basic Research (Grant No. 2019B030302001).
	L.W. thanks the support from the National Natural Science Foundation of China through grant 21BAA00619, 12073090 and 12233013, and the one-hundred-talent project of Sun Yat-sen University, the Fundamental Research Funds for the Central Universities, Sun Yat-sen University (22hytd09).
	A. T. is supported by JSPS KAKENHI grant Nos. 19K03907. A.A.T. acknowledges support from JSPS KAKENHI Grant Number 21K13914 and from the European Union’s Horizon 2020 research and innovation program under the Marie Sk\l{}odowska-Curie grant agreement no. 847523 \textsc{interactions} and from a Marie Sk\l{}odowska-Curie Individual Fellowship.

	\section*{DATA availability}
	The data in this paper were generated by the software \textsc{petar}, which is available in GitHub
	\href{https://github.com/lwang-astro/PeTar}{https://github.com/lwang-astro/PeTar}. The stellar evolution code \textsc{bseemp} is also available in GitHub
	\href{https://github.com/atrtnkw/bseemp}{https://github.com/atrtnkw/bseemp}. The initial conditions of star cluster models are generated by the software
	\textsc{mcluster}, which is written by Andreas H. W. Kupper et al. and is available in GitHub (modified version) \href{https://github.com/lwang-ast ro/mcluster}{https://github.com/lwang-astro/mcluster}. The
	simulation data will be shared via private communication with reasonable requests.



	\bibliographystyle{mnras}
	\bibliography{ref}

\begin{thebibliography}{}
\makeatletter
\relax
\def\mn@urlcharsother{\let\do\@makeother \do\$\do\&\do\#\do\^\do\_\do\%\do\~}
\def\mn@doi{\begingroup\mn@urlcharsother \@ifnextchar [ {\mn@doi@}
  {\mn@doi@[]}}
\def\mn@doi@[#1]#2{\def\@tempa{#1}\ifx\@tempa\@empty \href
  {http://dx.doi.org/#2} {doi:#2}\else \href {http://dx.doi.org/#2} {#1}\fi
  \endgroup}
\def\mn@eprint#1#2{\mn@eprint@#1:#2::\@nil}
\def\mn@eprint@arXiv#1{\href {http://arxiv.org/abs/#1} {{\tt arXiv:#1}}}
\def\mn@eprint@dblp#1{\href {http://dblp.uni-trier.de/rec/bibtex/#1.xml}
  {dblp:#1}}
\def\mn@eprint@#1:#2:#3:#4\@nil{\def\@tempa {#1}\def\@tempb {#2}\def\@tempc
  {#3}\ifx \@tempc \@empty \let \@tempc \@tempb \let \@tempb \@tempa \fi \ifx
  \@tempb \@empty \def\@tempb {arXiv}\fi \@ifundefined
  {mn@eprint@\@tempb}{\@tempb:\@tempc}{\expandafter \expandafter \csname
  mn@eprint@\@tempb\endcsname \expandafter{\@tempc}}}

\bibitem[\protect\citeauthoryear{Abbott et~al.}{Abbott
  et~al.}{2017}]{LIGOScientific:2016wof}
Abbott B.~P.,  et~al., 2017, \mn@doi [Class. Quant. Grav.]
  {10.1088/1361-6382/aa51f4}, 34, 044001

\bibitem[\protect\citeauthoryear{Abbott et~al.}{Abbott
  et~al.}{2019}]{LIGOScientific:2019ysc}
Abbott B.~P.,  et~al., 2019, \mn@doi [Phys. Rev. D]
  {10.1103/PhysRevD.100.064064}, 100, 064064

\bibitem[\protect\citeauthoryear{Abbott et~al.}{Abbott
  et~al.}{2020a}]{LIGOScientific:2020iuh}
Abbott R.,  et~al., 2020a, \mn@doi [Phys. Rev. Lett.]
  {10.1103/PhysRevLett.125.101102}, 125, 101102

\bibitem[\protect\citeauthoryear{{Abbott} et~al.,}{{Abbott}
  et~al.}{2020b}]{2020ApJ...900L..13A}
{Abbott} R.,  et~al., 2020b, \mn@doi [\apjl] {10.3847/2041-8213/aba493}, \href
  {https://ui.adsabs.harvard.edu/abs/2020ApJ...900L..13A} {900, L13}

\bibitem[\protect\citeauthoryear{Abbott et~al.}{Abbott
  et~al.}{2022}]{LIGOScientific:2021tfm}
Abbott R.,  et~al., 2022, \mn@doi [Astron. Astrophys.]
  {10.1051/0004-6361/202141452}, 659, A84

\bibitem[\protect\citeauthoryear{Abbott et~al.}{Abbott
  et~al.}{2023}]{KAGRA:2021duu}
Abbott R.,  et~al., 2023, \mn@doi [Phys. Rev. X] {10.1103/PhysRevX.13.011048},
  13, 011048

\bibitem[\protect\citeauthoryear{Abbott et~al.}{Abbott
  et~al.}{2024}]{LIGOScientific:2021usb}
Abbott R.,  et~al., 2024, \mn@doi [Phys. Rev. D] {10.1103/PhysRevD.109.022001},
  109, 022001

\bibitem[\protect\citeauthoryear{Ade et~al.}{Ade et~al.}{2016}]{Planck:2015fie}
Ade P. A.~R.,  et~al., 2016, \mn@doi [Astron. Astrophys.]
  {10.1051/0004-6361/201525830}, 594, A13

\bibitem[\protect\citeauthoryear{Amaro-Seoane \& Freitag}{Amaro-Seoane \&
  Freitag}{2006}]{Amaro-Seoane:2006zlf}
Amaro-Seoane P.,  Freitag M.,  2006, \mn@doi [Astrophys. J. Lett.]
  {10.1086/510405}, 653, L53

\bibitem[\protect\citeauthoryear{Amaro-Seoane, Gair, Freitag, Coleman~Miller,
  Mandel, Cutler  \& Babak}{Amaro-Seoane et~al.}{2007}]{Amaro-Seoane:2007osp}
Amaro-Seoane P.,  Gair J.~R.,  Freitag M.,  Coleman~Miller M.,  Mandel I.,
  Cutler C.~J.,   Babak S.,  2007, \mn@doi [Class. Quant. Grav.]
  {10.1088/0264-9381/24/17/R01}, 24, R113

\bibitem[\protect\citeauthoryear{Amaro-Seoane, Miller  \& Freitag}{Amaro-Seoane
  et~al.}{2009}]{Amaro-Seoane:2009iyc}
Amaro-Seoane P.,  Miller M.~C.,   Freitag M.,  2009, \mn@doi [Astrophys. J.
  Lett.] {10.1088/0004-637X/692/1/L50}, 692, L50

\bibitem[\protect\citeauthoryear{{Amaro-Seoane} et~al.,}{{Amaro-Seoane}
  et~al.}{2017}]{2017arXiv170200786A}
{Amaro-Seoane} P.,  et~al., 2017, \mn@doi [arXiv e-prints]
  {10.48550/arXiv.1702.00786}, \href
  {https://ui.adsabs.harvard.edu/abs/2017arXiv170200786A} {p. arXiv:1702.00786}

\bibitem[\protect\citeauthoryear{{Antognini}}{{Antognini}}{2015}]{2015MNRAS.452.3610A}
{Antognini} J.~M.~O.,  2015, \mn@doi [\mnras] {10.1093/mnras/stv1552}, \href
  {https://ui.adsabs.harvard.edu/abs/2015MNRAS.452.3610A} {452, 3610}

\bibitem[\protect\citeauthoryear{Antonini, Murray  \& Mikkola}{Antonini
  et~al.}{2014}]{Antonini:2013tea}
Antonini F.,  Murray N.,   Mikkola S.,  2014, \mn@doi [Astrophys. J.]
  {10.1088/0004-637X/781/1/45}, 781, 45

\bibitem[\protect\citeauthoryear{Antonini, Gieles  \& Gualandris}{Antonini
  et~al.}{2019}]{Antonini:2018auk}
Antonini F.,  Gieles M.,   Gualandris A.,  2019, \mn@doi [Mon. Not. Roy.
  Astron. Soc.] {10.1093/mnras/stz1149}, 486, 5008

\bibitem[\protect\citeauthoryear{{Arca Sedda} \& {Mastrobuono-Battisti}}{{Arca
  Sedda} \& {Mastrobuono-Battisti}}{2019}]{2019arXiv190605864A}
{Arca Sedda} M.,  {Mastrobuono-Battisti} A.,  2019, \mn@doi [arXiv e-prints]
  {10.48550/arXiv.1906.05864}, \href
  {https://ui.adsabs.harvard.edu/abs/2019arXiv190605864A} {p. arXiv:1906.05864}

\bibitem[\protect\citeauthoryear{Arca-Sedda, Amaro-Seoane  \& Chen}{Arca-Sedda
  et~al.}{2021}]{Arca-Sedda:2020lso}
Arca-Sedda M.,  Amaro-Seoane P.,   Chen X.,  2021, \mn@doi [Astron. Astrophys.]
  {10.1051/0004-6361/202037785}, 652, A54

\bibitem[\protect\citeauthoryear{{Ba{\~n}ados} et~al.,}{{Ba{\~n}ados}
  et~al.}{2018}]{Banados2018}
{Ba{\~n}ados} E.,  et~al., 2018, \mn@doi [\nat] {10.1038/nature25180}, \href
  {https://ui.adsabs.harvard.edu/abs/2018Natur.553..473B} {553, 473}

\bibitem[\protect\citeauthoryear{Barai \& de Gouveia Dal~Pino}{Barai \&
  de~Gouveia Dal~Pino}{2019}]{Barai:2018iyd}
Barai P.,  de Gouveia Dal~Pino E.~M.,  2019, \mn@doi [Mon. Not. Roy. Astron.
  Soc.] {10.1093/mnras/stz1616}, 487, 5549

\bibitem[\protect\citeauthoryear{{Baumgardt}}{{Baumgardt}}{2017}]{2017MNRAS.464.2174B}
{Baumgardt} H.,  2017, \mn@doi [\mnras] {10.1093/mnras/stw2488}, \href
  {https://ui.adsabs.harvard.edu/abs/2017MNRAS.464.2174B} {464, 2174}

\bibitem[\protect\citeauthoryear{{Baumgardt} et~al.,}{{Baumgardt}
  et~al.}{2019}]{2019MNRAS.488.5340B}
{Baumgardt} H.,  et~al., 2019, \mn@doi [\mnras] {10.1093/mnras/stz2060}, \href
  {https://ui.adsabs.harvard.edu/abs/2019MNRAS.488.5340B} {488, 5340}

\bibitem[\protect\citeauthoryear{{Belczynski}, {Bulik}  \&
  {Rudak}}{{Belczynski} et~al.}{2004}]{2004ApJ...608L..45B}
{Belczynski} K.,  {Bulik} T.,   {Rudak} B.,  2004, \mn@doi [\apjl]
  {10.1086/422172}, \href
  {https://ui.adsabs.harvard.edu/abs/2004ApJ...608L..45B} {608, L45}

\bibitem[\protect\citeauthoryear{Belczynski et~al.}{Belczynski
  et~al.}{2016}]{Belczynski:2016jno}
Belczynski K.,  et~al., 2016, \mn@doi [Astron. Astrophys.]
  {10.1051/0004-6361/201628980}, 594, A97

\bibitem[\protect\citeauthoryear{{Belczynski}, {Ryu}, {Perna}, {Berti},
  {Tanaka}  \& {Bulik}}{{Belczynski} et~al.}{2017}]{2017MNRAS.471.4702B}
{Belczynski} K.,  {Ryu} T.,  {Perna} R.,  {Berti} E.,  {Tanaka} T.~L.,
  {Bulik} T.,  2017, \mn@doi [\mnras] {10.1093/mnras/stx1759}, \href
  {https://ui.adsabs.harvard.edu/abs/2017MNRAS.471.4702B} {471, 4702}

\bibitem[\protect\citeauthoryear{{Binney} \& {Tremaine}}{{Binney} \&
  {Tremaine}}{1987}]{1987gady.book.....B}
{Binney} J.,  {Tremaine} S.,  1987, {Galactic dynamics}

\bibitem[\protect\citeauthoryear{{Boekholt}, {Portegies Zwart}  \&
  {Valtonen}}{{Boekholt} et~al.}{2020}]{2020MNRAS.493.3932B}
{Boekholt} T.~C.~N.,  {Portegies Zwart} S.~F.,   {Valtonen} M.,  2020, \mn@doi
  [\mnras] {10.1093/mnras/staa452}, \href
  {https://ui.adsabs.harvard.edu/abs/2020MNRAS.493.3932B} {493, 3932}

\bibitem[\protect\citeauthoryear{{Bovy}}{{Bovy}}{2015}]{2015ApJS..216...29B}
{Bovy} J.,  2015, \mn@doi [\apjs] {10.1088/0067-0049/216/2/29}, \href
  {https://ui.adsabs.harvard.edu/abs/2015ApJS..216...29B} {216, 29}

\bibitem[\protect\citeauthoryear{{Chabrier}}{{Chabrier}}{2003}]{2003PASP..115..763C}
{Chabrier} G.,  2003, \mn@doi [\pasp] {10.1086/376392}, \href
  {https://ui.adsabs.harvard.edu/abs/2003PASP..115..763C} {115, 763}

\bibitem[\protect\citeauthoryear{Chon, Omukai  \& Schneider}{Chon
  et~al.}{2021a}]{Chon:2021jlx}
Chon S.,  Omukai K.,   Schneider R.,  2021a, \mn@doi [Mon. Not. Roy. Astron.
  Soc.] {10.1093/mnras/stab2497}, 508, 4175

\bibitem[\protect\citeauthoryear{{Chon}, {Omukai}  \& {Schneider}}{{Chon}
  et~al.}{2021b}]{2021MNRAS.508.4175C}
{Chon} S.,  {Omukai} K.,   {Schneider} R.,  2021b, \mn@doi [\mnras]
  {10.1093/mnras/stab2497}, \href
  {https://ui.adsabs.harvard.edu/abs/2021MNRAS.508.4175C} {508, 4175}

\bibitem[\protect\citeauthoryear{{Costa}, {Mapelli}, {Iorio}, {Santoliquido},
  {Escobar}, { Klessen}  \& {Bressan}}{{Costa}
  et~al.}{2023}]{2023MNRAS.525.2891C}
{Costa} G.,  {Mapelli} M.,  {Iorio} G.,  {Santoliquido} F.,  {Escobar} G.~J.,
  { Klessen} R.~S.,   {Bressan} A.,  2023, \mn@doi [\mnras]
  {10.1093/mnras/stad2443}, \href
  {https://ui.adsabs.harvard.edu/abs/2023MNRAS.525.2891C} {525, 2891}

\bibitem[\protect\citeauthoryear{{D'Orazio} \& {Samsing}}{{D'Orazio} \&
  {Samsing}}{2018}]{2018MNRAS.481.4775D}
{D'Orazio} D.~J.,  {Samsing} J.,  2018, \mn@doi [\mnras]
  {10.1093/mnras/sty2568}, \href
  {https://ui.adsabs.harvard.edu/abs/2018MNRAS.481.4775D} {481, 4775}

\bibitem[\protect\citeauthoryear{Datta, Gupta, Kastha, Arun  \&
  Sathyaprakash}{Datta et~al.}{2021}]{Datta:2020vcj}
Datta S.,  Gupta A.,  Kastha S.,  Arun K.~G.,   Sathyaprakash B.~S.,  2021,
  \mn@doi [Phys. Rev. D] {10.1103/PhysRevD.103.024036}, 103, 024036

\bibitem[\protect\citeauthoryear{Deme, Hoang, Naoz  \& Kocsis}{Deme
  et~al.}{2020}]{Deme:2020ewx}
Deme B.,  Hoang B.-M.,  Naoz S.,   Kocsis B.,  2020, \mn@doi [Astrophys. J.]
  {10.3847/1538-4357/abafa3}, 901, 125

\bibitem[\protect\citeauthoryear{{Duch{\^e}ne} \& {Kraus}}{{Duch{\^e}ne} \&
  {Kraus}}{2013}]{2013ARA&A..51..269D}
{Duch{\^e}ne} G.,  {Kraus} A.,  2013, \mn@doi [\araa]
  {10.1146/annurev-astro-081710-102602}, \href
  {https://ui.adsabs.harvard.edu/abs/2013ARA&A..51..269D} {51, 269}

\bibitem[\protect\citeauthoryear{Emami \& Loeb}{Emami \&
  Loeb}{2020}]{Emami:2019bss}
Emami R.,  Loeb A.,  2020, \mn@doi [Mon. Not. Roy. Astron. Soc.]
  {10.1093/mnras/staa1200}, 495, 536

\bibitem[\protect\citeauthoryear{Farrell et~al.}{Farrell
  et~al.}{2012}]{Farrell:2011rj}
Farrell S.,  et~al., 2012, \mn@doi [Astrophys. J. Lett.]
  {10.1088/2041-8205/747/1/L13}, 747, L13

\bibitem[\protect\citeauthoryear{{Fishbach} \& {Holz}}{{Fishbach} \&
  {Holz}}{2020}]{2020ApJ...904L..26F}
{Fishbach} M.,  {Holz} D.~E.,  2020, \mn@doi [\apjl]
  {10.3847/2041-8213/abc827}, \href
  {https://ui.adsabs.harvard.edu/abs/2020ApJ...904L..26F} {904, L26}

\bibitem[\protect\citeauthoryear{{Fragione} \& {Kocsis}}{{Fragione} \&
  {Kocsis}}{2019}]{2019MNRAS.486.4781F}
{Fragione} G.,  {Kocsis} B.,  2019, \mn@doi [\mnras] {10.1093/mnras/stz1175},
  \href {https://ui.adsabs.harvard.edu/abs/2019MNRAS.486.4781F} {486, 4781}

\bibitem[\protect\citeauthoryear{Fragione \& Loeb}{Fragione \&
  Loeb}{2023}]{Fragione:2022ams}
Fragione G.,  Loeb A.,  2023, \mn@doi [Astrophys. J.]
  {10.3847/1538-4357/acb34e}, 944, 81

\bibitem[\protect\citeauthoryear{Fragione \& Silk}{Fragione \&
  Silk}{2020}]{Fragione:2020nib}
Fragione G.,  Silk J.,  2020, \mn@doi [Mon. Not. Roy. Astron. Soc.]
  {10.1093/mnras/staa2629}, 498, 4591

\bibitem[\protect\citeauthoryear{Fragione, Loeb, Kremer  \& Rasio}{Fragione
  et~al.}{2020}]{Fragione:2020rmf}
Fragione G.,  Loeb A.,  Kremer K.,   Rasio F.~A.,  2020, \mn@doi [Astrophys.
  J.] {10.3847/1538-4357/ab94b2}, 897, 46

\bibitem[\protect\citeauthoryear{Fragione, Kocsis, Rasio  \& Silk}{Fragione
  et~al.}{2022a}]{Fragione:2021nhb}
Fragione G.,  Kocsis B.,  Rasio F.~A.,   Silk J.,  2022a, \mn@doi [Astrophys.
  J.] {10.3847/1538-4357/ac5026}, 927, 231

\bibitem[\protect\citeauthoryear{Fragione, Loeb, Kocsis  \& Rasio}{Fragione
  et~al.}{2022b}]{Fragione:2022avp}
Fragione G.,  Loeb A.,  Kocsis B.,   Rasio F.~A.,  2022b, \mn@doi [Astrophys.
  J.] {10.3847/1538-4357/ac75d0}, 933, 170

\bibitem[\protect\citeauthoryear{Fregeau, Larson, Miller, O'Shaughnessy  \&
  Rasio}{Fregeau et~al.}{2006}]{Fregeau:2006yz}
Fregeau J.~M.,  Larson S.~L.,  Miller M.~C.,  O'Shaughnessy R.~W.,   Rasio
  F.~A.,  2006, \mn@doi [Astrophys. J. Lett.] {10.1086/507106}, 646, L135

\bibitem[\protect\citeauthoryear{Freitag, Gurkan  \& Rasio}{Freitag
  et~al.}{2006}]{Freitag:2005yd}
Freitag M.,  Gurkan M.~A.,   Rasio F.~A.,  2006, \mn@doi [Mon. Not. Roy.
  Astron. Soc.] {10.1111/j.1365-2966.2006.10096.x}, 368, 141

\bibitem[\protect\citeauthoryear{Fryer, Belczynski, Wiktorowicz, Dominik,
  Kalogera  \& Holz}{Fryer et~al.}{2012}]{Fryer:2011cx}
Fryer C.~L.,  Belczynski K.,  Wiktorowicz G.,  Dominik M.,  Kalogera V.,   Holz
  D.~E.,  2012, \mn@doi [Astrophys. J.] {10.1088/0004-637X/749/1/91}, 749, 91

\bibitem[\protect\citeauthoryear{{Gaia Collaboration} et~al.,}{{Gaia
  Collaboration} et~al.}{2023}]{2023A&A...674A..34G}
{Gaia Collaboration} et~al., 2023, \mn@doi [\aap]
  {10.1051/0004-6361/202243782}, \href
  {https://ui.adsabs.harvard.edu/abs/2023A&A...674A..34G} {674, A34}

\bibitem[\protect\citeauthoryear{Gair, Mandel, Miller  \& Volonteri}{Gair
  et~al.}{2011}]{Gair:2010dx}
Gair J.~R.,  Mandel I.,  Miller M.~C.,   Volonteri M.,  2011, \mn@doi [Gen.
  Rel. Grav.] {10.1007/s10714-010-1104-3}, 43, 485

\bibitem[\protect\citeauthoryear{{Garg}, {Derdzinski}, {Zwick}, { Capelo}  \&
  {Mayer}}{{Garg} et~al.}{2022}]{2022MNRAS.517.1339G}
{Garg} M.,  {Derdzinski} A.,  {Zwick} L.,  { Capelo} P.~R.,   {Mayer} L.,
  2022, \mn@doi [\mnras] {10.1093/mnras/stac2711}, \href
  {https://ui.adsabs.harvard.edu/abs/2022MNRAS.517.1339G} {517, 1339}

\bibitem[\protect\citeauthoryear{Gebhardt, Rich  \& Ho}{Gebhardt
  et~al.}{2002}]{Gebhardt:2002in}
Gebhardt K.,  Rich R.~M.,   Ho L.~C.,  2002, \mn@doi [Astrophys. J. Lett.]
  {10.1086/342980}, 578, L41

\bibitem[\protect\citeauthoryear{Gebhardt, Rich  \& Ho}{Gebhardt
  et~al.}{2005}]{Gebhardt:2005cy}
Gebhardt K.,  Rich R.~M.,   Ho L.~C.,  2005, \mn@doi [Astrophys. J.]
  {10.1086/497023}, 634, 1093

\bibitem[\protect\citeauthoryear{Gerssen, van~der Marel, Gebhardt,
  Guhathakurta, Peterson  \& Pryor}{Gerssen et~al.}{2002}]{Gerssen:2002iq}
Gerssen J.,  van~der Marel R.~P.,  Gebhardt K.,  Guhathakurta P.,  Peterson
  R.~C.,   Pryor C.,  2002, \mn@doi [Astron. J.] {10.1086/344584}, 124, 3270

\bibitem[\protect\citeauthoryear{Gerssen, van~der Marel, Gebhardt,
  Guhathakurta, Peterson  \& Pryor}{Gerssen et~al.}{2003}]{Gerssen:2002sd}
Gerssen J.,  van~der Marel R.~P.,  Gebhardt K.,  Guhathakurta P.,  Peterson
  R.~C.,   Pryor C.,  2003, \mn@doi [Astron. J.] {10.1086/345574}, 125, 376

\bibitem[\protect\citeauthoryear{{Giersz}, {Leigh}, {Hypki}, {L{ \"u}tzgendorf}
   \& {Askar}}{{Giersz} et~al.}{2015}]{2015MNRAS.454.3150G}
{Giersz} M.,  {Leigh} N.,  {Hypki} A.,  {L{ \"u}tzgendorf} N.,   {Askar} A.,
  2015, \mn@doi [\mnras] {10.1093/mnras/stv2162}, \href
  {https://ui.adsabs.harvard.edu/abs/2015MNRAS.454.3150G} {454, 3150}

\bibitem[\protect\citeauthoryear{Gonz\'alez, Kremer, Chatterjee, Fragione,
  Rodriguez, Weatherford, Ye  \& Rasio}{Gonz\'alez
  et~al.}{2021}]{Gonzalez:2020xah}
Gonz\'alez E.,  Kremer K.,  Chatterjee S.,  Fragione G.,  Rodriguez C.~L.,
  Weatherford N.~C.,  Ye C.~S.,   Rasio F.~A.,  2021, \mn@doi [Astrophys. J.
  Lett.] {10.3847/2041-8213/abdf5b}, 908, L29

\bibitem[\protect\citeauthoryear{Greene}{Greene}{2012}]{Greene:2012gk}
Greene J.~E.,  2012, \mn@doi [Nature Commun.] {10.1038/ncomms2314}, 3, 1304

\bibitem[\protect\citeauthoryear{{Greene}, {Strader}  \& {Ho}}{{Greene}
  et~al.}{2020}]{2020ARA&A..58..257G}
{Greene} J.~E.,  {Strader} J.,   {Ho} L.~C.,  2020, \mn@doi [\araa]
  {10.1146/annurev-astro-032620-021835}, \href
  {https://ui.adsabs.harvard.edu/abs/2020ARA&A..58..257G} {58, 257}

\bibitem[\protect\citeauthoryear{Gultekin, Miller  \& Hamilton}{Gultekin
  et~al.}{2004}]{Gultekin:2004pm}
Gultekin K.,  Miller M.~C.,   Hamilton D.~P.,  2004, \mn@doi [Astrophys. J.]
  {10.1086/424809}, 616, 221

\bibitem[\protect\citeauthoryear{{Hamers}}{{Hamers}}{2021}]{2021RNAAS...5..275H}
{Hamers} A.~S.,  2021, \mn@doi [Research Notes of the American Astronomical
  Society] {10.3847/2515-5172/ac3d98}, \href
  {https://ui.adsabs.harvard.edu/abs/2021RNAAS...5..275H} {5, 275}

\bibitem[\protect\citeauthoryear{{Hamers} \& {Safarzadeh}}{{Hamers} \&
  {Safarzadeh}}{2020}]{2020ApJ...898...99H}
{Hamers} A.~S.,  {Safarzadeh} M.,  2020, \mn@doi [\apj]
  {10.3847/1538-4357/ab9b27}, \href
  {https://ui.adsabs.harvard.edu/abs/2020ApJ...898...99H} {898, 99}

\bibitem[\protect\citeauthoryear{{Hartwig}, {Volonteri}, {Bromm}, { Klessen},
  {Barausse}, {Magg}  \& { Stacy}}{{Hartwig}
  et~al.}{2016}]{2016MNRAS.460L..74H}
{Hartwig} T.,  {Volonteri} M.,  {Bromm} V.,  { Klessen} R.~S.,  {Barausse} E.,
  {Magg} M.,   { Stacy} A.,  2016, \mn@doi [\mnras] {10.1093/mnrasl/slw074},
  \href {https://ui.adsabs.harvard.edu/abs/2016MNRAS.460L..74H} {460, L74}

\bibitem[\protect\citeauthoryear{{Hayashi}, {Trani}  \& {Suto}}{{Hayashi}
  et~al.}{2022}]{2022ApJ...939...81H}
{Hayashi} T.,  {Trani} A.~A.,   {Suto} Y.,  2022, \mn@doi [\apj]
  {10.3847/1538-4357/ac8f48}, \href
  {https://ui.adsabs.harvard.edu/abs/2022ApJ...939...81H} {939, 81}

\bibitem[\protect\citeauthoryear{Hild et~al.}{Hild et~al.}{2011}]{Hild:2010id}
Hild S.,  et~al., 2011, \mn@doi [Class. Quant. Grav.]
  {10.1088/0264-9381/28/9/094013}, 28, 094013

\bibitem[\protect\citeauthoryear{Hobbs, Lorimer, Lyne  \& Kramer}{Hobbs
  et~al.}{2005}]{Hobbs:2005yx}
Hobbs G.,  Lorimer D.~R.,  Lyne A.~G.,   Kramer M.,  2005, \mn@doi [Mon. Not.
  Roy. Astron. Soc.] {10.1111/j.1365-2966.2005.09087.x}, 360, 974

\bibitem[\protect\citeauthoryear{Hurley}{Hurley}{2007}]{Hurley:2007as}
Hurley J.~R.,  2007, \mn@doi [Mon. Not. Roy. Astron. Soc.]
  {10.1111/j.1365-2966.2007.11912.x}, 379, 93

\bibitem[\protect\citeauthoryear{{Hurley}, {Pols}  \& {Tout}}{{Hurley}
  et~al.}{2000}]{Hurley2000}
{Hurley} J.~R.,  {Pols} O.~R.,   {Tout} C.~A.,  2000, \mn@doi [\mnras]
  {10.1046/j.1365-8711.2000.03426.x}, \href
  {https://ui.adsabs.harvard.edu/abs/2000MNRAS.315..543H} {315, 543}

\bibitem[\protect\citeauthoryear{{Hurley}, {Tout}  \& {Pols}}{{Hurley}
  et~al.}{2002}]{Hurley2002}
{Hurley} J.~R.,  {Tout} C.~A.,   {Pols} O.~R.,  2002, \mn@doi [\mnras]
  {10.1046/j.1365-8711.2002.05038.x}, \href
  {https://ui.adsabs.harvard.edu/abs/2002MNRAS.329..897H} {329, 897}

\bibitem[\protect\citeauthoryear{Husa, Khan, Hannam, P\"urrer, Ohme,
  Jim\'enez~Forteza  \& Boh\'e}{Husa et~al.}{2016}]{Husa:2015iqa}
Husa S.,  Khan S.,  Hannam M.,  P\"urrer M.,  Ohme F.,  Jim\'enez~Forteza X.,
  Boh\'e A.,  2016, \mn@doi [Phys. Rev. D] {10.1103/PhysRevD.93.044006}, 93,
  044006

\bibitem[\protect\citeauthoryear{{Inayoshi}, {Hirai}, {Kinugawa}  \&
  {Hotokezaka}}{{Inayoshi} et~al.}{2017}]{2017MNRAS.468.5020I}
{Inayoshi} K.,  {Hirai} R.,  {Kinugawa} T.,   {Hotokezaka} K.,  2017, \mn@doi
  [\mnras] {10.1093/mnras/stx757}, \href
  {https://ui.adsabs.harvard.edu/abs/2017MNRAS.468.5020I} {468, 5020}

\bibitem[\protect\citeauthoryear{Inayoshi, Visbal  \& Haiman}{Inayoshi
  et~al.}{2020}]{Inayoshi:2019fun}
Inayoshi K.,  Visbal E.,   Haiman Z.,  2020, \mn@doi [Ann. Rev. Astron.
  Astrophys.] {10.1146/annurev-astro-120419-014455}, 58, 27

\bibitem[\protect\citeauthoryear{{Inayoshi}, {Kashiyama}, {Visbal}  \& {
  Haiman}}{{Inayoshi} et~al.}{2021}]{2021ApJ...919...41I}
{Inayoshi} K.,  {Kashiyama} K.,  {Visbal} E.,   { Haiman} Z.,  2021, \mn@doi
  [\apj] {10.3847/1538-4357/ac106d}, \href
  {https://ui.adsabs.harvard.edu/abs/2021ApJ...919...41I} {919, 41}

\bibitem[\protect\citeauthoryear{Iwasawa, Tanikawa, Hosono, Nitadori, Muranushi
   \& Makino}{Iwasawa et~al.}{2016}]{Iwasawa:2016zci}
Iwasawa M.,  Tanikawa A.,  Hosono N.,  Nitadori K.,  Muranushi T.,   Makino J.,
   2016, \mn@doi [Publ. Astron. Soc. Jap.] {10.1093/pasj/psw053}, 68, 54

\bibitem[\protect\citeauthoryear{{Iwasawa}, {Oshino}, {Fujii}  \&
  {Hori}}{{Iwasawa} et~al.}{2017}]{2017PASJ...69...81I}
{Iwasawa} M.,  {Oshino} S.,  {Fujii} M.~S.,   {Hori} Y.,  2017, \mn@doi [\pasj]
  {10.1093/pasj/psx073}, \href
  {https://ui.adsabs.harvard.edu/abs/2017PASJ...69...81I} {69, 81}

\bibitem[\protect\citeauthoryear{{Iwasawa}, {Namekata}, {Nitadori}, {Nomura},
  {Wang}, {Tsubouchi}  \& {Makino}}{{Iwasawa}
  et~al.}{2020}]{2020PASJ...72...13I}
{Iwasawa} M.,  {Namekata} D.,  {Nitadori} K.,  {Nomura} K.,  {Wang} L.,
  {Tsubouchi} M.,   {Makino} J.,  2020, \mn@doi [\pasj] {10.1093/pasj/psz133},
  \href {https://ui.adsabs.harvard.edu/abs/2020PASJ...72...13I} {72, 13}

\bibitem[\protect\citeauthoryear{Jani, Shoemaker  \& Cutler}{Jani
  et~al.}{2019}]{Jani:2019ffg}
Jani K.,  Shoemaker D.,   Cutler C.,  2019, \mn@doi [Nature Astron.]
  {10.1038/s41550-019-0932-7}, 4, 260

\bibitem[\protect\citeauthoryear{Kawamura et~al.}{Kawamura
  et~al.}{2011}]{Kawamura:2011zz}
Kawamura S.,  et~al., 2011, \mn@doi [Class. Quant. Grav.]
  {10.1088/0264-9381/28/9/094011}, 28, 094011

\bibitem[\protect\citeauthoryear{Khan, Husa, Hannam, Ohme, P\"urrer,
  Jim\'enez~Forteza  \& Boh\'e}{Khan et~al.}{2016}]{Khan:2015jqa}
Khan S.,  Husa S.,  Hannam M.,  Ohme F.,  P\"urrer M.,  Jim\'enez~Forteza X.,
  Boh\'e A.,  2016, \mn@doi [Phys. Rev. D] {10.1103/PhysRevD.93.044007}, 93,
  044007

\bibitem[\protect\citeauthoryear{{King}}{{King}}{1966}]{1966AJ.....71...64K}
{King} I.~R.,  1966, \mn@doi [\aj] {10.1086/109857}, \href
  {https://ui.adsabs.harvard.edu/abs/1966AJ.....71...64K} {71, 64}

\bibitem[\protect\citeauthoryear{{Kinugawa}, {Inayoshi}, {Hotokezaka},
  {Nakauchi}  \& {Nakamura}}{{Kinugawa} et~al.}{2014}]{2014MNRAS.442.2963K}
{Kinugawa} T.,  {Inayoshi} K.,  {Hotokezaka} K.,  {Nakauchi} D.,   {Nakamura}
  T.,  2014, \mn@doi [\mnras] {10.1093/mnras/stu1022}, \href
  {https://ui.adsabs.harvard.edu/abs/2014MNRAS.442.2963K} {442, 2963}

\bibitem[\protect\citeauthoryear{{Kinugawa}, {Nakamura}  \&
  {Nakano}}{{Kinugawa} et~al.}{2020}]{2020MNRAS.498.3946K}
{Kinugawa} T.,  {Nakamura} T.,   {Nakano} H.,  2020, \mn@doi [\mnras]
  {10.1093/mnras/staa2511}, \href
  {https://ui.adsabs.harvard.edu/abs/2020MNRAS.498.3946K} {498, 3946}

\bibitem[\protect\citeauthoryear{{Kinugawa}, {Nakamura}  \&
  {Nakano}}{{Kinugawa} et~al.}{2021a}]{2021MNRAS.501L..49K}
{Kinugawa} T.,  {Nakamura} T.,   {Nakano} H.,  2021a, \mn@doi [\mnras]
  {10.1093/mnrasl/slaa191}, \href
  {https://ui.adsabs.harvard.edu/abs/2021MNRAS.501L..49K} {501, L49}

\bibitem[\protect\citeauthoryear{{Kinugawa}, {Nakamura}  \&
  {Nakano}}{{Kinugawa} et~al.}{2021b}]{2021MNRAS.504L..28K}
{Kinugawa} T.,  {Nakamura} T.,   {Nakano} H.,  2021b, \mn@doi [\mnras]
  {10.1093/mnrasl/slab032}, \href
  {https://ui.adsabs.harvard.edu/abs/2021MNRAS.504L..28K} {504, L28}

\bibitem[\protect\citeauthoryear{{Kinugawa}, {Nakamura}  \&
  {Nakano}}{{Kinugawa} et~al.}{2021c}]{2021PTEP.2021b1E01K}
{Kinugawa} T.,  {Nakamura} T.,   {Nakano} H.,  2021c, \mn@doi [Progress of
  Theoretical and Experimental Physics] {10.1093/ptep/ptaa176}, \href
  {https://ui.adsabs.harvard.edu/abs/2021PTEP.2021b1E01K} {2021, 021E01}

\bibitem[\protect\citeauthoryear{Kozai}{Kozai}{1962}]{Kozai:1962zz}
Kozai Y.,  1962, \mn@doi [Astron. J.] {10.1086/108790}, 67, 591

\bibitem[\protect\citeauthoryear{{Kremer} et~al.,}{{Kremer}
  et~al.}{2019}]{2019PhRvD..99f3003K}
{Kremer} K.,  et~al., 2019, \mn@doi [\prd] {10.1103/PhysRevD.99.063003}, \href
  {https://ui.adsabs.harvard.edu/abs/2019PhRvD..99f3003K} {99, 063003}

\bibitem[\protect\citeauthoryear{Kremer et~al.,}{Kremer
  et~al.}{2020}]{Kremer:2020wtp}
Kremer K.,  et~al., 2020, \mn@doi [Astrophys. J.] {10.3847/1538-4357/abb945},
  903, 45

\bibitem[\protect\citeauthoryear{{Kroupa}}{{Kroupa}}{2001}]{2001MNRAS.322..231K}
{Kroupa} P.,  2001, \mn@doi [\mnras] {10.1046/j.1365-8711.2001.04022.x}, \href
  {https://ui.adsabs.harvard.edu/abs/2001MNRAS.322..231K} {322, 231}

\bibitem[\protect\citeauthoryear{Kroupa, Subr, Jerabkova  \& Wang}{Kroupa
  et~al.}{2020}]{Kroupa:2020sru}
Kroupa P.,  Subr L.,  Jerabkova T.,   Wang L.,  2020, \mn@doi [Mon. Not. Roy.
  Astron. Soc.] {10.1093/mnras/staa2276}, 498, 5652

\bibitem[\protect\citeauthoryear{{Lalande} \& {Trani}}{{Lalande} \&
  {Trani}}{2022}]{2022ApJ...938...18L}
{Lalande} F.,  {Trani} A.~A.,  2022, \mn@doi [\apj] {10.3847/1538-4357/ac8eab},
  \href {https://ui.adsabs.harvard.edu/abs/2022ApJ...938...18L} {938, 18}

\bibitem[\protect\citeauthoryear{Latif, Whalen  \& Khochfar}{Latif
  et~al.}{2022}]{Latif:2021xad}
Latif M.~A.,  Whalen D.,   Khochfar S.,  2022, \mn@doi [Astrophys. J.]
  {10.3847/1538-4357/ac3916}, 925, 28

\bibitem[\protect\citeauthoryear{{Lidov}}{{Lidov}}{1962}]{1962P&SS....9..719L}
{Lidov} M.~L.,  1962, \mn@doi [\planss] {10.1016/0032-0633(62)90129-0}, \href
  {https://ui.adsabs.harvard.edu/abs/1962P&SS....9..719L} {9, 719}

\bibitem[\protect\citeauthoryear{{Liu} \& {Bromm}}{{Liu} \&
  {Bromm}}{2020a}]{2020MNRAS.495.2475L}
{Liu} B.,  {Bromm} V.,  2020a, \mn@doi [\mnras] {10.1093/mnras/staa1362}, \href
  {https://ui.adsabs.harvard.edu/abs/2020MNRAS.495.2475L} {495, 2475}

\bibitem[\protect\citeauthoryear{{Liu} \& {Bromm}}{{Liu} \&
  {Bromm}}{2020b}]{2020ApJ...903L..40L}
{Liu} B.,  {Bromm} V.,  2020b, \mn@doi [\apjl] {10.3847/2041-8213/abc552},
  \href {https://ui.adsabs.harvard.edu/abs/2020ApJ...903L..40L} {903, L40}

\bibitem[\protect\citeauthoryear{{Liu}, {Meynet}  \& {Bromm}}{{Liu}
  et~al.}{2021}]{2021MNRAS.501..643L}
{Liu} B.,  {Meynet} G.,   {Bromm} V.,  2021, \mn@doi [\mnras]
  {10.1093/mnras/staa3671}, \href
  {https://ui.adsabs.harvard.edu/abs/2021MNRAS.501..643L} {501, 643}

\bibitem[\protect\citeauthoryear{Liu, Zhu, Hu, Zhang  \& Ji}{Liu
  et~al.}{2022}]{Liu:2021yoy}
Liu S.,  Zhu L.-G.,  Hu Y.-M.,  Zhang J.-d.,   Ji M.-J.,  2022, \mn@doi [Phys.
  Rev. D] {10.1103/PhysRevD.105.023019}, 105, 023019

\bibitem[\protect\citeauthoryear{Luo et~al.}{Luo
  et~al.}{2016}]{TianQin:2015yph}
Luo J.,  et~al., 2016, \mn@doi [Class. Quant. Grav.]
  {10.1088/0264-9381/33/3/035010}, 33, 035010

\bibitem[\protect\citeauthoryear{Mapelli, Huwyler, Mayer, Jetzer  \&
  Vecchio}{Mapelli et~al.}{2010}]{Mapelli:2010ht}
Mapelli M.,  Huwyler C.,  Mayer L.,  Jetzer P.,   Vecchio A.,  2010, \mn@doi
  [Astrophys. J.] {10.1088/0004-637X/719/2/987}, 719, 987

\bibitem[\protect\citeauthoryear{Mapelli et~al.}{Mapelli
  et~al.}{2021}]{Mapelli:2021syv}
Mapelli M.,  et~al., 2021, \mn@doi [Mon. Not. Roy. Astron. Soc.]
  {10.1093/mnras/stab1334}, 505, 339

\bibitem[\protect\citeauthoryear{Mapelli, Bouffanais, Santoliquido, Sedda  \&
  Artale}{Mapelli et~al.}{2022}]{Mapelli:2021gyv}
Mapelli M.,  Bouffanais Y.,  Santoliquido F.,  Sedda M.~A.,   Artale M.~C.,
  2022, \mn@doi [Mon. Not. Roy. Astron. Soc.] {10.1093/mnras/stac422}, 511,
  5797

\bibitem[\protect\citeauthoryear{{Mardling} \& {Aarseth}}{{Mardling} \&
  {Aarseth}}{2001}]{2001MNRAS.321..398M}
{Mardling} R.~A.,  {Aarseth} S.~J.,  2001, \mn@doi [\mnras]
  {10.1046/j.1365-8711.2001.03974.x}, \href
  {https://ui.adsabs.harvard.edu/abs/2001MNRAS.321..398M} {321, 398}

\bibitem[\protect\citeauthoryear{Mayer, Kazantzidis, Escala  \&
  Callegari}{Mayer et~al.}{2010}]{Mayer:2009qr}
Mayer L.,  Kazantzidis S.,  Escala A.,   Callegari S.,  2010, \mn@doi [Nature]
  {10.1038/nature09294}, 466, 1082

\bibitem[\protect\citeauthoryear{Mayer, Fiacconi, Bonoli, Quinn, Roskar, Shen
  \& Wadsley}{Mayer et~al.}{2015}]{Mayer:2014nva}
Mayer L.,  Fiacconi D.,  Bonoli S.,  Quinn T.,  Roskar R.,  Shen S.,   Wadsley
  J.,  2015, \mn@doi [Astrophys. J.] {10.1088/0004-637X/810/1/51}, 810, 51

\bibitem[\protect\citeauthoryear{McKernan, Ford, Lyra  \& Perets}{McKernan
  et~al.}{2012}]{McKernan:2012rf}
McKernan B.,  Ford K. E.~S.,  Lyra W.,   Perets H.~B.,  2012, \mn@doi [Mon.
  Not. Roy. Astron. Soc.] {10.1111/j.1365-2966.2012.21486.x}, 425, 460

\bibitem[\protect\citeauthoryear{McKernan, Ford, Kocsis, Lyra  \&
  Winter}{McKernan et~al.}{2014}]{McKernan:2014oxa}
McKernan B.,  Ford K. E.~S.,  Kocsis B.,  Lyra W.,   Winter L.~M.,  2014,
  \mn@doi [Mon. Not. Roy. Astron. Soc.] {10.1093/mnras/stu553}, 441, 900

\bibitem[\protect\citeauthoryear{{Mezcua}}{{Mezcua}}{2017}]{2017IJMPD..2630021M}
{Mezcua} M.,  2017, \mn@doi [International Journal of Modern Physics D]
  {10.1142/S021827181730021X}, \href
  {https://ui.adsabs.harvard.edu/abs/2017IJMPD..2630021M} {26, 1730021}

\bibitem[\protect\citeauthoryear{Mezcua, Roberts, Lobanov  \& Sutton}{Mezcua
  et~al.}{2015}]{Mezcua:2015pra}
Mezcua M.,  Roberts T.~P.,  Lobanov A.~P.,   Sutton A.~D.,  2015, \mn@doi [Mon.
  Not. Roy. Astron. Soc.] {10.1093/mnras/stv143}, 448, 1893

\bibitem[\protect\citeauthoryear{{Michie}}{{Michie}}{1963}]{1963MNRAS.125..127M}
{Michie} R.~W.,  1963, \mn@doi [\mnras] {10.1093/mnras/125.2.127}, \href
  {https://ui.adsabs.harvard.edu/abs/1963MNRAS.125..127M} {125, 127}

\bibitem[\protect\citeauthoryear{Michimura et~al.}{Michimura
  et~al.}{2020}]{Michimura:2020xnj}
Michimura Y.,  et~al., 2020, \mn@doi [Phys. Rev. D]
  {10.1103/PhysRevD.102.022008}, 102, 022008

\bibitem[\protect\citeauthoryear{{Mikkola} \& {Tanikawa}}{{Mikkola} \&
  {Tanikawa}}{1999a}]{1999CeMDA..74..287M}
{Mikkola} S.,  {Tanikawa} K.,  1999a, \mn@doi [Celestial Mechanics and
  Dynamical Astronomy] {10.1023/A:1008368322547}, \href
  {https://ui.adsabs.harvard.edu/abs/1999CeMDA..74..287M} {74, 287}

\bibitem[\protect\citeauthoryear{{Mikkola} \& {Tanikawa}}{{Mikkola} \&
  {Tanikawa}}{1999b}]{1999MNRAS.310..745M}
{Mikkola} S.,  {Tanikawa} K.,  1999b, \mn@doi [\mnras]
  {10.1046/j.1365-8711.1999.02982.x}, \href
  {https://ui.adsabs.harvard.edu/abs/1999MNRAS.310..745M} {310, 745}

\bibitem[\protect\citeauthoryear{Miller \& Hamilton}{Miller \&
  Hamilton}{2002}]{Miller:2001ez}
Miller M.~C.,  Hamilton D.~P.,  2002, \mn@doi [Mon. Not. Roy. Astron. Soc.]
  {10.1046/j.1365-8711.2002.05112.x}, 330, 232

\bibitem[\protect\citeauthoryear{Mouri \& Taniguchi}{Mouri \&
  Taniguchi}{2002}]{Mouri:2002mc}
Mouri H.,  Taniguchi Y.,  2002, \mn@doi [Astrophys. J. Lett.] {10.1086/339472},
  566, L17

\bibitem[\protect\citeauthoryear{{Naoz}}{{Naoz}}{2016}]{2016ARA&A..54..441N}
{Naoz} S.,  2016, \mn@doi [\araa] {10.1146/annurev-astro-081915-023315}, \href
  {https://ui.adsabs.harvard.edu/abs/2016ARA&A..54..441N} {54, 441}

\bibitem[\protect\citeauthoryear{{Naoz}, {Kocsis}, {Loeb}  \& {Yunes}}{{Naoz}
  et~al.}{2013}]{2013ApJ...773..187N}
{Naoz} S.,  {Kocsis} B.,  {Loeb} A.,   {Yunes} N.,  2013, \mn@doi [\apj]
  {10.1088/0004-637X/773/2/187}, \href
  {https://ui.adsabs.harvard.edu/abs/2013ApJ...773..187N} {773, 187}

\bibitem[\protect\citeauthoryear{{Navarro}, {Frenk}  \& {White}}{{Navarro}
  et~al.}{1996}]{1996ApJ...462..563N}
{Navarro} J.~F.,  {Frenk} C.~S.,   {White} S. D.~M.,  1996, \mn@doi [\apj]
  {10.1086/177173}, \href
  {https://ui.adsabs.harvard.edu/abs/1996ApJ...462..563N} {462, 563}

\bibitem[\protect\citeauthoryear{Nitz \& Capano}{Nitz \&
  Capano}{2021}]{Nitz:2020mga}
Nitz A.~H.,  Capano C.~D.,  2021, \mn@doi [Astrophys. J. Lett.]
  {10.3847/2041-8213/abccc5}, 907, L9

\bibitem[\protect\citeauthoryear{Noyola, Gebhardt  \& Bergmann}{Noyola
  et~al.}{2008}]{Noyola:2008kt}
Noyola E.,  Gebhardt K.,   Bergmann M.,  2008, \mn@doi [Astrophys. J.]
  {10.1086/529002}, 676, 1008

\bibitem[\protect\citeauthoryear{Oshino, Funato  \& Makino}{Oshino
  et~al.}{2011}]{Oshino:2011ja}
Oshino S.,  Funato Y.,   Makino J.,  2011, \mn@doi [Publ. Astron. Soc. Jap.]
  {10.1093/pasj/63.4.881}, 63, 881

\bibitem[\protect\citeauthoryear{{Peters}}{{Peters}}{1964}]{1964PhRv..136.1224P}
{Peters} P.~C.,  1964, \mn@doi [Physical Review] {10.1103/PhysRev.136.B1224},
  \href {https://ui.adsabs.harvard.edu/abs/1964PhRv..136.1224P} {136, 1224}

\bibitem[\protect\citeauthoryear{{Peters} \& {Mathews}}{{Peters} \&
  {Mathews}}{1963}]{1963PhRv..131..435P}
{Peters} P.~C.,  {Mathews} J.,  1963, \mn@doi [Physical Review]
  {10.1103/PhysRev.131.435}, \href
  {https://ui.adsabs.harvard.edu/abs/1963PhRv..131..435P} {131, 435}

\bibitem[\protect\citeauthoryear{Portegies~Zwart \& McMillan}{Portegies~Zwart
  \& McMillan}{2002}]{PortegiesZwart:2002iks}
Portegies~Zwart S.~F.,  McMillan S. L.~W.,  2002, \mn@doi [Astrophys. J.]
  {10.1086/341798}, 576, 899

\bibitem[\protect\citeauthoryear{Portegies~Zwart, Baumgardt, Hut, Makino  \&
  McMillan}{Portegies~Zwart et~al.}{2004}]{PortegiesZwart:2004ggg}
Portegies~Zwart S.~F.,  Baumgardt H.,  Hut P.,  Makino J.,   McMillan S. L.~W.,
   2004, \mn@doi [Nature] {10.1038/nature02448}, 428, 724

\bibitem[\protect\citeauthoryear{Portegies~Zwart, Baumgardt, McMillan, Makino,
  Hut  \& Ebisuzaki}{Portegies~Zwart et~al.}{2006}]{PortegiesZwart:2005zp}
Portegies~Zwart S.~F.,  Baumgardt H.,  McMillan S. L.~W.,  Makino J.,  Hut P.,
   Ebisuzaki T.,  2006, \mn@doi [Astrophys. J.] {10.1086/500361}, 641, 319

\bibitem[\protect\citeauthoryear{{Portegies Zwart}, {Boekholt}, {Por}, {Hamers}
   \& {McMillan}}{{Portegies Zwart} et~al.}{2022}]{2022A&A...659A..86P}
{Portegies Zwart} S.~F.,  {Boekholt} T.~C.~N.,  {Por} E.~H.,  {Hamers} A.~S.,
  {McMillan} S.~L.~W.,  2022, \mn@doi [\aap] {10.1051/0004-6361/202141789},
  \href {https://ui.adsabs.harvard.edu/abs/2022A&A...659A..86P} {659, A86}

\bibitem[\protect\citeauthoryear{{Portegies Zwart}, {Boekholt}  \& {
  Heggie}}{{Portegies Zwart} et~al.}{2023}]{2023MNRAS.tmp.2576P}
{Portegies Zwart} S.~F.,  {Boekholt} T. C.~N.,   { Heggie} D.~C.,  2023,
  \mn@doi [\mnras] {10.1093/mnras/stad2654}, \href
  {https://ui.adsabs.harvard.edu/abs/2023MNRAS.tmp.2576P} {}

\bibitem[\protect\citeauthoryear{Punturo et~al.}{Punturo
  et~al.}{2010}]{Punturo:2010zz}
Punturo M.,  et~al., 2010, \mn@doi [Class. Quant. Grav.]
  {10.1088/0264-9381/27/19/194002}, 27, 194002

\bibitem[\protect\citeauthoryear{Rasskazov, Fragione  \& Kocsis}{Rasskazov
  et~al.}{2020}]{Rasskazov:2019tgb}
Rasskazov A.,  Fragione G.,   Kocsis B.,  2020, \mn@doi [Astrophys. J.]
  {10.3847/1538-4357/aba2f4}, 899, 149

\bibitem[\protect\citeauthoryear{Reines \& Comastri}{Reines \&
  Comastri}{2016}]{Reines:2016kej}
Reines A.,  Comastri A.,  2016, \mn@doi [Publ. Astron. Soc. Austral.]
  {10.1017/pasa.2016.46}, 33, e054

\bibitem[\protect\citeauthoryear{Reinoso, Schleicher, Fellhauer, Klessen  \&
  Boekholt}{Reinoso et~al.}{2018}]{Reinoso:2018bfv}
Reinoso B.,  Schleicher D. R.~G.,  Fellhauer M.,  Klessen R.~S.,   Boekholt T.
  C.~N.,  2018, \mn@doi [Astron. Astrophys.] {10.1051/0004-6361/201732224},
  614, A14

\bibitem[\protect\citeauthoryear{Reitze et~al.}{Reitze
  et~al.}{2019}]{Reitze:2019iox}
Reitze D.,  et~al., 2019, Bull. Am. Astron. Soc., 51, 035

\bibitem[\protect\citeauthoryear{{Rizzuto} et~al.,}{{Rizzuto}
  et~al.}{2021}]{2021MNRAS.501.5257R}
{Rizzuto} F.~P.,  et~al., 2021, \mn@doi [\mnras] {10.1093/mnras/staa3634},
  \href {https://ui.adsabs.harvard.edu/abs/2021MNRAS.501.5257R} {501, 5257}

\bibitem[\protect\citeauthoryear{Robson, Cornish  \& Liu}{Robson
  et~al.}{2019}]{Robson:2018ifk}
Robson T.,  Cornish N.~J.,   Liu C.,  2019, \mn@doi [Class. Quant. Grav.]
  {10.1088/1361-6382/ab1101}, 36, 105011

\bibitem[\protect\citeauthoryear{{Rodriguez} \& {Antonini}}{{Rodriguez} \&
  {Antonini}}{2018}]{2018ApJ...863....7R}
{Rodriguez} C.~L.,  {Antonini} F.,  2018, \mn@doi [\apj]
  {10.3847/1538-4357/aacea4}, \href
  {https://ui.adsabs.harvard.edu/abs/2018ApJ...863....7R} {863, 7}

\bibitem[\protect\citeauthoryear{Rose, Naoz, Sari  \& Linial}{Rose
  et~al.}{2022}]{Rose:2021ftz}
Rose S.~C.,  Naoz S.,  Sari R.,   Linial I.,  2022, \mn@doi [Astrophys. J.
  Lett.] {10.3847/2041-8213/ac6426}, 929, L22

\bibitem[\protect\citeauthoryear{Ruan, Guo, Cai  \& Zhang}{Ruan
  et~al.}{2020}]{Ruan:2018tsw}
Ruan W.-H.,  Guo Z.-K.,  Cai R.-G.,   Zhang Y.-Z.,  2020, \mn@doi [Int. J. Mod.
  Phys. A] {10.1142/S0217751X2050075X}, 35, 2050075

\bibitem[\protect\citeauthoryear{Sakurai, Yoshida, Fujii  \& Hirano}{Sakurai
  et~al.}{2017a}]{Sakurai:2017opi}
Sakurai Y.,  Yoshida N.,  Fujii M.~S.,   Hirano S.,  2017a, \mn@doi [Mon. Not.
  Roy. Astron. Soc.] {10.1093/mnras/stx2044}, 472, 1677

\bibitem[\protect\citeauthoryear{{Sakurai}, {Yoshida}, {Fujii}  \& {
  Hirano}}{{Sakurai} et~al.}{2017b}]{2017MNRAS.472.1677S}
{Sakurai} Y.,  {Yoshida} N.,  {Fujii} M.~S.,   { Hirano} S.,  2017b, \mn@doi
  [\mnras] {10.1093/mnras/stx2044}, \href
  {https://ui.adsabs.harvard.edu/abs/2017MNRAS.472.1677S} {472, 1677}

\bibitem[\protect\citeauthoryear{{Sana} et~al.,}{{Sana}
  et~al.}{2012}]{2012Sci...337..444S}
{Sana} H.,  et~al., 2012, \mn@doi [Science] {10.1126/science.1223344}, \href
  {https://ui.adsabs.harvard.edu/abs/2012Sci...337..444S} {337, 444}

\bibitem[\protect\citeauthoryear{{Santoliquido}, {Mapelli}, {Iorio}, {Costa},
  {Glover}, { Hartwig}, {Klessen}  \& {Merli}}{{Santoliquido}
  et~al.}{2023}]{2023MNRAS.524..307S}
{Santoliquido} F.,  {Mapelli} M.,  {Iorio} G.,  {Costa} G.,  {Glover} S. C.~O.,
   { Hartwig} T.,  {Klessen} R.~S.,   {Merli} L.,  2023, \mn@doi [\mnras]
  {10.1093/mnras/stad1860}, \href
  {https://ui.adsabs.harvard.edu/abs/2023MNRAS.524..307S} {524, 307}

\bibitem[\protect\citeauthoryear{{Sesana}, {Haardt}, {Madau}  \&
  {Volonteri}}{{Sesana} et~al.}{2005}]{2005CQGra..22S.363S}
{Sesana} A.,  {Haardt} F.,  {Madau} P.,   {Volonteri} M.,  2005, \mn@doi
  [Classical and Quantum Gravity] {10.1088/0264-9381/22/10/030}, \href
  {https://ui.adsabs.harvard.edu/abs/2005CQGra..22S.363S} {22, S363}

\bibitem[\protect\citeauthoryear{Silk}{Silk}{2017}]{Silk:2017yai}
Silk J.,  2017, \mn@doi [Astrophys. J. Lett.] {10.3847/2041-8213/aa67da}, 839,
  L13

\bibitem[\protect\citeauthoryear{{Silsbee} \& {Tremaine}}{{Silsbee} \&
  {Tremaine}}{2017}]{2017ApJ...836...39S}
{Silsbee} K.,  {Tremaine} S.,  2017, \mn@doi [\apj] {10.3847/1538-4357/aa5729},
  \href {https://ui.adsabs.harvard.edu/abs/2017ApJ...836...39S} {836, 39}

\bibitem[\protect\citeauthoryear{{Skinner} \& {Wise}}{{Skinner} \&
  {Wise}}{2020}]{2020MNRAS.492.4386S}
{Skinner} D.,  {Wise} J.~H.,  2020, \mn@doi [\mnras] {10.1093/mnras/staa139},
  \href {https://ui.adsabs.harvard.edu/abs/2020MNRAS.492.4386S} {492, 4386}

\bibitem[\protect\citeauthoryear{Spera \& Mapelli}{Spera \&
  Mapelli}{2017}]{Spera:2017fyx}
Spera M.,  Mapelli M.,  2017, \mn@doi [Mon. Not. Roy. Astron. Soc.]
  {10.1093/mnras/stx1576}, 470, 4739

\bibitem[\protect\citeauthoryear{{Stacy}, {Bromm}  \& {Lee}}{{Stacy}
  et~al.}{2016}]{2016MNRAS.462.1307S}
{Stacy} A.,  {Bromm} V.,   {Lee} A.~T.,  2016, \mn@doi [\mnras]
  {10.1093/mnras/stw1728}, \href
  {https://ui.adsabs.harvard.edu/abs/2016MNRAS.462.1307S} {462, 1307}

\bibitem[\protect\citeauthoryear{{Suto}}{{Suto}}{1991}]{1991PASJ...43L...9S}
{Suto} Y.,  1991, \pasj, \href
  {https://ui.adsabs.harvard.edu/abs/1991PASJ...43L...9S} {43, L9}

\bibitem[\protect\citeauthoryear{Taniguchi, Shioya, Tsuru  \&
  Ikeuchi}{Taniguchi et~al.}{2000}]{Taniguchi:2000mp}
Taniguchi Y.,  Shioya Y.,  Tsuru T.~G.,   Ikeuchi S.,  2000, \mn@doi [Publ.
  Astron. Soc. Jap.] {10.1093/pasj/52.3.533}, 52, 533

\bibitem[\protect\citeauthoryear{{Tanikawa}, {Yoshida}, {Kinugawa}, {Takahashi}
   \& {Umeda}}{{Tanikawa} et~al.}{2020}]{2020MNRAS.495.4170T}
{Tanikawa} A.,  {Yoshida} T.,  {Kinugawa} T.,  {Takahashi} K.,   {Umeda} H.,
  2020, \mn@doi [\mnras] {10.1093/mnras/staa1417}, \href
  {https://ui.adsabs.harvard.edu/abs/2020MNRAS.495.4170T} {495, 4170}

\bibitem[\protect\citeauthoryear{{Tanikawa}, {Kinugawa}, {Yoshida}, {Hijikawa}
  \& {Umeda}}{{Tanikawa} et~al.}{2021a}]{2021MNRAS.505.2170T}
{Tanikawa} A.,  {Kinugawa} T.,  {Yoshida} T.,  {Hijikawa} K.,   {Umeda} H.,
  2021a, \mn@doi [\mnras] {10.1093/mnras/stab1421}, \href
  {https://ui.adsabs.harvard.edu/abs/2021MNRAS.505.2170T} {505, 2170}

\bibitem[\protect\citeauthoryear{{Tanikawa}, {Susa}, {Yoshida}, { Trani}  \&
  {Kinugawa}}{{Tanikawa} et~al.}{2021b}]{2021ApJ...910...30T}
{Tanikawa} A.,  {Susa} H.,  {Yoshida} T.,  { Trani} A.~A.,   {Kinugawa} T.,
  2021b, \mn@doi [\apj] {10.3847/1538-4357/abe40d}, \href
  {https://ui.adsabs.harvard.edu/abs/2021ApJ...910...30T} {910, 30}

\bibitem[\protect\citeauthoryear{{Tanikawa}, {Yoshida}, {Kinugawa}, {Trani},
  {Hosokawa}, {Susa}  \& {Omukai}}{{Tanikawa}
  et~al.}{2022}]{2022ApJ...926...83T}
{Tanikawa} A.,  {Yoshida} T.,  {Kinugawa} T.,  {Trani} A.~A.,  {Hosokawa} T.,
  {Susa} H.,   {Omukai} K.,  2022, \mn@doi [\apj] {10.3847/1538-4357/ac4247},
  \href {https://ui.adsabs.harvard.edu/abs/2022ApJ...926...83T} {926, 83}

\bibitem[\protect\citeauthoryear{Torres-Orjuela, Huang, Liang, Liu, Wang, Ye,
  Hu  \& Mei}{Torres-Orjuela et~al.}{2024}]{Torres-Orjuela:2023hfd}
Torres-Orjuela A.,  Huang S.-J.,  Liang Z.-C.,  Liu S.,  Wang H.-T.,  Ye C.-Q.,
   Hu Y.-M.,   Mei J.,  2024, \mn@doi [Sci. China Phys. Mech. Astron.]
  {10.1007/s11433-023-2308-x}, 67, 259511

\bibitem[\protect\citeauthoryear{Trani \& Spera}{Trani \&
  Spera}{2020}]{Trani:2020hwc}
Trani A.~A.,  Spera M.,  2020, \mn@doi [IAU Symp.] {10.1017/S1743921322001818},
  362, 404

\bibitem[\protect\citeauthoryear{{Trani}, {Fujii}  \& {Spera}}{{Trani}
  et~al.}{2019}]{2019ApJ...875...42T}
{Trani} A.~A.,  {Fujii} M.~S.,   {Spera} M.,  2019, \mn@doi [\apj]
  {10.3847/1538-4357/ab0e70}, \href
  {https://ui.adsabs.harvard.edu/abs/2019ApJ...875...42T} {875, 42}

\bibitem[\protect\citeauthoryear{Trani, Rastello, Di~Carlo, Santoliquido,
  Tanikawa  \& Mapelli}{Trani et~al.}{2022}]{Trani:2021tan}
Trani A.~A.,  Rastello S.,  Di~Carlo U.~N.,  Santoliquido F.,  Tanikawa A.,
  Mapelli M.,  2022, \mn@doi [Mon. Not. Roy. Astron. Soc.]
  {10.1093/mnras/stac122}, 511, 1362

\bibitem[\protect\citeauthoryear{Trani, Leigh, Boekholt  \& Zwart}{Trani
  et~al.}{2024}]{Trani:2024xwq}
Trani A.~A.,  Leigh N. W.~C.,  Boekholt T. C.~N.,   Zwart S.~P.,  2024, \mn@doi
  [Astron. Astrophys.] {10.1051/0004-6361/202449862}, 689, A24

\bibitem[\protect\citeauthoryear{Tremou et~al.}{Tremou
  et~al.}{2018}]{Tremou:2018rvq}
Tremou E.,  et~al., 2018, \mn@doi [Astrophys. J.] {10.3847/1538-4357/aac9b9},
  862, 16

\bibitem[\protect\citeauthoryear{Volonteri}{Volonteri}{2010}]{Volonteri:2010wz}
Volonteri M.,  2010, \mn@doi [Astron. Astrophys. Rev.]
  {10.1007/s00159-010-0029-x}, 18, 279

\bibitem[\protect\citeauthoryear{Wang et~al.}{Wang et~al.}{2019}]{Wang:2019ryf}
Wang H.-T.,  et~al., 2019, \mn@doi [Phys. Rev. D]
  {10.1103/PhysRevD.100.043003}, 100, 043003

\bibitem[\protect\citeauthoryear{{Wang}, {Nitadori}  \& {Makino}}{{Wang}
  et~al.}{2020a}]{2020MNRAS.493.3398W}
{Wang} L.,  {Nitadori} K.,   {Makino} J.,  2020a, \mn@doi [\mnras]
  {10.1093/mnras/staa480}, \href
  {https://ui.adsabs.harvard.edu/abs/2020MNRAS.493.3398W} {493, 3398}

\bibitem[\protect\citeauthoryear{{Wang}, {Iwasawa}, {Nitadori}  \& {
  Makino}}{{Wang} et~al.}{2020b}]{2020MNRAS.497..536W}
{Wang} L.,  {Iwasawa} M.,  {Nitadori} K.,   { Makino} J.,  2020b, \mn@doi
  [\mnras] {10.1093/mnras/staa1915}, \href
  {https://ui.adsabs.harvard.edu/abs/2020MNRAS.497..536W} {497, 536}

\bibitem[\protect\citeauthoryear{Wang, Tanikawa  \& Fujii}{Wang
  et~al.}{2021a}]{Wang:2021fxs}
Wang L.,  Tanikawa A.,   Fujii M.~S.,  2021a, \mn@doi [Mon. Not. Roy. Astron.
  Soc.] {10.1093/mnras/stab3255}, 509, 4713

\bibitem[\protect\citeauthoryear{{Wang} et~al.,}{{Wang}
  et~al.}{2021b}]{2021ApJ...907L...1W}
{Wang} F.,  et~al., 2021b, \mn@doi [\apjl] {10.3847/2041-8213/abd8c6}, \href
  {https://ui.adsabs.harvard.edu/abs/2021ApJ...907L...1W} {907, L1}

\bibitem[\protect\citeauthoryear{{Wang}, {Tanikawa}  \& {Fujii}}{{Wang}
  et~al.}{2022}]{Wang2022b}
{Wang} L.,  {Tanikawa} A.,   {Fujii} M.,  2022, \mn@doi [\mnras]
  {10.1093/mnras/stac2043}, \href
  {https://ui.adsabs.harvard.edu/abs/2022MNRAS.515.5106W} {515, 5106}

\bibitem[\protect\citeauthoryear{{Willems}, {Kalogera}, {Vecchio}, {Ivanova},
  {Rasio}, {Fregeau}  \& {Belczynski}}{{Willems}
  et~al.}{2007}]{2007ApJ...665L..59W}
{Willems} B.,  {Kalogera} V.,  {Vecchio} A.,  {Ivanova} N.,  {Rasio} F.~A.,
  {Fregeau} J.~M.,   {Belczynski} K.,  2007, \mn@doi [\apjl] {10.1086/521049},
  \href {https://ui.adsabs.harvard.edu/abs/2007ApJ...665L..59W} {665, L59}

\bibitem[\protect\citeauthoryear{{Winter-Granic}, {Petrovich}, {Pe{\~n}
  a-Donaire}  \& {Hamilton}}{{Winter-Granic}
  et~al.}{2023}]{2023arXiv231217319W}
{Winter-Granic} M.,  {Petrovich} C.,  {Pe{\~n} a-Donaire} V.,   {Hamilton} C.,
  2023, \mn@doi [arXiv e-prints] {10.48550/arXiv.2312.17319}, \href
  {https://ui.adsabs.harvard.edu/abs/2023arXiv231217319W} {p. arXiv:2312.17319}

\bibitem[\protect\citeauthoryear{{Wu} et~al.,}{{Wu} et~al.}{2015}]{Wu2015}
{Wu} X.-B.,  et~al., 2015, \mn@doi [\nat] {10.1038/nature14241}, \href
  {https://ui.adsabs.harvard.edu/abs/2015Natur.518..512W} {518, 512}

\bibitem[\protect\citeauthoryear{Yagi}{Yagi}{2012}]{Yagi:2012gb}
Yagi K.,  2012, \mn@doi [Class. Quant. Grav.] {10.1088/0264-9381/29/7/075005},
  29, 075005

\bibitem[\protect\citeauthoryear{{Zocchi}, {Gieles}  \&
  {H{\'e}nault-Brunet}}{{Zocchi} et~al.}{2017}]{2017MNRAS.468.4429Z}
{Zocchi} A.,  {Gieles} M.,   {H{\'e}nault-Brunet} V.,  2017, \mn@doi [\mnras]
  {10.1093/mnras/stx316}, \href
  {https://ui.adsabs.harvard.edu/abs/2017MNRAS.468.4429Z} {468, 4429}

\bibitem[\protect\citeauthoryear{van~der Marel, Gerssen, Guhathakurta, Peterson
   \& Gebhardt}{van~der Marel et~al.}{2002}]{vanderMarel:2002ip}
van~der Marel R.~P.,  Gerssen J.,  Guhathakurta P.,  Peterson R.,   Gebhardt
  K.,  2002, \mn@doi [Astron. J.] {10.1086/344583}, 124, 3255

\makeatother
\end{thebibliography}




	\bsp	
	\label{lastpage}
\end{CJK}
\end{document}